\documentclass[10pt,oneside]{article}
\usepackage{amsmath,amssymb,color,tikz,MnSymbol,stmaryrd}
\usepackage[english]{babel}
\usepackage[mathcal]{euscript}
\usetikzlibrary{matrix,arrows}
\DeclareMathAlphabet{\mathpzc}{OT1}{pzc}{m}{it}
\usepackage{dsfont,url}
\usepackage{bbm}
\usepackage{pstricks,mathtools,pst-plot,pst-node}
\usepackage[latin1]{inputenc}

\usepackage{mathrsfs}
\usepackage[T1]{fontenc}
\usepackage{andsupp}
\usepackage{amscd}

\newtheorem{theorem}{Theorem}[section]

\newtheorem{proposition}[theorem]{Proposition}
\newtheorem{lemma}[theorem]{Lemma}
\newtheorem{corollary}[theorem]{Corollary}
\newtheorem{example}{Example}[section]
\newtheorem{definition}[example]{Definition}
\newtheorem{problem}[theorem]{Problem}
\newtheorem{remark}[example]{Remark}

\def\beaq{\begin{eqnarray}}
\def\eeaq{\end{eqnarray}}

\newcommand{\eq}[1]{Eqn.~(\ref{#1})}

\newcommand{\beq}{\begin{equation}}
\newcommand{\eeq}{\end{equation}}
\newcommand{\bea}{\begin{eqnarray}}
\newcommand{\eea}{\end{eqnarray}}
\newcommand{\dd}{\mathrm{d}}

\newcommand{\Res}{\mathop{\,\rm Res\,}}
\newcommand{\bs}[1]{\ensuremath{\boldsymbol{#1}}}

\definecolor{rouge}{rgb}{0.84,0.18,0.07}
\definecolor{bleu}{rgb}{0.22,0.41,0.74}
\definecolor{vertf}{rgb}{0.08,0.46,0.07}
\usepackage{hyperref}
\hypersetup{colorlinks,urlcolor=vertf,citecolor=rouge,linkcolor=bleu,filecolor=black}

\renewcommand{\epsilon}{\varepsilon}
\renewcommand{\rho}{\varrho}
\renewcommand{\phi}{\varphi}

\newcommand{\Bosonnormalord}{\stackrel{\textstyle \circ}{ \circ}}

\newcommand{\im}{\mathop{\fam0 Im}\nolimits}

\newcommand{\tr}{\mathop{\fam0 Tr}\nolimits}

\newcommand{\Lie}[1]{\mbox{$\mathfrak #1$}}

\newcommand{\Id}{\mathop{\fam0 Id}\nolimits}

\newcommand{\ra}{\mathop{\fam0 \rightarrow}\nolimits}

\def\Res{\operatornamewithlimits{Res}}

\def\tr{\operatorname{tr}}

\def\:{:}

\def\Bosonnormalordconstruction#1{\vcenter{\hbox{\ooalign{%
\raise.8ex\hbox{$#1\circ$}\crcr\lower.8ex\hbox{$#1\circ$}}}}}

\DeclareMathOperator{\End}{End}

\newcommand{\bt}{\begin{theorem}}
\newcommand{\et}{\end{theorem}}
\newcommand{\br}{\begin{remark}}
\newcommand{\er}{\end{remark}}

\def\Bosonnormalord{\,\lower.8ex \hbox{$\circ$} \llap{\raise.8ex\hbox{$\circ$}} \,}
\def\normalord{\,\lower.8ex \hbox{$\cdot$} \llap{\raise.8ex\hbox{$\cdot$}} \,}

\def\ra{\rightarrow}

\def\Res{\mathop{\rm Res}\limits}

\def\tr{\mathop{\rm Tr}\nolimits}

\def\Id{\mathop{\rm Id}\nolimits}

\let\operatorname=\mathrm
\let\text=\mathrm

\def\ln{\operatorname{ln}}

\def\beq{\begin{equation}}
\def\eeq{\end{equation}}
\def\bea{\begin{eqnarray}}
\def\eea{\end{eqnarray}}

\newcommand{\cpict}[3]{
\dimen1=#1\advance\dimen1 by-\hsize\divide\dimen1 by-2 \vtop to #2{
\noindent\hskip\dimen1{\special{em:graph #3.bmp}} \vfil}\hskip-2cm }

\newcommand{\be}{\begin{equation}}
\newcommand{\ee}{\end{equation}}

\def\Col#1{
\begin{pspicture}(0.2,1)
\multiput(0,0)(0,0.2){#1}{\psframe(0,0)(0.2,0.2)}
\end{pspicture}
}

\def\ColOne#1{
\begin{pspicture}(0.2,1)
\multiput(0,0)(0,0.2){#1}{\psframe(0,0)(0.2,0.2)}
\rput(0,-0.2){\makebox(0,0)[lb]{\tiny$1$}}
\end{pspicture}
}

\def\ColP#1{
\begin{pspicture}(0.2,1)
\multiput(0,0)(0,0.2){#1}{\psframe(0,0)(0.2,0.2)}
\rput(0,-0.2){\makebox(0,0)[lb]{\tiny $p$}}
\end{pspicture}
}

\def\Colg#1{
\begin{pspicture}(0.2,1)
\multiput(0,0)(0,0.2){#1}{\psframe[fillstyle=solid,fillcolor=lightgray](0,0)(0.2,0.2)}
\end{pspicture}
}

\def\ColgOne#1{
\begin{pspicture}(0.2,1)
\multiput(0,0)(0,0.2){#1}{\psframe[fillstyle=solid,fillcolor=lightgray](0,0)(0.2,0.2)}
\rput(0,-0.2){\makebox(0,0)[lb]{\tiny$1$}}
\end{pspicture}
}

\def\ColgP#1{
\begin{pspicture}(0.2,1)
\multiput(0,0)(0,0.2){#1}{\psframe[fillstyle=solid,fillcolor=lightgray](0,0)(0.2,0.2)}
\rput(0,-0.2){\makebox(0,0)[lb]{\tiny $p$}}
\end{pspicture}
}

\def\Cols#1{
\begin{pspicture}(0.2,0.6)
\multiput(0,0)(0,0.2){#1}{\psframe(0,0)(0.2,0.2)}
\end{pspicture}
}

\def\ColsOne#1{
\begin{pspicture}(0.2,0.6)
\multiput(0,0)(0,0.2){#1}{\psframe(0,0)(0.2,0.2)}
\rput(0,-0.2){\makebox(0,0)[lb]{\tiny$1$}}
\end{pspicture}
}

\def\ColsP#1{
\begin{pspicture}(0.2,0.6)
\multiput(0,0)(0,0.2){#1}{\psframe(0,0)(0.2,0.2)}
\rput(0,-0.2){\makebox(0,0)[lb]{\tiny $p$}}
\end{pspicture}
}

\def\Colsg#1{
\begin{pspicture}(0.2,0.6)
\multiput(0,0)(0,0.2){#1}{\psframe[fillstyle=solid,fillcolor=lightgray](0,0)(0.2,0.2)}
\end{pspicture}
}

\def\ColsgOne#1{
\begin{pspicture}(0.2,0.6)
\multiput(0,0)(0,0.2){#1}{\psframe[fillstyle=solid,fillcolor=lightgray](0,0)(0.2,0.2)}
\rput(0,-0.2){\makebox(0,0)[lb]{\tiny$1$}}
\end{pspicture}
}

\def\ColsgP#1{
\begin{pspicture}(0.2,0.6)
\multiput(0,0)(0,0.2){#1}{\psframe[fillstyle=solid,fillcolor=lightgray](0,0)(0.2,0.2)}
\rput(0,-0.2){\makebox(0,0)[lb]{\tiny $p$}}
\end{pspicture}
}

\usepackage{scrextend}
\deffootnote{2em}{0em}{\thefootnotemark\quad}

\begin{document}

\sloppy

\pagestyle{empty}
\addtolength{\baselineskip}{0.20\baselineskip}
\begin{center}

\vspace{26pt}

{\Large \textbf{The ABCD of topological recursion}}

\vspace{26pt}

\textsl{J\o{}rgen Ellegaard Andersen}\footnote{\noindent Centre for Quantum Geometry of Moduli Spaces, Department of Mathematics, Ny Munkegade 118, 8000 Aarhus C, Denmark.\\  \textit{Current address:} Center for Quantum Mathematics, Danish Institute for Advanced Study, SDU, Campusvej 55, 5230 Odense, Denmark \\ \href{mailto:jea.qgm@gmail.com}{\texttt{jea@sdu.dk}}}, \textsl{Ga\"etan Borot}\footnote{Max Planck Institut f\"ur Mathematik, Vivatsgasse 7, 53111 Bonn, Germany. \\  \textit{Current address:} Humboldt-Universit\"at zu Berlin, Institut f\"ur Mathematik \& Institut f\"ur Physik, Unter den Linden 6, 10099 Berlin, Germany. \\ \href{mailto:gaetan.borot@hu-berlin.de}{\texttt{gaetan.borot@hu-berlin.de}}}, \textsl{Leonid O. Chekhov}\footnote{Steklov Mathematical Institute, Gubkin 8, 119991, Moscow, Russia. \\ Niels Bohr Institute, Blegdamsvej 17, 2100 Copenhagen, Denmark. \\ Department of Mathematics, Michigan State University, East Lansing, USA.  \\ \href{mailto:chekhov@msu.edu}{\texttt{chekhov@msu.edu}}}, \textsl{Nicolas Orantin}\footnote{\'{E}cole Polytechnique F\'{e}d\'{e}rale de Lausanne, D\'epartement de Math\'ematiques, 1015 Lausanne, Switzerland. \\ \textit{Current address:} Section de Math\'ematiques, Universit\'e de Gen\'{e}ve, Uni Dufour, 24, rue du G\'{e}n\'{e}ral Dufour, Case postale 64, 1211 Gen\'{e}ve 4, Switzerland. \\ \href{mailto:nicolas.orantin@unige.ch}{\texttt{nicolas.orantin@unige.ch}}}
\end{center}

\vspace{20pt}

%

\begin{center}
\textbf{Abstract}
\end{center}

\vspace{0.2cm}

Kontsevich and Soibelman reformulated and slightly generalised the topological recursion of \cite{EOFg}, seeing it as a quantisation of certain quadratic Lagrangians in $T^*V$ for some vector space $V$. KS topological recursion is a procedure which takes as initial data a quantum Airy structure --- a family of at most quadratic differential operators on $V$ satisfying some axioms --- and gives as outcome a formal series of functions on $V$ (the partition function) simultaneously annihilated by these operators. Finding and classifying quantum Airy structures  modulo the gauge group action, is by itself an interesting problem which we study here. We provide some elementary, Lie-algebraic tools to address this problem, and give some elements of the classification for ${\rm dim}\,V = 2$. We also describe four more interesting classes of quantum Airy structures, coming from respectively Frobenius algebras (here we retrieve the 2d TQFT partition function as a special case), non-commutative Frobenius algebras, loop spaces of Frobenius algebras and a $\mathbb{Z}_{2}$-invariant version of the latter. This $\mathbb{Z}_{2}$-invariant version in the case of a semi-simple Frobenius algebra corresponds to the topological recursion of \cite{EOFg}.



\vspace{26pt}
\pagestyle{plain}
\setcounter{page}{1}

\vfill

\vspace{0.5cm}

\newpage

\tableofcontents

\newpage

\section{Introduction}

\subsection{A new point of view on topological recursion}

The topological recursion (TR) is a formalism developed by Eynard, Orantin \cite{EOFg,EORev} and Chekhov \cite{CEO06} which has in recent years found  many applications in random matrices \cite{E1MM,BEO}, enumerative geometry \cite{BMconj,EMS,Ebook,ACEH,BDBKS}, intersection theory on the moduli space of curves \cite{Mirza1,EOwp,EInter}, integrable systems \cite{BEInt,MDHitchin,BeEMgenus}, topological strings \cite{BKMP,EOBKMP,MCCLiuBKMP}, quantum field theories \cite{BEknots,BESeifert,ABO1}, see \cite{EynardICM} for a recent overview. In its simplest version, it takes as input a spectral curve $\Sigma$ embedded as a Lagrangian in $(\mathbb{C} \times \mathbb{C},\dd x \wedge \dd y)$, and returns a collection of meromorphic forms $\omega_{g,n}$ defined on ${\rm Sym}^n \Sigma$, indexed by integers $g \geq 0$ and $n \geq 1$. It also returns scalars $F_{g} = \omega_{g,0}$ for $g \geq 0$, which enjoy a property of symplectic invariance \cite{EO2MM,EOxy}.
The argument leading to symplectic invariance assumes $\Sigma$ is compact, the embedding algebraic, and it is a computational tour de force: it does not explain why this property is true and does not allow an easy generalisation to weaker conditions. These applications hint at interpreting the topological recursion as a quantisation procedure, but a thorough understanding of its underlying (symplectic) geometric nature is still incomplete.

Kontsevich and Soibelman \cite{KSTR} recently proposed a new point of view and setting for TR which generalises the TR of \cite{EOFg}. We refer to it as KS-TR. Their starting point is the notion of classical Airy structure, \textit{i.e.} a Lagrangian defined by quadratic equations in a symplectic vector space $T^*V$. The initial data for KS-TR is a lift of the former to a sub-Lie algebra of the Weyl algebra of $V$, which they call a \emph{quantum Airy structure} (Definition~\ref{Def:QAiry}). Quantum Airy structures are equivalently determined by their coefficients, collected in four tensors $(A,B,C,D)$, which must satisfy relations \eqref{SymA}-\eqref{Drel} coming from the ``Lie subalgebra'' condition. The outcome of KS-TR is a formal function on $V$ of the form $$Z = \exp\big(\sum_{g \geq 0} \hbar^{g - 1}S_{g}\big)$$ annihilated by the differential operators determining the quantum Airy structure (Proposition~\ref{Propmain}). The $n$-th order Taylor coefficients $F_{g,n}$ of $S_{g}$ are computed as in TR by induction on $2g - 2 + n$ using $(A,B,C,D)$, and encode the same information as the $\omega_{g,n}$ did in TR. More precisely, the data of a spectral curve can be used to produce a quantum Airy structure, such that the $F_{g,n}$ computed by KS-TR are the coefficients of the decomposition of $\omega_{g,n}$ computed by TR in a suitable basis of meromorphic differentials (Section 3.5 in \cite{KSTR}, and Section 6.1 here).

Kontsevich and Soibelman emphasise in \cite{KSTR} the geometry of Lagrangians in $T^*V$ and the relations between KS-TR and deformation quantisation. The present work is complementary to \cite{KSTR}. It focuses on the study of the relations defining quantum Airy structures, with the aim to exhibiting initial data for KS-TR.

\subsection{Outline}

Let us summarise the content of the article.

In Section~\ref{S2} we concisely present the KS-TR formalism. We write down explicitly in Section~\ref{S22} the relations satisfied by $(A,B,C,D)$, for which we give a graphical interpretation as three coupled IHX-like relations (Figure~\ref{3RelIHX}). The existence of the partition function (Proposition~\ref{Propmain}) is proved in \cite{KSTR} by general holonomicity arguments. We prove it in Section~\ref{S24} by direct computations. Section~\ref{SS3} shows that the partition function can be explicitly computed when some of the tensors $(A,B,C)$ are zero.
 
In Section~\ref{SS3bis}, we give an equivalent characterisation of classical and quantum Airy structures in terms of ``torsion-free'' symplectic representations of Lie algebras $\rho_{1}\,;\,V \rightarrow \mathfrak{sp}(T^*V)$, together with the data of a Lagrangian linear embedding $\mathcal{I}\,:\,V \rightarrow T^*V$. As a result, we describe in Section~\ref{SS4} an action of the group of at most quadratic differential operators on quantum Airy structures (the analog of a gauge group) and their partition function. Therefore, we are especially interested in quantum Airy structure modulo the action of this group. One can define in this way the moduli space of quantum Airy structures (Section~\ref{SModspace}), and deformation theory of quantum Airy structures is governed by twisted Lie algebra cohomology. We also define (Section~\ref{S34}) an action of commuting flows corresponding to translations in $V$. It means that, from a given quantum Airy structure $(L_i)_{i \in I}$, we can obtain a deformed Airy structure $(L_i^{(t)})_{i \in I}$ parametrised by $t$ in a formal neighbourhood of $0$ in $V$. The action of translation is non-linear even at the infinitesimal level, thus non-trivial modulo the gauge group action.

The remaining of the paper is devoted to exhibiting examples of quantum Airy structures. In Section~\ref{SS5}, we study general properties of the above symplectic representations, and apply them to prove some results aiming towards a classification finite-dimensional quantum Airy structures forming semi-simple Lie algebras. In particular, representation theory allows us to construct a quantum Airy structure forming the Lie algebra $\mathfrak{sl}_{2}(\mathbb{C})$. In Section~\ref{S5}, we give a classification of abelian quantum Airy structures in dimension two and three, a partial classification of two-dimensional quantum Airy structures, and an example of a non-trivial quantum Airy structure for a non semi-simple three-dimensional Lie algebra.

We then progress towards more geometric examples. In Section~\ref{S4}, we describe four classes of quantum Airy structures, associated respectively to Frobenius algebras, non-commutative Frobenius algebras, the loop space of Frobenius algebras, and a $\mathbb{Z}_{2}$-invariant version of the latter. Our proposal for the Frobenius algebra class (Section~\ref{S4Frob}) satisfies rather trivially the axioms of a quantum Airy structure, and contains as a special case the enumeration of the trivalent graphs underlying TR and the partition functions of 2d TQFTs (Lemma~\ref{TQFTpart}). For the three other classes, checking that our proposal is a quantum Airy structure is a computation. For the Frobenius and non-commutative Frobenius algebra class, we are able to give an explicit formula --- in the form of a finite-dimensional path integral, well-defined at the level of formal power series --- for the partition function in full generality (Section~\ref{S4NCFrob}). For the loop space of Frobenius algebras class (Section~\ref{S4Loop}-\ref{S444}), the partition function necessarily has $S_{0} = 0$, but $(S_{g})_{g \geq 1}$ can be non-trivial. For the $\mathbb{Z}_{2}$-invariant version, a priori all $S_{g}$ can be non-trivial.

The class of quantum Airy structures we describe in Proposition~\ref{ThmLoopZ2TQFT} for $\mathbb{Z}_{2}$-invariant loop spaces of Frobenius algebras are in correspondence with local spectral curves, and KS-TR gets identified with TR in this case. Section~\ref{comparTR} explains this correspondence to TR in more detail. In Section~\ref{SJUING}, we explain how the gauge group action on quantum Airy structures relates to Givental group action on Lagrangian cones. This brings our understanding of the correspondence between TR and correlation functions of semi-simple cohomological field theories established in \cite{DBOSS} closer to original spirit of Givental quantisation procedure \cite{GiventalQuad}. Independently, Section~\ref{SYoung} interpretes the recursion for quantum Airy structure on loop spaces as a dynamic on Young diagrams.

We conclude in Section~\ref{SConclu} with a list of open problems raised throughout the article.

\subsection{List of Airy structures}

For convenience of the reader we compile the list of Airy structures/partition functions constructed in this article and refer to the text for details and notations.

\subsubsection{With known enumerative interpretation}
 
\noindent $\bullet$ The $1$-dimensional Airy structures: $L = \hbar \partial_x - \frac{1}{2}Ax^2 - B x \hbar \partial_x - \frac{1}{2}C\hbar^2\partial_x^2 - \hbar D$. For arbitrary $A,B,C,D$, the partition function $Z$ is a Whittaker function $M_{\mu,1/4}$ for $\mu$ specified by $A,B,C,D$. When $B^2 = AC$ it degenerates to an Airy function. $Z$ is the weighted count of the number of graphs in topological recursion. See Section~\ref{1dZ}.

\noindent $\bullet$ Airy structures based on Frobenius algebras $\mathbb{A}$
\begin{equation}
\begin{split}
A & = \varphi(\theta_A e^* \otimes e^* \otimes e^*), \\
B& = \varphi(\theta_B e^* \otimes e^*  \otimes e), \\
C & = \varphi(\theta_C e^* \otimes e \otimes e) \\
\end{split}
\end{equation}
for arbitrary $\theta_A,\theta_B,\theta_C \in \mathbb{A}$ and $D \in \mathbb{A}^*$. Its partition function $Z$ is expressible in terms of the 2d TQFT attached to $\mathbb{A}$ (Proposition~\ref{FrobABCD}). It can also be written as a finite-dimensional path integral (Propositions~\ref{Cequal0Frob}--\ref{WhitZFrob}).

\noindent $\bullet$ Airy structures based on $\mathbb{A}[\![z^2]\!]$
\bea
A & = & \varphi(\theta e^* \otimes \dd e^* \otimes \dd e^*), \nonumber \\
B & = & \varphi(\theta e^* \otimes \dd e^* \otimes e) \nonumber \\
C & = & \varphi(\theta e^* \otimes e \otimes e) \nonumber
\eea
with $\theta \in z^{-2}\mathbb{A}[\![z^2]\!](\dd z)^{-1}$, see Propositions~\ref{Loop2} and \ref{ThmLoopZ2TQFT}. These Airy structures correspond to smooth spectral curves with simple ramifications in Chekhov--Eynard--Orantin theory, see Section~\ref{TRcomp}. Writing $\theta = O(z^2)$, we have $Z = 1$ as consequence of Lemma~\ref{Lemm2}. When $\theta_{\alpha} = t_{\alpha} z^{2s_{\alpha}}$ with $t_{\alpha}$ invertible and $s_{\alpha} \in \{-1,0\}$ for any $\alpha$, $Z$ is expressible in terms of intersection theory on $\overline{\mathcal{M}}_{g,n}$ in view of \cite{EInter,NorburyDo,Chekhovnor,negativerspin} (see also \cite[Section 7.5]{BKS21} for a summary).

\subsubsection{With unknown enumerative interpretation}

\noindent $\bullet$ Five continuous families of $2$-dimensional Airy structures based on the Lie algebra ${\rm Aff}(\mathbb{R}^2)$ (Proposition~\ref{d2Leonid}). In four cases $Z$ is an elementary function, in the last case it is expressible in terms of the Bessel function $J_{\nu}$, see Equation~\eqref{d2LeonidZ}.

\noindent $\bullet$ A $3$-dimensional Airy structure based on the Lie algebra $\mathfrak{sl}_2$ is constructed in Proposition~\ref{sl2th}. $Z$ is expressible in terms of the Hankel function of the second kind $\widetilde{H}^{(2)}_{\frac{1}{5}}$.

\noindent $\bullet$ A $3$-dimensional Airy structure based on a Bianchi VI Lie algebra is constructed in Proposition~\ref{Ldim3}. We did not compute its partition function.

\noindent $\bullet$ $2$- and $3$-dimensional Airy structures based on abelian Lie algebras. (Section~\ref{Secabelian}). In all cases the partition function either trivial or given by an elementary function.

\noindent $\bullet$ Airy structures based on non-commutative Frobenius algebras $\mathbb{A}$
\bea
A & = & \varphi(\lambda_A \,e \otimes  \{e \mathop{,}^{\otimes}e\}), \nonumber \\
B & = & \varphi(\lambda_B \,[e\mathop{,}^{\otimes}e] \otimes e), \nonumber \\
C & =  & \varphi(\lambda_C\, e \otimes \{e\mathop{,}^{\otimes}e\}), \nonumber
\eea 
with central $\lambda_A,\lambda_B,\lambda_C$ satisfying $\lambda_B^2 + \lambda_A \lambda_C = 0$, and $\lambda_B D \in [\mathbb{A},\mathbb{A}]^{\bot}$, see Proposition~\ref{NCFrob}.  The partition function $Z$ is expressible as a finite-dimensional integral (Corollary~\ref{co89}).

\noindent $\bullet$ Airy structures based on $\mathbb{A}[\![z]\!]$, where $\mathbb{A}$ is a Frobenius algebra
\begin{equation}
\begin{split}
A & = \varphi(\theta\, e^* \otimes \dd e^* \otimes \dd e^*), \\
B & = \varphi(\theta\, e^* \otimes \dd e^* \otimes e), \\
C & = \varphi(\theta\, e^* \otimes e \otimes e),
\end{split}
\end{equation} 
with $\theta \in z^{-1}\mathbb{A}[\![z]\!](\dd z)^{-1}$, see Propositions~\ref{Loop1} and \ref{ThmLoopTQFT}. When $\theta(z) \sim t z^{-1}(\dd z)^{-1}$ for $t \in \mathbb{A}$ invertible, we have $Z = 1$ as consequence of Lemma~\ref{Lemm1}.

\subsubsection{General properties of $Z$ in special cases}

\noindent $\bullet$ When $C = 0$, $Z$ is computed explicitly in Proposition~\ref{Cequal0}, and $F_{g,n} = 0$ for $g \geq 2$ and all $n$. 

\noindent $\bullet$ When $A = 0$, $F_{0,n} = 0$ for all $n$ (Lemma~\ref{Aequal0genus0}). For $A = B = 0$ only $F_{g,n} = 0$ for $n \geq 2$ and $F_{g,1}$ is a sum over rooted trivalent trees (Proposition~\ref{ABequal0}).

\subsection{Comments}

We stress that KS-TR is not only a reformulation of TR. It comes with new non-trivial examples of initial data, \textit{e.g.} having a finite-dimensional $V$ (Sections~\ref{S4Frob}-\ref{S4NCFrob}), and the case where $V$ is infinite-dimensional and attached to a curve without reference to a local involution (Proposition~\ref{ThmLoopTQFT}). The latter may be used to propose a TR for spectral curves without ramification points, see Section~\ref{SWithout}. Although the motivations mainly come from geometry, KS-TR can be presented only resorting to multilinear algebra and combinatorics, without complex analysis. A short survey on applications of topological recursion in geometry from the perspective of Airy structures was given in \cite{BToulouse}. The beginners or non-geometers interested in the theory of topological recursion --- \textit{e.g.} for the enumeration of maps \cite{Ebook} --- may find the simplicity of this new framework (concentrated in Section~\ref{S2}) appealing.

Since the first release of this article, many other works have brought forward the theory of quantum Airy structures taking \cite{KSTR} and the present work as starting point. A general strategy to construct Airy structures from vertex operator algebras having a free field representation and establish their equivalence with a spectral curve formulation was developed in \cite{HigherAiry,BouchardMastel,BKS21,Airyideal}. The extension to super Airy structures and supersymmetric VOAs was studied in \cite{SuperAiry1,SuperAiry2}. It gave a conceptual approach to the computations of Section~\ref{S4}, provided many other classes of Airy structures notably based on $W$-algebras, resolved symmetry questions in Bouchard--Eynard topological recursion \cite{BHLMR,BEthink}, and was applied to construct Whittaker vectors for $W$-algebras and Nekrasov partition function of four-dimensional pure $\mathcal{N} = 2$ supersymmetric gauge theory \cite{Whittaker,SuperAiry3}. The latter in fact provides a geometric application to the topological recursion without branched covering of Section~\ref{SWithout}. The classification of Airy structures based on simple Lie algebras initiated in Section~\ref{SS5} has been completed in \cite{RHAiry}. Developing further ideas from the work of Kontsevich and Soibelman \cite{KSTR}, Lagrangian foliations of symplectic surfaces and  corresponding families of Airy structures were used to study the geometry of the deformation space of global spectral curves \cite{NorburyAiry,ChaimaAiry}, generalising what was previously known in the setting of cotangent bundles of curves. A reformulation of Section~\ref{TRcomp} as Airy structure on the space of generalised cycles of a spectral curve was proposed in \cite{EynardAiry}. The ABCD (rather than spectral curve) perspective on topological recursion permitted the geometric refinement of topological recursion proposed in \cite{GRpaper}.

\subsubsection*{Acknowledgments}

We thank M.~Kontsevich and Y.~Soibelman for communicating preliminary versions of their work with us, P.~Biane, M.~Shapiro, P.~Teichner and F.~Wagemann for discussions, M.~Karev, D.~Noshchenko, B.~Ruba and Y.~Sch\"uler for comments and corrections. We also thank the organisers of the AMS Symposium \emph{Topological recursion and its applications}, Charlotte in July 2016 --- where this collaboration was initiated --- and the organisers of the thematic month on \emph{Topological recursion and modularity} at the Matrix, Creswick in December 2016 --- where a preliminary version of this work was presented. J.E.A. was funded in part by the Danish National Research Foundation grant DNRF95 (Centre for Quantum Geometry of Moduli Spaces, QGM) and by the ERC Synergy grant Recursive and Exact New Quantum Theory (ReNewQuantum) which receives funding from the European Research Council (ERC) under the European Union's Horizon 2020 research and innovation programme under grant agreement No 810573. The work of G.B. benefited from the support of the Max-Planck-Gesellschaft. The work of L.C. was supported by the Russian Foundation for Basic Research (Grant No. 15-01-99504a) and the ERC Advance Grant 291092 ``Exploring the Quantum Universe'' (EQU).

\section{Kontsevich-Soibelman approach to topological recursion}
\label{S2}

\subsection{Setting}
\label{S21}
Let $V$ be a vector space over $\mathbb{C}$. It could be finite or infinite-dimensional. We will mostly work in a basis $(e_i)_{i \in I}$ of $V$, and with its dual basis $(x_i)_{i \in I}$ which forms a set of linear coordinates on $V$. In the cases where $\dim V = \infty$, convergence issues will not play a role in this article. For general discussions it will be implicitly assumed that all seemingly infinite sums are actually finite or make sense after introducing if necessary suitable filtrations or completions. For specific examples where $\dim\,V = \infty$ we will justify that the sums contain only finitely many non-zero terms. We equip $T^*V$ with its canonical symplectic structure, and consider its Weyl algebra

$$
\mathcal{W}_{V}^{\hbar} = \mathbb{C}[\hbar]\big\langle (x_i,\partial_{i})_{i \in I})\big\rangle/\langle[\partial_i,x_i] = 1 \rangle.
$$

 Kontsevich and Soibelman \cite{Soibeltalk,KSTR} proposed the following setting, motivated by the problem of quantisation of Lagrangians in $T^*V$ defined by quadratic equations.
 
 By convention, $i,j,k,\ldots$ are fixed indices, while indices $a,b,c,\ldots$ should be summed over $I$. For instance, $A^i_{a,b}x_ax_b := \sum_{a,b \in I} A^i_{a,b}x_ax_b$. We warn the reader that the position of indices (upper or lower) does not respect Einstein's convention.

\begin{definition}
\label{Def:QAiry} A \emph{quantum Airy structure} on $V$ is a sequence $(L_i)_{i \in I}$ of elements of $\mathcal{W}_{V}^{\hbar}$ of the form
\beq
\label{Lform} L_i = \hbar \partial_{i} - \tfrac{1}{2}A^i_{a,b}x_ax_b - \hbar B^i_{a,b}x_a\partial_{b} - \tfrac{\hbar^2}{2}C^i_{a,b}\partial_a\partial_b - \hbar D^i,
\eeq
where $\hbar$ is a formal parameter and $A^i_{j,k},B^i_{j,k},C^i_{j,k}$ and $D^i$ are scalars, which form a Lie subalgebra of $\mathcal{W}_{V}^{\hbar}$, \textit{i.e.}
\beq
\label{Liealg} [L_i,L_j] = \hbar\,f_{i,j}^a L_a
\eeq
for some scalars $f_{i,j}^k$.
\end{definition}
In this definition, we can always assume that $A^i_{j,k} = A^i_{k,j}$ and $C^i_{j,k} = C^i_{k,j}$. The coefficients defining a quantum Airy structure can be rearranged in a basis-free way
\beq
\label{tensorini} A \in {\rm Hom}(V^{\otimes 3},\mathbb{C}),\qquad B \in {\rm Hom}(V^{\otimes 2},V),\qquad C \in {\rm Hom}(V,V^{\otimes 2}),\qquad D \in {\rm Hom}(V,\mathbb{C}),
\eeq
by the assignments
$$
A(e_i \otimes e_j \otimes e_k) = A^i_{j,k},\qquad B(e_i \otimes e_j) = B^i_{j,a}e_{a},\qquad C(e_i) = C^i_{a,b}e_{a} \otimes e_{b},\qquad D(e_i) = D^i.
$$
Equation \eqref{Liealg} puts strong constraints on $A,B,C,D$. They will be studied in Section~\ref{S22} in a pedestrian way, and in Section~\ref{SS3bis} in a more abstract way. We remark that for any choice of $A$ and $D$, $(A,B = C = 0,D)$ defines a (rather trivial) quantum Airy structure. The justification for the name ``Airy structures'' will appear in the examples provided in Section~\ref{S4Frob}. The notion of classical Airy structure will only be presented in Section~\ref{SSymp}, as it does not play a central role here although it served as motivation in \cite{KSTR}.

The Weyl algebra $\mathcal{W}_{V}^{\hbar}$ naturally acts by differential operators on functions on $V$. Equation \eqref{Liealg} is a sufficient condition for the existence of a function $Z$ on $V$ which is a common solution to $L_i\cdot Z = 0$ for all $i \in I$. More precisely, we have
\begin{proposition}
\label{Propmain}There exists a unique formal series of the form
\beq
\label{Zpart} Z = \exp\bigg(\sum_{\substack{g \geq 0 \\ n \geq 1}} \frac{\hbar^{g - 1}}{n!} \sum_{i_1,\ldots,i_n \in I} F_{g,n}(i_1,\ldots,i_n)\,x_{i_1}\cdots x_{i_n}\bigg),
\eeq
where $F_{g,n}(i_1,\ldots,i_n)$ are scalars, invariant under permutation of the $(i_m)_{m = 1}^n$, such that $F_{0,1}(i) = F_{0,2}(i,j) = 0$ for all $i,j$, and
$$
\forall i \in I\qquad L_i\cdot Z = 0.
$$
More precisely,
\beq
\label{F03} F_{0,3}(i,j,k) = A^i_{j,k},\qquad F_{1,1}(i) = D^i,
\eeq
and for $2g - 2 + n \geq 2$
\bea
\label{TRForm} F_{g,n}(i_1,J) & = & \sum_{m = 2}^n B^{i_1}_{i_m,a} F_{g,n - 1}(a,J \setminus \{i_m\}) \\
&& + \tfrac{1}{2}C^{i_1}_{a,b}\bigg(F_{g - 1,n + 1}(a,b,J) + \sum_{\substack{J' \sqcup J'' = J \\ h' + h'' = g}} F_{h',1 + |J'|}(a,J') F_{h'',1 + |J''|}(b,J'')\bigg), \nonumber
\eea
where $J = \{i_2,\ldots,i_n\}$ is a $(n-1)$-uple of indices in $I$.
\end{proposition}
\noindent\textbf{Proof.} The uniqueness is obvious. We take $i_1 \in I$, insert \eqref{Zpart} into the equation $L_{i_1}\cdot Z = 0$, and for each $g \geq 0$, $n \geq 1$ and $(i_2,\ldots,i_n) \in I^{n - 1}$ we collect the coefficient of $\hbar^{g - 1} \frac{x_{i_2}\cdots x_{i_n}}{(n - 1)!}$. The equations for $(g,n) = (0,1),(0,2),(0,3)$ are
\begin{eqnarray}
0 & = & -F_{0,1}(i_1) + C_{a,b}^i F_{0,1}(a)F_{0,1}(b), \nonumber \\
0 & = & - F_{0,2}(i_1,i_2) + B_{i_2,a}^{i_1} F_{0,1}(a) + C^{i_1}_{a,b} F_{0,1}(a) F_{0,2}(b,i_2), \nonumber \\
0 & = & - F_{0,3}(i_1,i_2,i_3) + A_{i_2,i_3}^{i_1} + \big(B_{i_2,a}^{i_1} F_{0,2}(a,i_3) + B_{i_3,a}^{i_1} F_{0,2}(a,i_2)\big) \nonumber \\
&& +\tfrac{1}{2} C^{i_1}_{a,b}\big(F_{0,1}(a) F_{0,3}(b,i_2,i_3) + F_{0,2}(a,i_2)F_{0,2}(b,i_3)\big).
\end{eqnarray}
As we take $F_{0,1}(i) = F_{0,2}(i,j) = 0$, the two first equations are automatically satisfied. The third equation yields $F_{0,3}(i_1,i_2,i_3) = A^{i_1}_{i_2,i_3}$. For $(g,n) = (1,1)$ we find $F_{1,1}(i_1) = D^{i_1}$. In general, isolating the term $F_{g,n}(i_1,\ldots,i_n)$ readily gives \eqref{TRForm}. In this equation, the order of the indices in the set $J'$ and $J''$ does not matter as the $F_{g',n'}(j_1,\ldots,j_n)$ were assumed symmetric under permutation of $(j_m)_{m = 1}^n$.

Conversely, we shall see that the existence is guaranteed by the constraints \eqref{Liealg}. Alternatively, and effectively, we can define $F_{g,n}$ by formula \eqref{TRForm} inductively on $2g - 2 + n$, provided we justify that the result is symmetric when $i_1$ is permuted with the other $i_m$s. We will show this is true later in Proposition~\ref{Explicitsym}, by direct computations involving the relations between $(A,B,C,D)$ following from \eqref{Liealg}. We see for instance that the symmetry of $F_{0,3}$ imposes that $A^{i}_{j,k} = A^j_{i,k}$, hence $A$ must be fully symmetric in its three indices. It will indeed be a consequence (see Section~\ref{S24}) of \eqref{Liealg} for operators of the form \eqref{Lform}. \hfill $\Box$

\begin{center}
\begin{figure}[h!]
\centering
\includegraphics[width=\textwidth]{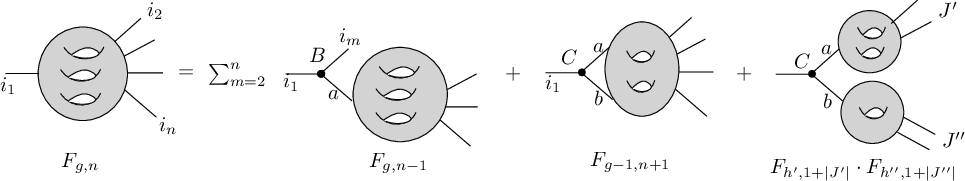}
\caption{\label{TRgraph} $F_{g,n}(i_1,\ldots,i_n)$ is represented as a surface of genus $g$ with $n$ boundaries, carrying the labels $i_1,\ldots,i_n$. In this graphical language, the terms appearing in the recursion \eqref{TRForm} are all the topologies resulting from the removal of a pair of pants $P$ bounding the first boundary. The weight of $P$ is a $B$ or a $C$ depending on whether it has one or two external boundary components.}
\end{figure}
\end{center}

Formula \eqref{TRForm} has a graphical interpretation (Figure~\ref{TRgraph}), which contains the same kind of terms as the topological recursion introduced in \cite{EOFg}. One can therefore propose, following Kontsevich and Soibelman, an elementary definition of the topological recursion.
\begin{itemize}
\item[$\bullet$] the initial data is a quantum Airy structure, \textit{i.e.} $(A,B,C,D)$ as in \eqref{tensorini} satisfying the relations we will write in Section~\ref{S2}.
\item[$\bullet$] the outcome are symmetric tensors $F_{g,n} \in {\rm Hom}(V^{\otimes n},\mathbb{C})$ indexed by $2g - 2 + n > 0$, which we consider as the coefficients of the Taylor expansion at $0$ of a formal series/function on $V$, whose exponential is denoted $Z$ and called the partition function.
\end{itemize}
The topological recursion of \cite{EOFg} rather takes as initial data a spectral curve, \textit{i.e.} a simple branched cover $x\,:\,\Sigma \rightarrow \Sigma_0$ between two smooth complex curves, together with a meromorphic $1$-form $\omega_{0,1}$ on $\Sigma$, and a fundamental meromorphic bidifferential of the second kind $\omega_{0,2}$ on $\Sigma^2$. We will show in Section~\ref{TRcomp} that this data determines a quantum Airy structure based on $V = H^0(U,K_U)/x^*H^0(U_0,K_{U_0})$ where $U$ is a (small enough) neighbourhood of the ramification points (zeroes of $\dd x$) in $\Sigma$, and $U_0 := x(U)$. The $F_{g,n}$ computed by \eqref{TRForm} are then coefficients of the decompositions of the meromorphic $n$-differentials $\omega_{g,n}$ defined by \cite{EOFg} in a suitable basis of meromorphic forms (Proposition~\ref{FWid}). Therefore, Kontsevich--Soibelman topological recursion can be seen as a generalisation of \cite{EOFg}.

\begin{center}
\begin{figure}[h!]
\centering
\includegraphics[width=0.8\textwidth]{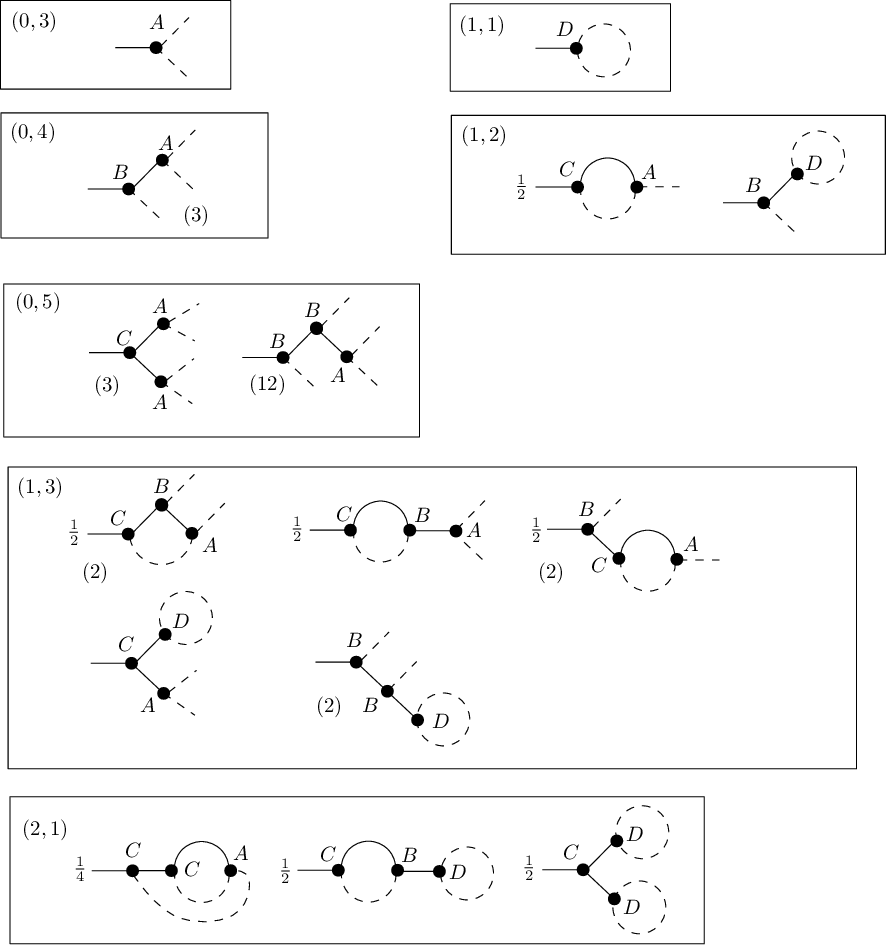}
\caption{\label{TRGraph} Unfolding \eqref{TRForm} gives \cite[Section 3]{EORev} a set $\mathfrak{G}_{g,n}(1)$ of pairs $(G,T)$ where $G$ is a trivalent graph with first Betti number $g$ and $n$ leaves, with cyclic order at each vertex, and $T$ is a spanning tree rooted at the first leaf, having the extra property that edges which are not in $T$ must connect vertices $v$ and $v'$ which are \emph{parent}. This means that the geodesic in $T$ from the root to $v$ contains (or is contained) in the geodesic from the root to $v'$. Vertices incident to a loop are assigned a $D$, vertices incident to one external leg are $B$s, vertices incident to two external legs are $A$s. Internal vertices can be $A,B,C$ as prescribed by the recursive construction of the graph --- which is remembered by the spanning tree rooted at the first leg. We have listed these graphs for low values of $(g,n)$. The $\tfrac{1}{2^p}$ is the symmetry factor which arises from the repetition of factor of $\tfrac{1}{2}$ in the $C$-term of \eqref{TRForm}. A $(k)$ indicates that there are $k$ such graphs, which differ by the labeling $2,\ldots,n$ of the legs. When two such graphs are related by an exchange of the two legs outgoing from a $C$, the two graphs give the same contribution to $F_{g,n}$, and we listed it as a single graph with a factor of $\tfrac{1}{2}$ less.}
\end{figure}
\end{center}

\subsection{The relations between \texorpdfstring{$(A,B,C,D)$}{(A,B,C,D)}}
\label{S22}
Let $L := (L_i)_{i \in I}$ be differential operators of the form \eqref{Lform}. We now describe the necessary and sufficient conditions on $(A,B,C,D)$ for $L$ to be a quantum Airy structure, \textit{i.e.} to satisfy \eqref{Liealg}. Evaluating the commutator between the first terms with a pure single derivative and the $B$-terms we again obtain terms with pure single derivatives. Because this commutator is the only source of such  terms in the right-hand side, we immediately conclude that the structure constants $f^k_{i,j}$ are determined by the $B$-terms alone
\beq
\forall i,j,k \in I,\qquad f^k_{i,j}=B^i_{j,k}-B^j_{i,k}.
\label{f2B}
\eeq
Evaluating now the commutator between $L_i$ and $L_j$ and comparing with the right-hand side in Equation~\eqref{Liealg} we obtain further constraints on $(A,B,C,D)$. First, the absence of a linear term in $x_i$ immediately implies the full symmetry of the coefficients $A$
\beq
\forall i,j,k \in I,\qquad A^i_{j,k}=A^j_{i,k},
\label{SymA}
\eeq
as anticipated for the symmetry of $F_{0,3}$ in \eqref{F03}. We obtain three more relations matching the coefficients of the terms $x_kx_\ell$, $\partial_{k}\partial_{\ell}$, $x_k\partial_{\ell}$, for any $i,j,k,\ell \in I$
\begin{eqnarray}
\label{BBACeq} B^i_{j,a}B^a_{k,\ell} + B^i_{k,a}B^j_{a,\ell} + C^i_{\ell,a}A^j_{a,k} & = & (i \leftrightarrow j), \\
\label{BCeq} B^i_{j,a}C^a_{k,\ell} + C^i_{k,a}B^j_{a,\ell} + C^i_{\ell,a}B^j_{a,k} & = & (i \leftrightarrow j),\\
\label{BAeq} B^i_{j,a}A^a_{k,\ell} + B^i_{k,a}A^j_{a,\ell} + B^i_{\ell,a}A^j_{a,k} & = & (i \leftrightarrow j).
\end{eqnarray}
And matching the coefficient of $\hbar.1$ we find, for all $i,j \in I$
\beq
\label{Drel} B^i_{j,a}D^a + \tfrac{1}{2}C^i_{a,b}A^j_{a,b} = (i \leftrightarrow j).
\eeq
Consequently we have the lemma
\begin{lemma}
$(L_i)_{i \in I}$ is a quantum Airy structure if and only if $(A,B,C,D)$ satisfy the torsion-free conditions \eqref{f2B}-\eqref{SymA}, the \textbf{BB-CA} relation \eqref{BBACeq}, the \textbf{BC} relation \eqref{BCeq}, the \textbf{BA} relation \eqref{BAeq}, and the \textbf{D} relation \eqref{Drel}. \hfill $\Box$
\end{lemma}
Equation \eqref{f2B} can be taken as a definition of the structure constants $f$, and one can check by direct computation that the above relations imply the Jacobi identity for $f$. The full symmetry of $A$ could be added to the axioms of quantum Airy structures. We call \eqref{f2B}-\eqref{SymA} ``torsion-free condition" for a reason explained in Section~\ref{S32}. The three relations for $(A,B,C)$ are rather non-trivial. If $d = {\rm dim}\,V$, let us count the number of unknowns and \textit{a priori} independent equations determining quantum Airy structures. $A$ has $\tfrac{d(d + 1)(d + 2)}{6}$ independent coefficients, $B$ has $d^3$ coefficients, and $C$ has $\frac{d^2(d + 1)}{2}$ coefficients. The \textbf{BB-CA} relation is antisymmetric in $i,j$, so give $\tfrac{d(d - 1)}{2}\cdot d^2$ constraints. The \textbf{BC} and the \textbf{AC} relations are antisymmetric in $i,j$, symmetric in $k,\ell$, so give $2 \cdot \frac{d(d - 1)}{2}\cdot \tfrac{d(d + 1)}{2}$ constraints. So, as far as $(A,B,C)$ are concerned, we have $\tfrac{d}{3}(5d^2 + 3d + 1)$ unknowns, and $\tfrac{d^2(d - 1)(2d + 1)}{2}$ constraints. The first values are

$$\begin{array}{|c||c|c|c|c|c|c|}
\hline
d & 1 & 2 & 3 & 4 & 5 & 6 \\
\hline
\hline
\#{\rm unknowns} & 3 & 18 & 55 & 124 & 235 & 398  \\
\hline
\#{\rm constraints} & 0 & 10 & 63 & 216 & 550 & 1170 \\
\hline
\end{array}
$$
For $d \geq 3$, we find that the three relations form an overdetermined system. Therefore, it is \textit{a priori} not obvious that non-zero solutions for $(A,B,C)$ can be found at all. If $(A,B,C)$ is a solution, the set of allowed $D$ satisfying \eqref{Drel} is an affine space, hence easier to describe. We will however show in Sections~\ref{S5} and \ref{S4} that many non-trivial solutions can be found.

\begin{lemma}\label{trL}
\label{basrem} Assume $(A,B,C)$ solves the \textbf{BB-CA}, \textbf{BC} and \textbf{BA} relations, as well as \eqref{f2B} and \eqref{SymA}.
\begin{itemize}
\item[$(i)$] If $V$ is an abelian Lie algebra (namely $f_{i,j}^k = 0$ for all $i,j,k \in I$), any choice of $D$ gives a quantum Airy structure.
\item[$(ii)$] If ${\rm tr}\,B^i := \sum_{a} B^i_{a,a}$ exists (for instance, it is always the case when $\dim V\,< \infty$), then
$$
D^i_{{\rm ref}} = \tfrac{1}{2}\,{\rm tr}\,B^i
$$
completes $(A,B,C)$ into a quantum Airy structure. Further, $(A,B,C,D)$ completes $(A,B,C)$ into a quantum Airy structure if and only if for any $v_1,v_2 \in V$, $(D - D_{{\rm ref}})([v_1,v_2]) = 0$  where we consider $D \in V^*$.
\end{itemize}
\end{lemma}
In particular, when $V$ is finite-dimensional and semi-simple, $[V,V] = V$ thus, for given $(A,B,C)$ satisfying the relations, there is a unique way to complete into a quantum Airy structure $(A,B,C,D)$.

\noindent \textbf{Proof.} The claim follows from the observation that, summing \eqref{BBACeq} over $k = \ell \in I$, we find
$$
f^i_{j,a} {\rm Tr}\,B^a + C^i_{a,b}A^j_{a,b} - C^j_{a,b}A^i_{a,b} = 0.
$$
\hfill $\Box$

\vspace{0.2cm}
Note that
$$
\hbar B^i_{a,b} x_a\partial_{b} = \tfrac{\hbar}{2} B^i_{a,b}(x_a\partial_b + \partial_b x_a) - \tfrac{\hbar}{2}\,{\rm tr}\,B^i
$$
when it makes sense. So, the solution exhibited in $(ii)$ corresponds to choosing the democratic ordering of $x_a$ and $\partial_b$, rather than the normal ordering $x_a\partial_b$. When $V$ is infinite-dimensional, we will  in Section~\ref{S4Loop} see cases where ${\rm tr}\,B^i$ is not well-defined, but we can nevertheless find solutions for $D$.

\subsection{Graphical interpretation of the relations}

\begin{center}
\begin{figure}[h!]
\centering
\includegraphics[width=0.6\textwidth]{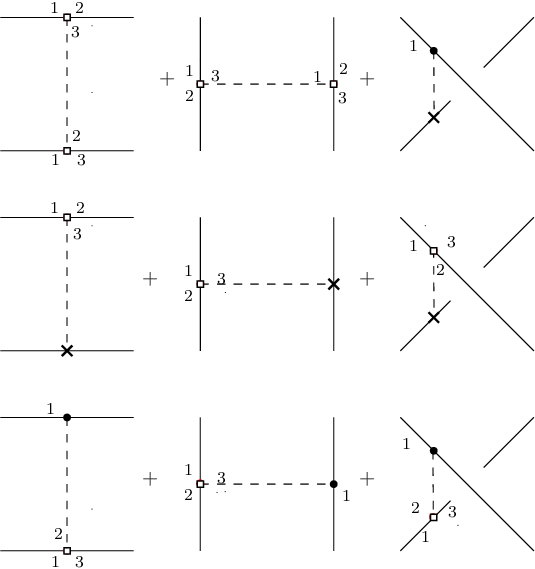}
\caption{\label{3RelIHX} A $\Box$ indicates a $B^{i_1}_{i_2,i_3}$ with incident edges labeling as indicated by the numbers. A $\boldsymbol{\times}$ indicates an $A^{*}_{**}$. A $\bullet$ represents a $C^{i_1}_{**}$, with first incident edge carrying the upper index. The three relations are that these combinations are symmetric with respect to permutation of the two left legs.}
\end{figure}
\end{center}

\begin{center}
\begin{figure}[h!]
\centering
\includegraphics[width=0.6\textwidth]{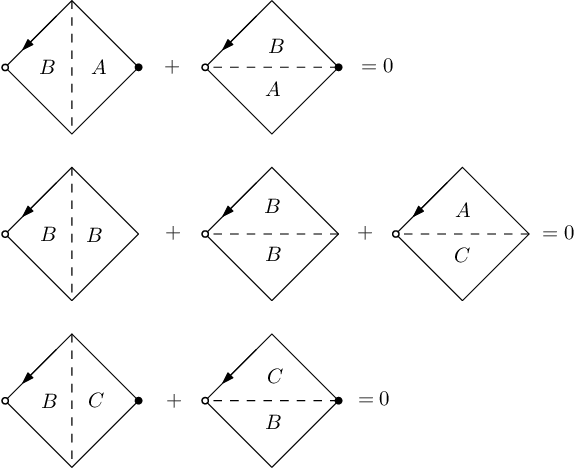}
\caption{\label{3REL} Each edge carries an index. Indices carried by dashed edge are summed over. $\bullet$ means symmetrisation of the indices of the edge incident at that vertex, while $\circ$ means antisymmetrisation. The arrow indicates which index is placed first, and thus defines the order of composition of the two operators.}
\end{figure}
\end{center}

The structure of the indices $(i,j,k,l)$ and the summation index $a$ is the same in the three relations \textbf{BB-CA}, \textbf{BC} and \textbf{BA}. They can in fact be presented as a system of three IHX relations (Figure~\ref{3RelIHX}). Another graphical form is given in Figure~\ref{3REL}.

This IHX form is in fact not surprising.  The original IHX relation is the graphical interpretation of the Jacobi relation for a Lie algebra $\mathfrak{l}$, and the Jacobi relation itself expresses that the adjoint representation $\mathfrak{l} \rightarrow {\rm GL}(\mathfrak{l})$ is a homomorphism of Lie algebras. It is a special case, for the adjoint representation, of the STU relation expressing that one has a representation $\mathfrak{l} \rightarrow {\rm End}(\mathfrak{m})$, where $\mathfrak{m}$ is a module for $\mathfrak{l}$. We will see in Section~\ref{S32} that the three relations for $(A,B,C)$ are equivalent to requiring that the adjoint action ${\rm ad}_{L}\,:\,V \rightarrow {\rm End}(T^*_{\hbar}V)$ is a representation of the Lie algebra $(V,(f_{i,j}^k)_{i,j,k})$. So, these three relations come from specialising the STU relation to this representation which has the special form $[L_i,\cdot]$ with $L_i$ of the form \eqref{Lform}.

\vspace{0.5cm}

\subsection{Proof of symmetry}
\label{S24}
\begin{proposition}
\label{Explicitsym}
If $(L_i)_{i \in I}$ is a quantum Airy structure, $F_{g,n}(i_1,\ldots,i_n)$ defined recursively by \eqref{TRForm} for $2g - 2 + n > 0$ is symmetric under permutation of $(i_m)_{m = 1}^n$.
\end{proposition}

\begin{remark}
The converse of Proposition~\ref{Explicitsym} is not true. Indeed, if $A = 0$ and $D = 0$, one finds that $F_{g,n} = 0$ for all $g,n$, hence it is symmetric, whether or not $B$ and $C$ satisfy the --- still non-trivial --- relations \textbf{BB-CA} and \textbf{BC}.
\end{remark}

\noindent\textbf{Proof.} We already saw in Section~\ref{S21} that $F_{0,3}(i,j,k) = A^i_{j,k}$, which is fully symmetric due to \eqref{SymA}, and for $F_{1,1}$ there is nothing to check. We compute from \eqref{TRForm}
$$
F_{0,4}(i,j,k,\ell) = B_{j,a}^i A_{k,\ell}^{a} + {\rm cycl.}\,\,(j,k,\ell)
$$
which is fully symmetric thanks to \eqref{BAeq}, and
$$
F_{1,2}(i,j) = B_{j,a}^iD^a + \tfrac{1}{2}C^i_{a,b}A_{a,b}^j
$$
which is fully symmetric thanks to \eqref{Drel}. So the result holds for $2g - 2 + n \leq 2$. Take $(g,n)$ such that $2g -2 + n > 2$, and assume we have proved fully symmetry of $F_{g',n'}$ for all $2g' - 2 + n' < 2g - 2 + n$. Let $K = \{k_1,\ldots,k_n\}$. Let us define $F_{g,n + 2}(i,j,K)$ by applying \eqref{TRForm} with first index $i$. The resulting terms are in the range of the induction hypothesis, thus  completely symmetric under permutation of $(j,k_1,\ldots,k_n)$. So, we only need to prove it is symmetric under permutation of $i$ and $j$. For this purpose, we use again \eqref{TRForm} with first index $j$, except for the term involving $B^i_{j,a}F_{g,n + 1}(a,K)$, for which we use \eqref{TRForm} with first index $a$. Denote $K^{[k]} := K \setminus\{k\}$ and $K^{[k,\ell]} := K\setminus\{k,\ell\}$.  We also implicitly use the full symmetry of $A$ and the symmetry of $C$ in its two lower indices. We find that
\begin{eqnarray}
&& F_{g,n + 2}(i,j,K) \nonumber \\
 & = & B_{j,a}^iF_{g,n + 1}(a,K) + \sum_{k \in K} B_{k,a}^iF_{g,n + 1}(a,j,K^{[k]}) + \tfrac{1}{2}C_{a,b}^i F_{g - 1,n + 3}(a,b,j,K) \nonumber \\
&& + \sum_{\substack{h' + h'' = g \\ K' \sqcup K'' = K}} C_{a,b}^i F_{h',2 + |K'|}(a,j,K') F_{h'',1 + |K''|}(b,K'')  \nonumber \\
& = & B_{j,a}^i\bigg\{\sum_{k \in K} B_{k,b}^a F_{g,n}(b,K^{[k]}) + \tfrac{1}{2}C_{b,c}^aF_{g - 1,n + 2}(b,c,K) +\!\!\! \sum_{\substack{h' + h'' = g \\ K' \sqcup K'' = K}} \!\!\! \tfrac{1}{2}C_{b,c}^a F_{h',1+|K'|}(b,K')F_{h'',1 + |K''|}(c,K'')\bigg\} \nonumber \\
&& + \sum_{k \in K} B_{k,a}^i\bigg\{B_{a,b}^j F_{g,n}(b,K^{[k]}) + \sum_{\ell \in K^{[k]}} B_{\ell,b}^j F_{g,n}(b,a,K^{[k,\ell]}) + \tfrac{1}{2} C_{b,c}^j F_{g - 1, n + 2}(b,c,a,K^{[k]}) \nonumber \\
&& + \sum_{\substack{h' + h'' = g \\ K' \sqcup K'' = K^{[k]}}} \tfrac{1}{2} C^j_{b,c}F_{h',2 + |K'|}(a,b,K') F_{h'',1 + |K''|}(c,K'') + \delta_{g,0}\delta_{n,2} A_{a,k}^j\bigg\} \nonumber \\
&& + \tfrac{1}{2}C_{a,b}^i\bigg\{B_{a,c}^jF_{g - 1,n + 2}(b,c,K) + B^j_{b,c}F_{g - 1,n + 2}(a,c,K) \nonumber \\
&& + \sum_{k \in K} B_{k,c}^j F_{g - 1,n + 2}(a,b,c,K^{[k]}) + \tfrac{1}{2}C_{c,d}^j F_{g - 2,n + 4}(a,b,c,d,K) \nonumber \\
&& + \sum_{\substack{h' + h'' = g - 1 \\ K' \sqcup K'' = K}}  C_{c,d}^jF_{h',3+|K'|}(a,b,c,K')F_{h'',1 + |K''|}(d,K'') + C_{c,d}^j F_{h',2 + |K'|}(a,c,K')F_{h'',2 + |K''|}(b,d,K'') \nonumber \\
&& + \sum_{\substack{h' + h'' = g \\ K' \sqcup K'' = K}} C_{a,b}^i F_{h'',1 + |K''|}(b,K'')\bigg\{B_{a,c}^j F_{h',1 + |K'|}(c,K') + \sum_{k \in K'} B_{k,c}^j F_{h',1 + |K'|}(a,c,K^{'[k]}) \nonumber \\
&& + \tfrac{1}{2}C_{c,d}^j F_{h' - 1,3 + |K'|}(a,c,dK') + \sum_{\substack{s + s' = h' \\ L \sqcup L' = K'}} C_{c,d}^j F_{s,2 + |L|}(a,c,L)F_{s',1 + |L'|}(d,L') + \delta_{h',0}\delta_{|K'|,1}A_{a,k'}^j\bigg\}. \nonumber
\end{eqnarray}
We now collect the various terms
\begin{eqnarray}
&& F_{g,n + 2}(i,j,K) \nonumber \\
&= & \sum_{k \in K} \textcolor{red}{F_{g,n}(b,K^{[k]}) \Big(B_{j,a}^iB_{k,b}^a + B_{a,b}^jB_{k,a}^i + A_{a,k}^jC_{a,b}^i\Big)}  + \sum_{k \neq \ell \in K} F_{g,n}(a,b,K^{[k,\ell]})\,B_{k,a}^i B_{\ell,b}^j \nonumber \\
&& + \textcolor{red}{\tfrac{1}{2} F_{g - 1,n + 2}(b,c,K)\big(B_{j,a}^iC_{b,c}^a + B_{a,c}^jC_{a,b}^i + B^j_{b,c}C^i_{a,b}\big)} + \sum_{k \in K} \tfrac{1}{2} F_{g - 1,n + 2}(a,b,c,K^{[k]})\big(B^i_{k,a}C^j_{b,c} + B^j_{k,c}C^i_{a,b}\big) \nonumber \\ 
&& + \tfrac{1}{4} F_{g - 2,n + 4}(a,b,c,d,K)\,C_{a,b}^i C_{c,d}^j + \textcolor{red}{\sum_{\substack{h' + h'' = g \\ K' \sqcup K'' = K}} \tfrac{1}{2} F_{h',1+|K'|}(b,K')F_{h'',1+|K''|}(c,K'')\big(C_{b,c}^aB_{j,a}^i + 2C_{a,b}^iB_{a,c}^j\big)} \nonumber \\
&&+ \sum_{k \in K} \sum_{\substack{h' + h'' = g \\ K' \sqcup K'' = K^{[k]}}}  \tfrac{1}{2} F_{h',2+|K'|}(a,b,K')F_{h'',1 + |K''|}(c,K'')\big(B_{k,a}^iC_{b,c}^j + B^j_{k,b}C^i_{a,c}\big) \nonumber \\
&& + \sum_{\substack{h' + h'' = g - 1\\ K' \sqcup K'' = K}} \tfrac{1}{2}F_{h',3 + |K'|}(a,b,c,K') F_{h'',1 + |K''|}(d,K'')\big(C_{a,b}^i C_{c,d}^j+ C_{a,d}^iC_{c,b}^j\big) \nonumber \\
&& + \sum_{\substack{h' + h'' = g - 1\\ K' \sqcup K'' = K}} \tfrac{1}{2}F_{h',2 + |K'|}(a,c,K')F_{h'',2 + |K''|}(b,d,K'')\,C_{a,b}^iC_{c,d}^j \nonumber \\
&& + \sum_{\substack{h'' + s + s' = g \\ K'' \sqcup L \sqcup L' = K}} F_{h'',1 + |K''|}(b,K'')F_{s,2 + |L|}(a,c,L)F_{s,1 + |L'|}(d,L')C_{a,b}^iC_{c,d}^j. \nonumber
\end{eqnarray}
In the right-hand side, the third term of type $CA$ in the first sum comes from the last term in the previous equation, and to obtain the second term in the bracket in the fourth (resp. fifth) line we have renamed the dummy indices $b \leftrightarrow c$ (resp. $b \leftrightarrow d$). The non-red terms are obviously symmetric by permutation of $i$ and $j$. The three relations \eqref{BBACeq}-\eqref{BCeq}-\eqref{BAeq} allow the conclusion that the red terms are also symmetric. So, $F_{g,n}(i,j,K)$ is fully symmetric, and by induction this entails the claim. \hfill $\Box$

\section{Lie algebraic approach to Airy structures}
\label{SS3bis}

\subsection{Reformulation \textit{via} the adjoint representation}

\label{S32}
In $\mathcal{W}_{V}^{\hbar}$, we have two notions of degree: the $\hbar$-degree, and the variable degree assigning degree $1$ to $x_i$ and $\partial_{x_i}$. We denote $T^*_{\hbar}V = V^* \oplus V.\hbar$ and
consider the subspace $\mathcal{D}_{V} \subset \mathcal{W}_{V}^{\hbar}$
$$
\mathcal{D}_{V} = \bigoplus_{d = 0}^{2} \mathcal{D}_{V,d},\qquad \mathcal{D}_{V,0} = \mathbb{C}.\hbar,\qquad \mathcal{D}_{V,1} = T^*_{\hbar}V,\qquad \mathcal{D}_{V,2} = {\rm Sym}^2 T^*_{\hbar}V.
$$
Note that the copy of $V^*$ in $\mathcal{D}_{V}$ contains the linear functions $x_i$, while the copy of $V$ correspond to differential operators $\partial_{i}$. $\mathcal{D}_{V}$ is naturally a sub-Lie algebra of $\mathcal{W}_{V}^{\hbar}$. Let $\pi_1$ and $\pi_1^*$ be the linear projections to the subspaces $V^*$ and $V.\hbar$ of $\mathcal{D}_{V,1}$. We are interested in linear maps $V \rightarrow \mathcal{D}_{V}$, which we parametrise --- after a choice of basis $(e_i)_{i \in I}$ of $V$ --- in the form
\beq
\label{Lpreform} L_i = M^i_{a} \hbar \partial_{a} + N^i_{a} x_a - \bigg(\tfrac{1}{2}A^i_{a,b}x_ax_b + \hbar B^i_{a,b}x_a\partial_{b} + \tfrac{\hbar^2}{2} C^i_{a,b}\partial_{a}\partial_{b} + \hbar D^i\bigg).
\eeq

\begin{definition}
We say that a linear map $L\,:\,V \rightarrow \mathcal{D}_V$ has normal form 
if $M^i_{j} = \delta_{i,j}$ and $N^i_{j} = 0$. 
\end{definition}
We observe that the condition of normal form is independent of the choice of the basis $(e_i)_{i\in I}$ of $V$.
\begin{definition}
A \emph{quasi-Airy structure} on $V$ is the data of a Lie algebra structure on $V$, together with a homomorphism of Lie algebras $L\,:\,V \rightarrow \mathcal{D}_V$. In this case, we denote $\mathcal{Z}_{L}$ the space of solutions of 
$$
\forall v \in V,\qquad L(v)\cdot Z = 0
$$ 
of the form
$$
Z = \exp\bigg(\sum_{\substack{g \geq 0 \\ n \geq 1}} \frac{\hbar^{g - 1}}{n!}\,\mathcal{F}_{g,n}\bigg),\qquad \mathcal{F}_{g,n} \in {\rm Sym}^n V^*.
$$
Note that the relations imposed by $[L_i,L_j] = f_{i,j}^a L_{a}$ are \textit{a priori} different than those described in Section~\ref{S22} due to the presence of $M$ and $N \neq 0$.
\end{definition}
Tautologically, quasi-Airy structures having normal form are quantum Airy structures.

Let $L\,:\,V \rightarrow \mathcal{D}_V$ be a quasi-Airy structure. In particular, $V$ has the structure of a Lie algebra, and we can consider the adjoint representation ${\rm ad}_{L}\,:\,V \rightarrow {\rm End}(\mathcal{D}_V)$ defined by ${\rm ad}_{L}(v) = \hbar^{-1}[L(v),\cdot]$. Computation with \eqref{Lpreform} shows that ${\rm ad}_{L}$ has a block decomposition with respect to $\mathcal{D}_{V} = \bigoplus_{d = 0}^{2} \mathcal{D}_{V,d}$
$$
{\rm ad}_{L} = \left(\begin{array}{ccc} 0 & \rho_{0,1} & \rho_{0,2} \\ 0 & \rho_1 & \rho_{1,2} \\ 0 & 0 & \rho_2 \end{array}\right).
$$
If we choose a basis of $V$, since $L$ has the form \eqref{Lpreform} the central block has a further decomposition with respect to $\mathcal{D}_{V,1} = V^* \oplus V.\hbar$ of the form
\beq
\label{rho1} \rho_1(e_i) = \left(\begin{array}{cc} -B^i & A^i \\ -C^i & (B^i)^{T} \end{array}\right),
\eeq
where $X^i = (X^i_{j,k})_{j,k}$ is considered as a matrix for $X \in \{A,B,C\}$. Note that \eqref{rho1} is the general form of a matrix $Y$ such that $YS$ is symmetric, $S$ being the symplectic transformation
$$
S = \left(\begin{array}{cc} 0 & - 1 \\ 1 & 0 \end{array}\right).
$$
We note that $\rho_1$ is a representation of $V$ on $T_{\hbar}^*V$ if and only if $[\rho_1(e^i),\rho_1(e^j)] = f_{i,j}^a \rho_1(e^a)$. 
Further, with respect to the same decomposition of $\mathcal{D}_{V,1}$
\beq
\label{rho01ti} \rho_{0,1} = \big( M^i, N^i \big).
\eeq
We also define $\tilde{\rho}_{1}\,:\,V \rightarrow {\rm End}(\mathbb{C}.\hbar \oplus T^*_{\hbar}V)$ by
\beq
\label{rho1ti}\tilde{\rho}_{1} = \left(\begin{array}{cc} 0 & \rho_{0,1} \\ 0 & \rho_{1} \end{array}\right).
\eeq
We observe that if $L$ is of normal form and $V$ is the complexification of a real Lie algebra $V_{{\mathbb R}}$, then $\tilde{\rho}_1$ is the complexification of a real
$$ \tilde{\rho}_1^{{\mathbb R}} : V_{{\mathbb R}} \rightarrow {\rm End}(\mathbb{R}.\hbar \oplus T^*_{\hbar}V_{{\mathbb R}}).$$
This follows immediately since the standard form $(M^i, N^i)$ are in particular real.
An easy computation shows
\begin{lemma}
\label{LLL1}Let $L\,:\,V \rightarrow \mathcal{D}_V$ be a linear map of normal form \eqref{Lform}. Then $\rho_1$ defined by \eqref{rho1} is a representation of $V$ on $T^*_{\hbar}V$ if and only if the \textbf{BB-CA}, \textbf{BC} and \textbf{BA} relations are satisfied.
\end{lemma}
\begin{lemma}
\label{LLLfdsf} Let $L\,:\,V \rightarrow \mathcal{D}_V$ be a linear map of normal form, such that $\rho_1$ is a representation of $V$ on $T^*_{\hbar}V$. The three properties below are equivalent.
\begin{itemize}
\item[$(i)$] $\tilde{\rho}_{1}$ is a representation of $V$ on $\mathbb{C}.\hbar \oplus T^*_{\hbar} V$.
\item[$(ii)$] the two extra relations \eqref{f2B} and \eqref{SymA} are satisfied.
\item[$(iii)$] if $\mathcal{I}\,:\,V \rightarrow T^*_{\hbar} V$ is the natural inclusion, the representation $\rho_{1}$ satisfies
\beq
\label{commuts}\forall v_1,v_2 \in V,\qquad \rho_{1}(v_1)(\mathcal{I}(v_2)) - \rho_{1}(v_2)(\mathcal{I}(v_1)) = \mathcal{I}([v_1,v_2]).
\eeq
\end{itemize} 
\end{lemma}
The only property which may be missing in Lemma~\ref{LLLfdsf} for $L$ to be a quantum Airy structure is the \textbf{D} relation \eqref{Drel} --- compare however with Lemma~\ref{basrem}, $(ii)$. Indeed the $D^i$ contribute to the constant part of the operators, and thus are not seen at the level of the adjoint action. Condition $(iii)$ is very similar to a \emph{torsion-free condition} for connections in vector bundles, which is the reason why we adopted this name to refer to the linear relation \eqref{commuts}.

Remark that, if $(L_i)_{i \in I}$ are operators of normal form with $A = C = 0$, $\rho_1$ is a representation if and only if $\rho_1$ acting on $V$ is a representation --- given by the matrices $(B^i)^T$. In particular, if $(f_{i,j}^k)$ are the structure constants of a Lie algebra structure on $V$, the choice $B^i_{j,k} = \tfrac{1}{2}f_{i,j}^k$ makes $(V,\rho_1^{V})$ (where $\rho_1^{V}$ means $\rho_1$ acting on $V$) the adjoint representation, and $(V,\rho_{1}^{V^*})$ its dual. One then sees that the \textbf{BB-CA} relation is equivalent to the Jacobi relation that $f_{i,j}^k$ indeed satisfy. In this case, $(L_i)_{i \in I}$ is the well-known expression of the adjoint representation by differential operators. It is however not a quantum Airy structure because the first --- and very important --- term $\hbar\partial_i$ is missing. Or, equivalently, this representation $\rho_{1}$ in general violates the torsion-free condition \eqref{commuts}.

\subsection{Classical Airy structures from symplectic representations}
\label{SSymp}

This paragraph is a digression to relate Section~\ref{S32} to the notion of \emph{classical Airy structure} introduced in \cite{KSTR}, which is the Poisson analog of Definition~\ref{Def:QAiry}. Let $\mathcal{P}_{V,d}$ be the space of polynomial functions on $T^*V$ of degree $d$, and
$$
\mathcal{P}_{V} = \bigoplus_{d = 0}^2 \mathcal{P}_{V,d}
$$
equipped with its canonical Poisson bracket. We denote $p_{V^*}\,:\,\mathcal{P}_{V} \rightarrow V^* \subset \mathcal{P}_{V,1}$ the natural projection.

\begin{definition}
\label{clAiry}A \emph{classical Airy structure} is the data of a linear map $\Lambda\,:\,V \rightarrow \mathcal{P}_{V,1} \oplus \mathcal{P}_{V,2}$ such that $\mathsf{I} := p_{V^*}\circ \Lambda$ is an isomorphism, and such that ${\rm Im}\,\Lambda$ is closed under Poisson bracket.
\end{definition}

If $(e_i)_{i \in I}$ is a basis of $V$, we denote $(x_i)_{i \in I}$ and $(y_i)_{i \in I}$ the corresponding linear coordinates on $T^*V$. By defining $\Lambda_i:= I^{-1}_{i,a}\Lambda(e_a)$, a classical Airy structure is uniquely determined by a family $(\Lambda_i)_{i \in I}$ of elements of $\mathcal{P}_{V}$ of the form
$$
\Lambda_i = y_i  - \tfrac{1}{2}A^i_{a,b}x_ax_b - B^i_{a,b}x_ay_b - \tfrac{1}{2}C^i_{a,b}y_ay_b,
$$
such that
\beq
\label{Poissoncom} \{\Lambda_i,\Lambda_j\} = f_{i,j}^a\Lambda_{a}
\eeq
for some scalars $f_{i,j}^k$. Note that \eqref{Poissoncom} implies that $\{(x,y) \in T^*V\,\,:\,\,\forall i \in I \quad \Lambda_i (x,y)= 0\}$ is Lagrangian subvariety (at least near $0$) in $T^*V$. This Lagrangian for non-zero $(A,B,C)$ is a perturbation of the canonical Lagrangian given by the zero section $\{(x,y) \in T^*V\,\,:\,\,\forall i \in I \quad y_i = 0\} \subset T^*V$.

The data of such $(\Lambda_i)_{i \in I}$ is the definition adopted by \cite{KSTR} for classical Airy structure. Here, we prefer to adopt the basis-independent Definition~\ref{clAiry}. The analysis we did for quantum Airy structures in the realm of the Weyl algebra can be transposed directly for classical Airy structures in the Poisson realm.
\begin{proposition}
\label{clasicProp}The data $(A,B,C)$ defines a classical Airy structure if and only if it satisfies \textbf{BB-CA}, \textbf{BC}, \textbf{BA} relations, as well as the torsion-free conditions \eqref{f2B}-\eqref{SymA}. This is also equivalent (Lemma~\ref{LLLfdsf}) to requiring that $\tilde{\varrho}_{1}$ is a representation of the Lie algebra $V$ defined by the structure constants $f_{i,j}^k$ in the basis $(e_i)_{i \in I}$. \hfill $\Box$
\end{proposition}
In other words, quantum Airy structures are just classical Airy structures together with the data $D$ satisfying the affine relation \eqref{Drel}.

This allows a more geometric perspective on Airy structures, which we are going to explain again but in a slightly different way. The keypoint, which is also put forward and exploited in \cite{KSTR}, is that suitable family of infinitesimal symplectomorphisms in $T^*V$ give rise to classical Airy structures.

Let us fix a Lie algebra structure on $V$. We then consider the vector space $W = V \oplus V^*$ with the canonical symplectic structure $\omega$, \textit{i.e.} $\omega((v_1,l_1),(v_2,l_2)) = l_2(v_1) - l_1(v_2)$. Let $ \pi_V : W \ra V$ be the projection along $V^*$. We denote $\mathfrak{sp}(W)$ the Lie algebra of infinitesimal symplectomorphisms of $W$. Combining Proposition~\ref{clasicProp} with Lemma~\ref{LLLfdsf} we find

\begin{lemma}
There is a one to one correspondence between classical Airy structures, and Lie algebra homomorphisms $\mathcal{L}\,:\,V \rightarrow \mathfrak{sp}(T^*V)$ together with a Lagrangian embedding $\mathcal{I}\,:\,V \rightarrow T^*V$ such that
\be\label{Con1}
\forall v_1,v_2 \in V,\qquad {\cal L}_{v_1} \mathcal{I}(v_2) - {\cal L}_{v_2}\mathcal{I}(v_1) = \mathcal{I}([v_1,v_2]).
\ee
The coefficients of the classical Airy structure are given in terms of the symplectic form and the action on $W$ of the linear symplectomorphisms determined by $\mathcal{L}$ in \eqref{ABCs} below.
\end{lemma}
The advantage of this formulation is that given a Lie algebra homomorphism of $V$ into $\mathfrak{sp}(T^*V)$, for instance coming from geometry, then one only need to construct $\mathcal{I}$ such that the \emph{linear} condition \eqref{Con1} is satisfied. We also remark that the Lie algebra structure on $V$ is completely specified by the $\mathcal{L}$ \textit{via} \eqref{Con1}.
 
\noindent {\bf Proof}. First we will assume that we are given a Lie algebra homomorphism $\mathcal{L}\,:\,V \rightarrow \mathfrak{sp}(T^*V)$ as above,  and from this we will construct a classical Airy structure. We pick a basis $(e_i)_{i \in I}$ of $V$. We let $e_i^*$ be the dual basis of $e_i$, then $((0,e_i^*),(e_i,0))_{i \in I}$ is a symplectic basis of $T^*V = W$, \textit{i.e.}
$$
\forall i,j \in I,\qquad \omega(e_i,e_j) = \omega(e_i^*,e_j^*) = 0,\qquad \omega(e_i,e_j^*) = \delta_{i,j} .
$$
We let $f_{i,j}^k$ be the structure constants of $V$ in the basis $(e_i)_{i \in I}$ of $V$. One then has the decomposition
$$
\forall w \in W,\qquad w =  -x_a e_a + y_a e_a^*,\qquad x_i = \omega(e_i^*,w),\qquad y_i = \omega(e_i,w) .
$$
Denoting ${\cal L}_i := {\cal L}_{e_i}$, we have by definition of a symplectic representation
$$
\forall (w_1,w_2) \in W^2,\qquad \omega({\cal L}_i w_1,w_2) + \omega(w_1, {\cal L}_i  w_2)  = 0.
$$
Hence ${\cal L}_i$ is represented in the symplectic basis of $T^*V \cong V^* \oplus V$ by the matrix
$$
H_i = \left(\begin{array}{cc}
-B^i & A^i \cr
C^i & (B^i)^{T} \cr
\end{array}
\right)
$$
where $A^i$ and $C^i$ are symmetric matrices with
\beq
\label{ABCs} A^i_{j,k} = -\omega(\mathcal{L}_ie_j,e_k),\qquad B^i_{j,k} = \omega(\mathcal{L}_{i}e_j,e_k^*),\qquad C^i_{j,k} = -\omega(\mathcal{L}_{i}e_j^*,e_k^*).
\eeq
With this choice of $(A,B,C)$, one defines the hamiltonians
$$
h_i(w) := \tfrac{1}{2}\omega(\mathcal{L}_{i}w,w) = -\Big(\tfrac{1}{2} A_{a,b}^i x_a x_b + B_{a,b}^i x_a y_b + \tfrac{1}{2} C_{a,b}^i y_a y_b\Big)
$$
for $w = -x_a e_a + y_a e_a^*$. Then, we can define $\Lambda\,:\,e_i \rightarrow (w \mapsto \omega(e_i,w) + h_i(w))$, and we claim this is a classical Airy structure. Indeed, one can check using the commutation relations for $\mathcal{L}$ that $\{w \in W\,\,:\,\,\forall i\,\,\,\ \omega(e_i,w) + h_i(w) = 0\}$ is a Lagrangian subvariety of $T^*V$ near $0$.

Conversely, a classical Airy structure gives the desired representation by just using \eqref{ABCs} to define ${\cal L}_i$ in terms of $(A,B,C)$. \hfill $\Box$

\section{Moduli spaces of Airy structures}
\label{SS4}

\subsection{Group action}

\label{S31}
The affine extended symplectic group $\mathcal{G}_{V} := \exp(\mathcal{D}_{V}/\hbar)$ acts by conjugation on its Lie algebra $\mathcal{D}_{V}$, hence inducing an action on the set of quasi-Airy structures, as well as on the partition function $Z$
$$
\mathcal{U} \in \mathcal{G}_{V},\qquad \tilde{L}_i = \mathcal{U}L_i\mathcal{U}^{-1},\qquad \tilde{Z} = \mathcal{U}\cdot Z.
$$
It contains --- and is generated by --- the Heisenberg subgroup $\exp\big((\mathcal{D}_{V,0} + \mathcal{D}_{V,1})/\hbar\big)$, the metaplectic group $\exp(\mathcal{D}_{V,2}/\hbar)$. This perspective makes it clear that we ought to study (quasi-)Airy structures only up to the action of ${\mathcal{G}}_{V}$.

Computations show that the subgroup of $\mathcal{G}_{V}$ which preserves the normal form \eqref{Lform} of quantum Airy structures only consists of multiplication by scalars (which are central and do not change the $L_i$), renormalisation $(A,B,C,D) \rightarrow (\lambda^3 A,\lambda B,\lambda^{-1}C,\lambda D)$ for $\lambda \in \mathbb{C}^*$ (which can be realised via $\mathcal{U} = \lambda^{x_a\partial_{a}}$) and differential operators of order two
\beq
\label{Udiff2} \mathcal{U} = \exp\big( \tfrac{\hbar}{2} u_{a,b} \partial_{a}\partial_{b}\big),\qquad u_{i,j} = u_{j,i} \in \mathbb{C}.
\eeq
If $(A,B,C,D)$ are the coefficients of a quantum Airy structure, the new quantum Airy structure obtained by the action of \eqref{Udiff2} has coefficients $(\tilde{A},\tilde{B},\tilde{C},\tilde{D})$ given by
\bea
\tilde{A}^i_{j,k} & = & A^i_{j,k}. \nonumber \\
\tilde{B}^i_{j,k} & = & B^i_{j,k} + u_{k,a}A^i_{j,a}.  \nonumber \\
\tilde{C}^i_{j,k} & = & C^i_{j,k} + u_{j,a}B^i_{a,k} + u_{k,a}B^i_{a,j} + u_{j,a}u_{k,b}A^i_{a,b}. \nonumber \\
 \label{SymU} \tilde{D}^i & = & D^i + \tfrac{1}{2}u_{a,b}A^i_{a,b}.
\eea
We have just proved
\begin{corollary}
If $(u_{i,j})_{i,j \in I}$ is a symmetric matrix, \eqref{SymU} is a symmetry of the relations of Section~\ref{S22}.
\end{corollary}
This can also be checked directly by inserting \eqref{SymU} in the relations. At the level of the partition functions, if ${\rm u} = (u_{i,j})_{i,j}$ is invertible, Wick's theorem shows that the action of \eqref{Udiff2} can be realised by a formal Gaussian convolution
$$
\exp\Big(\tfrac{\hbar}{2} u_{a,b} \partial_{a}\partial_{b} \Big)\cdot Z (x) = \int_{V} \frac{\mathbf{\dd} \xi}{\det(2\pi \hbar\,{\rm id})^{1/2}}\,\exp\bigg(-\frac{({\rm u}^{-1})_{a,b}\xi_a\xi_b}{2\hbar}\bigg)\,Z(x + \xi),
$$
where $\mathbb{\dd} \xi = \prod_{i \in I} \dd\xi_i$ is the Lebesgue measure on $V$.

\begin{remark}
Another easy transformation of the $L_i$ is the rescaling of $\hbar$. It transforms $L_i$ into $z^{-1}\exp(z\hbar\partial_{\hbar})L_i$, and thus $(A,B,C,D)$ into $(z^{-1}A,B,zC,D)$. We prefer not to include it in $\mathcal{G}_{V}$.
\end{remark}

\begin{lemma}
\label{pretoAiry} If $L$ is a quasi-Airy structure, its $\mathcal{G}_{V}$-orbit contains a quantum Airy structure if and only if $(\pi_1 \oplus \pi_1^*) \circ L\,:\,V \rightarrow T^*_{\hbar}V$ is a (linear) Lagrangian embedding of $V$.
\end{lemma}
\noindent\textbf{Proof.} If $L = (L_i)_{i \in I}$ is a quantum Airy structure, then $\pi_1^* \circ L$ is the isomorphism between $V$ and $V^*$ induced by the choice of a basis $(e_i)_{i \in I}$. In particular, $(\pi_1 \oplus \pi_1^*) \circ L$ is a Lagrangian embedding of $V$ into $T^*_{\hbar}V$. These properties remain true for $\tilde{L}$ in the $\mathcal{G}_{V}$-orbit of $L$. Conversely, let $L$ be a quasi-Airy structure such that $(\pi_1 \oplus \pi_1^*) \circ L\,:\,V \rightarrow T^*_{\hbar}V$ is a Lagrangian embedding. We can always compose it with a linear symplectomorphism of $T_{\hbar}^*V$ bringing this Lagrangian to $V^*$. It is well-known that $\mathcal{G}_{V}$ contains elements which can realise as automorphisms of the Weyl algebra the linear symplectomorphisms
$$
\tilde{x}_i = \alpha_{i,a}x_a + \beta_{i,a} \hbar\partial_{a},\qquad \hbar\tilde{\partial}_i = \gamma_{i,a}x_{a} + \epsilon_{i,a}\hbar\partial_{a}
$$
for arbitrary matrices $(\alpha,\beta,\gamma,\varepsilon)$ satisfying the symplectic conditions
$$
\left\{\begin{array}{l} \alpha\beta^{T}\,\,{\rm symmetric,} \\ (\gamma\epsilon^{T})\,\,{\rm symmetric,} \\ \alpha\epsilon^{T} - \beta\gamma^{T} = {\rm id}_{V}. \end{array}\right. 
$$
Therefore, we can find $\tilde{L}$ in the $\mathcal{G}_{V}$-orbit of $L$ such that $\pi_1 \circ L = 0$ and $\pi_1^* \circ L$ is an isomorphism. In other words, in a given basis, $\tilde{L}_i$ has the form \eqref{Lpreform} with $\tilde{N}^i_{j} = 0$ and ${\rm M} = (\tilde{M}^i_{j})_{i,j}$ invertible. So, putting $\check{L}_i = ({\rm M}^{-1})_{i,a}\tilde{L}_{a}$ gives an operator in normal form, \textit{i.e.} a quantum Airy structure. \hfill $\Box$

\begin{corollary}
A quasi-Airy structure $(L_i)_{i \in I}$ has a quantum Airy structure in its $\mathcal{G}_{V}$-orbit if and only if $\rho_{1}$ is a representation of the Lie algebra $V$ into $T_{\hbar}^*V$, and $\tilde{\rho}_{1}$ is a one-dimensional extension of this representation such that $v \mapsto (v,\rho_{0,1}(v))$ is a Lagrangian embedding of $V$ into $T^*_{\hbar}V$.
\end{corollary}
\textbf{Proof.} Comparing Lemmas~\ref{LLL1}-\ref{LLLfdsf} with the relations found in Section~\ref{S22} shows that $(L_i)_{i \in I}$ of normal form is a quantum Airy structure if and only if $\rho_1$ is a representation of $V$ in $T^*_{\hbar}V$ and $\tilde{\rho}_{1}$ is a one-dimensional extension of this representation such that ${\rho}_{0,1} = (\psi_{V*},0_{V})$ where $\psi_{V*}\,:\,V \rightarrow V^*$ is the isomorphism determined by the choice of basis in which $(L_i)_{i \in I}$ is defined. In this case, $\tilde{\rho}_{1}$ determines an exact sequence of $V$-modules:
$$
0 \longrightarrow \mathbb{C} \longrightarrow \mathbb{C}.\hbar \oplus T^*_{\hbar}V \longrightarrow (\mathbb{C}.\hbar \oplus T^*_{\hbar}V)/\mathbb{C} \longrightarrow 0,
$$
If $L_i \in \mathcal{D}$ does not have normal form, the block $\rho_{0,1}$ nevertheless gives the map $(\pi_1^* \oplus \pi_1) \circ L$, and the general claim is a consequence of Lemma~\ref{pretoAiry}. \hfill $\Box$

\subsection{Definition of moduli spaces}
\label{SModspace}
\label{S3def}

We shall now introduce various moduli spaces associated to Airy structures.

Let $V$ be a (finite-dimensional) Lie algebra. We denote $\mathfrak{A}_{V}^{{\rm cl}}$ (resp. $\mathfrak{A}_{V}$, $\mathfrak{A}_{V}^{{\rm q}}$) the set of classical (resp quasi-, quantum) Airy structures based on the Lie algebra $V$. Of course, each of them is a subset of the set of all Airy structures where we also vary the Lie algebra structure on the vector space $V$. However, we restrict here to study the set of Airy structures based on a fixed Lie algebra.
 
As a subset of $\mathbb{A}_{V} := ({\rm Sym}^3 V^*) \times (V^* \otimes V^* \otimes V) \times ({\rm Sym}^2 V^* \otimes V)$ cut out by the (finitely many) quadratic \textbf{BB-CA}, \textbf{BC} and \textbf{BA} relations and the linear relation \eqref{f2B} (we have included the relation \eqref{SymA} in the definition of $\mathbb{A}_{V}$), $\mathfrak{A}_{V}^{{\rm cl}}$ naturally has the structure of an affine algebraic variety. Likewise, $\mathfrak{A}_{V}$ is an affine algebraic variety. $\mathfrak{A}_{V}^{{\rm q}}$ is obviously a subvariety of $\mathfrak{A}_{V}$. It can also be seen as a subvariety of $V^* \times \mathbb{A}_{V}$ cut out by the extra \textbf{D} relation, where $D^i$ are the coordinates in the first factor. In fact, as the \textbf{D} relation is affine, it arises as the total space of an affine subbundle $\pi\,:\,\mathfrak{A}_{V}^{{\rm q}} \rightarrow \mathfrak{A}_{V}^{{\rm cl}}$ of the trivial vector bundle $V^* \times \mathbb{A}_{V} \rightarrow \mathbb{A}_{V}$ restricted to $\mathfrak{A}_{V}^{{\rm cl}}$. According to Lemma~\ref{basrem}, $\pi$ has a section given by 
$$
D_{{\rm ref}}^i := \tfrac{1}{2}\,{\rm Tr}\,B^i.
$$
Mapping $D$ to $(D - D_{{\rm ref}})$  turns $\mathfrak{A}_{V}^{{\rm q}}$ into a trivial vector bundle over $\mathfrak{A}_{V}^{{\rm cl}}$, with fiber
$$
(V')^{\bot} = \big\{\varphi \in V^*\,\,:\,\,\forall x,y \in V,\quad \varphi([x,y]) = 0\big\}.
$$

As we saw in Section~\ref{S31}, $\mathcal{G}_{V}$ acts algebraically on $\mathfrak{A}_{V}^{{\rm cl}}$ and $\mathfrak{A}_{V}$, and its algebraic subgroup $\tilde{\mathcal{G}}_{V}$ preserving Airy structures acts on $\mathfrak{A}_{V}^{{\rm q}}$. Since we are interested in Airy structures up to the action of $\mathcal{G}_{V}$, the appropriate algebraic way to proceed would be to consider the quotient stack or the GIT quotients $\mathfrak{A}_{V}^{{\rm cl}}/\!/\mathcal{G}_{V}$ and $\mathfrak{A}_{V}/\!/\mathcal{G}_{V}$. In this paper we shall just consider the set-theoretic quotients
$\mathfrak{M}_{V}^{{\rm cl}}  = \mathfrak{A}_{V}^{{\rm cl}}/\mathcal{G}_{V}$ and  $\mathfrak{M}_{V}=  \mathfrak{A}_{V}/\mathcal{G}_{V}$ with the induced topology from $\mathfrak{A}_{V}^{{\rm cl}}$ and $\mathfrak{A}_{V}$ respectively, and present some preliminary remarks about this quotient space, which we will call the moduli space of quasi-Airy structures. We caution the reader that this quotient space will in general not even be Hausdorff. The same comment applies to the moduli space of quantum Airy structures $\mathfrak{M}_{V}^{{\rm q}} = \mathfrak{A}_{V}^{{\rm q}}/\tilde{\mathcal{G}}_{V}$. The fibration $\pi\,:\,\mathfrak{A}_{V}^{{\rm q}} \rightarrow \mathfrak{A}_{V}^{{\rm cl}}$ is $\tilde{\mathcal{G}}_{V}$-equivariant, therefore we have a natural fibration $\mathfrak{M}_{V}^{{\rm q}} \rightarrow \mathfrak{M}_{V}$. The description of its fibers can in principle be obtained by looking at the action of $\mathcal{G}_{V}$ on $D$ via \eqref{SymU}.

\subsection{Deformation theory}

As the moduli space of classical Airy structures just consists of Lie algebra homomorphisms modulo inner automorphisms of the target Lie algebra $\mathcal{D}_{V}$, one can use the theory of deformation of algebraic structures to study their moduli space. The deformations of the Lie algebra homomorphism $L\,:\,V \rightarrow \mathcal{D}_{V}$ are governed by the differential graded algebra $E_{L}^{\bullet} := {\rm Hom}(\Lambda^{\bullet}V,\mathcal{D}_{V})$, which is equipped with the Cartan--Eilenberg differential
\begin{equation*}
\begin{split}
& \quad (\dd_{E_{L}^{\bullet}} \phi)(v_1\wedge \cdots v_{n + 1}) \\
& := \sum_{\ell = 1}^{n + 1} (-1)^{\ell + 1} \phi(v_1 \wedge \cdots \widehat{v_{\ell}} \cdots \wedge v_{n + 1}) L(v_{\ell}) + \sum_{\ell < m} (-1)^{\ell + m}\phi\big([v_k,v_{\ell}]_{V} \wedge v_1 \wedge \cdots \widehat{v_k} \cdots \widehat{v_{\ell}} \cdots \wedge v_{n + 1} \big) 
\end{split}
\end{equation*}
for $\varphi \in E_{L}^n$ and the bracket
$$
[\phi,\psi]_{E_{L}^{\bullet}}(v_1 \wedge \cdots \wedge v_{n + m}) := \sum_{\substack{\sigma \in \mathfrak{S}_{n + m}  \\ \sigma(1) < \cdots < \sigma(m) \\ \sigma(m + 1) < \cdots < \sigma(m + n)}} \varepsilon(\sigma) \big[\phi(v_{\sigma(1)}\wedge \cdots \wedge v_{\sigma(m)}),\psi(v_{\sigma(m + 1)}\wedge \cdots \wedge v_{\sigma(m + n)})\big]_{\mathcal{D}}
$$
for $\phi \in E^m_{L}$ and $\psi \in E^n_{L}$ where $v_1\wedge \cdots \wedge \widehat{v_{\ell}} \wedge \cdots \wedge v_n$ stands for  $v_1\wedge \cdots \wedge v_n$ with $v_{\ell}$ removed and  $[\cdot,\cdot]_{X}$ stands for the bracket in $X$. We denote $H_{L}^i(V,\mathcal{D}_{V})$ the cohomology of this complex, and $Z_{L}^{n}(V,\mathcal{D}_{V})$ the space of $n$-cocycles. In particular we get a quadratic map
$$\Omega \,:\, H^1_L(V,{\mathcal D}_{V}) \ra H^2_L(V,{\mathcal D}_{V}),$$
given by
$$\Omega(\phi) = [\phi,\phi]_{E^\bullet_L}.$$

We get the following proposition as a direct consequence of the results of \cite{Nijenhuis}. 

\begin{proposition}
\label{THDEF1}Let $L\, :\, V \ra \mathcal{D}_{V}$ be a classical Airy structure. If $H_{L}^1(V,\mathcal{D}_{V}) = 0$, then $L$ is rigid, \textit{i.e.} all continuous deformations of $L$ remain in its $\mathcal{G}_{V}$-orbit. In general there exists an open neighbourhood $U$ of zero in  $H_{L}^1(V,\mathcal{D}_{V})$ and an open neighbourhood $\mathfrak{U}$ of $[L]\in \mathfrak{M}^{{\rm cl}}_{V}$ such that 
$$ \mathfrak{U} \cong \left(U \cap \Omega^{-1}(0)\right)/\mathcal{G}_{V}(L)$$
where $\mathcal{G}_{V}(L)$ is the stabiliser of $L$ in $\mathcal{G}_{V}$.
\end{proposition}

\begin{corollary}
\label{THDEF2}If we have that $H^2_L(V,{\mathcal D}) = 0$ and that the action of $\mathcal{G}_{V}(L)$ on $H_{L}^1(V,\mathcal{D})$ factors through a finite group, then $M$ will have the structure of an orbifold near $[L]\in \mathfrak{M}_{V}^{{\rm cl}}$ and
$$ T_{[L]}\mathfrak{M}^{{\rm cl}}_{V} \cong H_{L}^1(V,\mathcal{D}_{V}).$$
\end{corollary}

We have seen by explicit computations that $\tilde{\varrho}_{1}\,:\,V \rightarrow {\rm End}(\mathbb{C}.\hbar \oplus T^*_{\hbar}V)$ is a Lie algebra homomorphism if and only if $\varrho\,:\,V \rightarrow \mathcal{D}_{V}$ is a Lie algebra homomorphism. So, we can in fact replace in Proposition~\ref{THDEF1} and Corollary~\ref{THDEF2} the module $\mathcal{D}_{V}$ by the module $\mathbb{C}.\hbar \oplus T^*_{\hbar}V$. However, if we only used the module $T^*_{\hbar}V$, we would miss the constraints imposed the torsion-free condition \eqref{commuts}.

\subsection{Translations}
\label{S34}
So far, the symmetries we have described act linearly on the coefficients of quantum Airy structures. Among them, translations $x_i \rightarrow x_i + t_i$ transform a quasi-Airy structure into operators $(\tilde{L}_i)_{i \in I}$ such that $(\tilde{L}^i - P^i)_{i \in I}$ is a quasi-Airy structure, for some constants $P^i$. The solution of $\tilde{L}_i\cdot \tilde{Z} = 0$ for all $i$  is $Z(t + x)$, which is the Taylor expansion of $Z(\xi)$ around the point $\xi = t$. If we write $Z(x) = \exp\big(\sum_{g \geq 0} \hbar^{g - 1}S_{g}(x)\big)$, and assume momentarily that $S_g$ has a non-zero radius of convergence uniformly in $g$,
$$
Z(t + x) = \exp\Big(\sum_{g \geq 0} \sum_{n \geq 1} \sum_{i_1,\ldots,i_n} \frac{\hbar^{g - 1}}{n!}\,F_{g,n}^{(t)}(i_1,\ldots,i_n)x_{i_1}\cdots x_{i_n}\Big),
$$
where now $F_{0,1}^{(t)}(i) = \partial_{i}S_{0}(t)$ and $F_{0,2}^{(t)}(i,j) = \partial_{i}\partial_{j}S_{0}(t)$ are a priori now zero. It is natural to suspect that
$$
\tilde{Z}(x) := \exp\big(-\tfrac{F_{0,1}^{(t)}(a)x_a}{\hbar} - \tfrac{F_{0,2}^{(t)}x_ax_b}{2\hbar}\big)Z(x + t)
$$
is the partition function of a new, $t$-dependent quantum Airy structure. The next theorem will confirm and make sense of this, using formal series in $t$. 

We introduce the graded vector space $\hat{K} = \mathbb{K}[\![(t_i)_{i \in I}]\!]$, by assigning degree $1$ to each $t_i$. We get the decomposition into homogeneous pieces
$$
\hat{K} = \prod_{n \geq 0} \hat{K}^{[n]}.
$$
Let $(A,B,C,D)$ be a quantum Airy structure. We first describe a formal replacement for $\partial_{i}S_{0}(t)$ and $\partial_{i}\partial_{j}S_0(t)$. These are elements of $\hat{K}$ whose homogeneous components are inductively defined
$$
G_{0,1}(i) = \sum_{n \geq 2} G_{0,1}^{[n]}(i),\qquad G_{0,2}(i,j) = \sum_{n \geq 1} G_{0,2}^{[n]}(i,j),
$$
with
\bea
\label{G01ini}G_{0,1}^{[2]}(i) & = & \tfrac{1}{2}A^i_{a,b}t_{a}t_{b}\,, \\
\label{G01rec} \forall n \geq 3,\quad G_{0,1}^{[n]}(i) & = & B^i_{a,b}t_aG_{0,1}^{[n - 1]}(b) + \tfrac{1}{2}C^i_{a,b}\sum_{\substack{n_1 + n_2 = n \\ n_1,n_2 \geq 2}} G_{0,1}^{[n_1]}(a)G_{0,1}^{[n_2]}(b)\,,
\eea
and
\bea
\label{G02ini} G_{0,2}^{[1]}(i,j) & = & A^i_{j,a}t_{a}\,, \\
\label{G02rec} \forall n \geq 2,\quad G_{0,2}^{[n]}(i,j) & = & B^i_{a,b}t_{a}G_{0,2}^{[n - 1]}(b,j) + B^i_{j,a}G_{0,1}^{[n]}(a) + C^i_{a,b} \sum_{\substack{n_1 + n_2 = n \\ n_1 \geq 2,\,\,n_2 \geq 1}} G_{0,1}^{[n_1]}(a)G_{0,2}^{[n_2]}(b,j)\,.
\eea
Then, we define $(\tilde{A},\tilde{B},\tilde{C},\tilde{D})$ with coefficients in $\hat{K}$, again inductively by their homogeneous components. The initial conditions are
\beq
\label{iniform} \tilde{X}^{[0]} = X,\qquad X \in \{A,B,C,D\}
\eeq
and the recursions read for $n \geq 1$
\bea
\label{Arecform} (\tilde{A}^{[n]})^{i}_{j,k} & = & B^i_{a,b}(\tilde{A}^{[n - 1]})^{b}_{j,k}t_{a} + B^i_{j,a}G_{0,2}^{[n]}(a,k) + B^i_{k,a}G_{0,2}^{[n]}(a,j) \\
&& +  C^i_{a,b}\bigg(\sum_{\substack{n_1 + n_2 = n \\ n_1 \geq 2\,\,n_2 \geq 0}} G_{0,1}^{[n_1]}(a)(\tilde{A}^{[n_2]})^{b}_{j,k} +  \sum_{\substack{n_1 + n_2 = n \\ n_1,n_2 \geq 1}} G_{0,2}^{[n_1]}(a,j)G_{0,2}^{[n_2]}(b,k)\bigg). \\
\label{Brecform} (\tilde{B}^{[n]})^i_{j,k} & = & B^i_{a,b}(\tilde{B}^{[n - 1]})^{b}_{j,k}t_{a} + C^i_{a,k}G_{0,2}^{[n]}(a,j) +  \sum_{\substack{n_1 + n_2 = n \\ n_1 \geq 2,\,\,n_2 \geq 0}} C^i_{a,b}G_{0,1}^{[n_1]}(a)(\tilde{B}^{[n_2]})_{j,k}^b.  \\
\label{Crecform} (\tilde{C}^{[n]})^i_{j,k} & = & B^i_{a,b}(\tilde{C}^{[n - 1]})^b_{j,k}t_a + \sum_{\substack{n_1 + n_2 = n \\ n_1 \geq 2,\,\,n_2 \geq 0}} C^i_{a,b}G_{0,1}^{[n_1]}(a)(\tilde{C}^{[n_2]})^b_{j,k}. \\
\label{Drecform} (\tilde{D}^{[n]})^i & = & B^i_{a,b}(\tilde{D}^{[n - 1]})^bt_a + C^i_{a,b} \sum_{\substack{n_1 + n_2 = n \\ n_1 \geq 2,\,\,n_2 \geq 0}} G_{0,1}^{[n_1]}(a)(\tilde{D}^{[n_2]})^b. 
\eea

\begin{proposition}
If $(A,B,C,D)$ is a quantum Airy structure whose partition function has Taylor coefficients $F_{g,n}(i_1,\ldots,i_n)$, then $(\tilde{A},\tilde{B},\tilde{C},\tilde{D})$ is a quantum Airy structure with coefficients in $\hat{K}$ whose partition function has Taylor coefficients
\beq
\label{Fgnshift} \tilde{F}_{g,n}(i_1,\ldots,i_n) = \sum_{m \geq 0} \frac{1}{m!} \sum_{j_1,\ldots,j_m} F_{g,n+m}(i_1,\ldots,i_n,j_1,\ldots,j_m)t_{j_1}\cdots t_{j_m} \in \hat{K}.
\eeq
\end{proposition}

The above formulas form a non-linear\footnote{We mean that $\tilde{X}$ does not depend linearly of $X$.}, infinitesimal symmetry of quantum Airy structures. In the convergent case, this symmetry is the one expected.

\begin{lemma}
Assume $V$ finite-dimensional. Let $Z = \exp\big(\sum_{g \geq 0} \hbar^{g - 1}S_{g}(x)\big)$ be the partition function of a quantum Airy structure, where $S_g \in \mathbb{K}[\![(x_i)_{i \in I}]\!]$. If $S_0$ has positive radius of convergence, then for any $g \geq 0$, $S_{g}$ has a radius of convergence bounded from below by a positive constant independent of $g$. We denote $\underline{S}_g$ the analytic function defined by those series at least in a neighbourhood of $0$ in $V$. The formal series defining $(\tilde{A},\tilde{B},\tilde{C},\tilde{D},G_{0,1},G_{0,2})$ have positive radius of convergence, and we also use underlined letters to denote the analytic functions of $t$ they define. Then
$$
\underline{G}_{0,1}(t;i) = \partial_{i} \underline{S}_0(t),\qquad \underline{G}_{0,2}(t;i,j) = \partial_{i}\partial_{j} \underline{S}_{0}(t),\qquad \underline{A}^i_{j,k}(t) = \partial_i\partial_j\partial_k\underline{S}_0(t),\qquad \underline{D}^i_{j,k}(t) = \partial_{i} \underline{S}_{1}(t).
$$
Further, for $x$ and $t$ in a neighbourhood of $0$,
$$
\underline{\tilde{S}}_{g}(t;x) = \underline{S}_{g}(x + t).
$$
\end{lemma}
\begin{remark} The results of \cite{RubaAiry} show that the assumption "$S_0$ has positive radius of convergence" is automatically satisfied.
\end{remark}

\noindent \textbf{Proof.} A computation shows that the translation $x \rightarrow x + t$, followed by conjugation by
$$
\mathcal{U} = \exp\Big(\hbar^{-1}\big(G_{0,1}(a)x_a + \tfrac{1}{2}G_{0,2}(a,b)x_ax_b\big)\Big)
$$
transforms the quantum Airy structure $(A,B,C,D)$ into the quasi-Airy structure $(\hat{L}_i - \hat{P}^i)_{i \in I}$ with coefficients (see \eqref{Lpreform} for notations, we replaced here $M^i_{j}$ by $M_{i,j}$ and $N^i_{j}$ by $N_{i,j}$ for convenience)
\bea
\hat{M}_{i,j} & = & \delta_{i,j} - B^i_{a,j}t_a - C^i_{j,a}G_{0,1}(a). \nonumber \\
\hat{N}_{i,j} & = & -A^i_{j,a}t_{a} + (\delta_{i,b} - B^i_{a,b}t_a - C^i_{a,b}G_{0,1}(a))G_{0,2}(b,j) - B^i_{j,a}G_{0,1}(a). \nonumber \\
\hat{P}^i & = & -\tfrac{1}{2}A^i_{a,b}t_{a}t_{b} - \tfrac{1}{2}C^i_{a,b}G_{0,1}(a)G_{0,1}(b) + (\delta_{i,b} - B^i_{a,b}t_{a})G_{0,1}(b). \nonumber \\
\hat{A}^i_{j,k} & = & A^i_{j,k} + B^i_{j,a}G_{0,2}(a,k) + B^i_{k,a}G_{0,2}(a,j) + C^i_{a,b}G_{0,2}(a,j)G_{0,2}(b,k). \nonumber \\
\hat{B}^i_{j,k} & = & B^i_{j,k} + C^i_{a,k}G_{0,2}(a,j). \nonumber \\
\hat{C}^i_{j,k} & = & C^i_{j,k}. \nonumber \\
\hat{D}^i & = & D^i + \frac{1}{2} C^i_{a,b}G_{0,2}(a,b). \nonumber
\eea 
We indeed remark that the operations of translation and conjugation by the exponential of a quadratic form preserve the Lie commutation relations, so $(\hat{L}_i - \hat{P}^i)_{i \in I}$ is indeed a quasi-Airy structure, with same structure constants. This determines a quantum Airy structure provided $\hat{M}$ is an invertible matrix, and provided one can choose $\hat{N}_{i,j} = \hat{P}^i = 0$ for all $i,j$. In this case, the quantum Airy structure is $\tilde{L}_i = (\hat{M}^{-1})_{i,a}\hat{L}_a$, \textit{i.e.} its coefficients are $\tilde{X}^i = (\hat{M}^{-1})_{i,a}\hat{X}^a$ for $X \in \{A_{j,k},B_{j,k},C_{j,k},D\}$. 

We can indeed solve the equation $\hat{P}^i = 0$ by choosing $G_{0,1}(i)$  as in \eqref{G01ini}-\eqref{G01rec}. Then, $G_{0,2}(i,j)$ is obtained by solving perturbatively $\hat{N}_{i,j} = 0$, leading to \eqref{G02ini}-\eqref{G02rec}. Inserting these series in the expression of the coefficients of $\tilde{L}_i$
$$
\tilde{X}^i = \hat{M}^{-1}_{i,a}\hat{X}^a,\qquad X \in \{A_{j,k},B_{j,k},C_{j,k},D\}
$$
leads to formulas \eqref{iniform} and \eqref{Arecform}-\eqref{Drecform}. The partition function for this new quantum Airy structure is
$$
\tilde{Z}(x) = \exp\big(- \tfrac{G_{0,1}(a)x_{a}}{\hbar} - \tfrac{G_{0,2}(a,b)x_ax_b}{2\hbar}\big)Z(x + t)
$$
and by consistency, we deduce that
\beq
\label{inininini}G_{0,1}(i) = \partial_{i} S_{0}(t),\qquad G_{0,2}(i,j) = \partial_{i}\partial_{j}S_0(t),\qquad \tilde{A}^i_{j,k} = \partial_{i}\partial_{j}\partial_{k} S_0(t),\qquad \tilde{D}^i = \partial_{i} S_1(t)
\eeq
and the Taylor coefficients of $\tilde{Z}$ are given by \eqref{Fgnshift}, both in the sense of formal series in $(t_i)_{i \in I}$. 

Now assume that $V$ is finite-dimensional and $S_0(t)$ has a non-zero radius of convergence. The equation $L_i\cdot Z = 0$ implies for $g \geq 1$
$$
\hat{M}_{i,a}(t)\partial_{a}S_g(t) = \delta_{g,1}D^i + \tfrac{1}{2}C^i_{a,b}\bigg(\partial_{a}\partial_{b} S_{g - 1}(t) + \sum_{\substack{g_1 + g_2 = g \\ g_1,g_2 \geq 1}} \partial_{a}S_{g_1}(t) \partial_{b}S_{g_2}(t)\bigg).
$$
We recall that $\hat{M}_{i,a}(t) = \delta_{i,a} - B^i_{a,b}t_{b} - C^i_{a,b}\partial_{b}S_0(t)$. As $S_0(t) = O(t)$, we have $\hat{M}(t) = {\rm Id} + O(t)$. Hence, the (finite-dimensional) matrix $\hat{M}(t)$ is invertible for $t$ small enough. So, we can prove by induction on $g \geq 1$ that $S_g(t)$, as a solution of the (compatible) system of linear ODEs with analytic coefficients in a neighbourhood of $t = 0$, 
$$
\partial_{i}S_g(t) = [\hat{M}^{-1}(t)]_{i,c}\bigg\{\delta_{g,1}D^c + \tfrac{1}{2}C^c_{a,b}\Big(\partial_a\partial_{b}S_{g - 1}(t) + \sum_{\substack{g_1 + g_2 = g \\ g_1,g_2 \geq 1}} \partial_{a}S_{g_1}(t) \partial_{b}S_{g_2}(t)\Big)\bigg\}
$$
is the formal Taylor series at $0$ of an analytic function $\underline{S}_{g}(t)$ in the neighbourhood $\Omega$ of $0$ on which $\hat{M}(t)$ is invertible and $S_0(t)$ is analytic. This contains a neighbourhood of $0$ independent of $g$.

Independently, as $S_0(t)$ is analytic in a neighbourhood of $0$, the equality
\beq
\label{G01S0} G_{0,1}(i) = \partial_{i} S_0(t)
\eeq
provides a definition of $\underline{G}_{0,1}$ as the analytic function $\partial_{i} \underline{S}_0(t)$, whose Taylor series at $0$ is $G_{0,1}(i)$, hence such that \eqref{G01S0} holds at the level of analytic functions. Then, the expression for the formal series $G_{0,2}(i,j)$ obtained by enforcing $N_{i,j} = 0$ above shows that it is the formal Taylor series at $0$ of an analytic function $\underline{G}_{0,2}(i,j)$ for $t$ in $\Omega$. And the expression of $(\tilde{A},\tilde{B},\tilde{C},\tilde{D})$ in terms of $S_0(t)$ and its first and second order derivatives shows they upgrade in the same way to analytic functions of $t \in \Omega$, in such a way that the equality between formal series at $0$ continue to hold at the level of analytic functions of $t \in \Omega$. \hfill $\Box$

\section{Formulas for the partition function in two simple cases}
\label{SS3}
\subsection{The case \texorpdfstring{$C = 0$}{C=0}}
\label{S35}
Quantum Airy structures with $C = 0$ give rise to a compatible system of linear ODEs for the partition function
$$
\forall i\qquad \hbar(\delta_{i,b}  - B^i_{a,b}x_a)\partial_{x_b}\ln\,Z = \tfrac{1}{2}A^i_{a,b}x_{a}x_{b} + \hbar D^i.
$$
The partition function in this case can be computed in exact form. Let us introduce matrices $\mathbf{B}_j = (B^a_{j,b})_{a,b}$ and column vectors $\mathbf{A}_{j,k} = (A_{j,k}^a)_{a}$ and $\mathbf{D} = (D^a)_{a}$, as well as the formal power series
\bea
\psi_0(z) & := & \tfrac{1}{2z^3}\big(-\ln(1 - z) - z - \tfrac{z^2}{2}\big) = \tfrac{1}{6} + O(z), \nonumber \\
\psi_1(z) & := & -\tfrac{1}{z}\,\ln(1 - z) = 1 + O(z), \nonumber
\eea
which we will apply to matrices.

\begin{proposition}\label{Cequal0}
The partition function of a quantum Airy structure having $C = 0$ reads
$$
Z = \exp\big(\hbar^{-1}S_0(x) + S_1(x)\big),
$$
where
$$
S_0(x) =\Big[\psi_0\Big(\sum_{j}x_j \mathbf{B}_j\Big)\cdot \mathbf{A}_{a,b}\Big]_{c}x_{a}x_{b}x_{c},\qquad S_1(x) = \Big[\psi_1\Big(\sum_{j} x_j\mathbf{B}_j\Big)\cdot \mathbf{D}\Big]_{a}x_a\,,
$$
and $\cdot$ is the multiplication between a matrix and a column vector.
\end{proposition}

\noindent \textbf{Proof.} We use the expression of the Taylor coefficients $F_{g,n}(i_1,\ldots,i_n)$ of the partition function as sums over trivalent graphs. Since $C = 0$, all trivalent vertices should be incident to one leaf (a $B$), two leaves (an $A$), or a loop (a $D$). This drastically simplifies the structure of the graphs which can contribute to the sum, in particular their genus is $0$ and $1$. Thus $F_{g,n} = 0$ for $g \geq 2$.

\begin{center}
\begin{figure}[h!]
\centering
\includegraphics[width=0.6\textwidth]{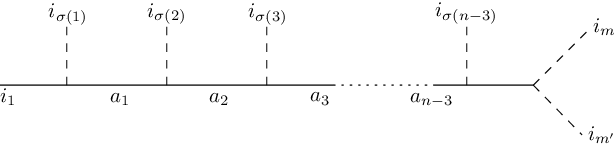}
\caption{Genus $0$ graphs without inner trivalent vertices. \label{Fig:genus0graph}}
\end{figure}
\end{center}

In genus $0$ (Figure~\ref{Fig:genus0graph}), the graphs are characterised by the sequence $\sigma(1),\ldots,\sigma(n - 3)$ of leaves successively attached when moving away from the root, and the pair $\{m,m'\}$ of leaves terminating the graph. Thus:
\bea
F_{0,n}(i_1,\ldots,i_n) & = & \sum_{\sigma}  \sum_{m < m'} B^{i_1}_{i_{\sigma(1)},a_1}B^{a_2}_{i_{\sigma(2)},a_2} \cdots B^{a_{n - 4}}_{i_{\sigma(n - 3)},a_{n - 3}} A^{a_{n - 3}}_{i_m,i_{m'}} \nonumber \\
& = & \sum_{\sigma} \sum_{m < m'} \big[\mathbf{B}_{i_{\sigma(1)}}\mathbf{B}_{i_{\sigma(2)}}\cdots \mathbf{B}_{i_{\sigma(n - 3)}}\cdot \mathbf{A}_{i_m,i_{m'}}\big]_{i_1} \nonumber
\eea
where the first sum ranges over bijections $\sigma\,:\,\{1,\ldots,n - 3\} \rightarrow \{2,\ldots,n\}\setminus \{m,m'\}$ and we recall that the indices $a_1,a_2,\ldots,a,b,c,\ldots $ are implicitly summed over. Then
$$
\sum_{i_1,\ldots,i_n \in I} F_{0,n}(i_1,\ldots,i_n)\,\frac{x_{i_1}\cdots x_{i_n}}{n!} = \frac{1}{n!} \sum_{\sigma} \sum_{m < m'} \Big[\big(x_d\mathbf{B}_{d}\big)^{n - 3}\cdot \mathbf{A}_{a,b}\Big]_{c}x_ax_bx_c.
$$
The set of $(\sigma,\{m,m'\})$ appearing in this sum is in bijection with permutations of $\{2,\ldots,n\}$, thus
$$
\sum_{i_1,\ldots,i_n \in I} F_{0,n}(i_1,\ldots,i_n)\,\frac{x_{i_1}\cdots x_{i_n}}{n!} = \frac{1}{n} \Big[\big(x_d\mathbf{B}_{d}\big)^{n - 3}\cdot \mathbf{A}_{a,b}\Big]_{c}x_ax_bx_c,
$$ 
and summing over $n \geq 3$ gives the announced expression.

\begin{center}
\begin{figure}[h!]
\centering
\includegraphics[width=0.6\textwidth]{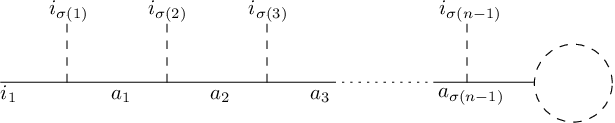}
\caption{Genus $1$ graphs without inner trivalent vertices. \label{Fig:genus1graph}}
\end{figure}
\end{center}

In genus $1$ (Figure~\ref{Fig:genus1graph}), the same graphs appear except that the terminal vertex is a loop instead of a pair of leaves. Thus for $n \geq 1$
\bea
F_{1,n}(i_1,\ldots,i_n) & = & \sum_{\sigma} B^{i_1}_{i_{\sigma(1)},a_1}B^{a_2}_{i_{\sigma(2)},a_2} \cdots B^{a_{n - 2}}_{i_{\sigma(n - 1)},a_{n - 1}} D^{a_{n - 1}} \nonumber \\
& = & \sum_{\sigma} \big[\mathbf{B}_{i_{\sigma(1)}}\mathbf{B}_{i\sigma(2)}\cdots\, \mathbf{B}_{i_{\sigma(n - 1)}}\cdot \mathbf{D}\big]_{i_1}\,, \nonumber
\eea
where $\sigma$ are bijections from $\{1,\ldots,n - 1\}$ to $\{2,\ldots,n\}$. Therefore
$$
\sum_{i_1,\ldots,i_n \in I} F_{1,n}(i_1,\ldots,i_n)\,\frac{x_{i_1}\cdots\, x_{i_n}}{n!} = \frac{1}{n}\, \Big[\big(x_b\mathbf{B}_{b}\big)^{n - 1}\cdot \mathbf{D}\big]_{a}x_{a},
$$
and summing over $n \geq 1$ gives the announced expression.  \hfill $\Box$

\begin{remark} If $(L_i)_{i \in I}$ forms an abelian quantum Airy structure, all matrices $\rho_1(e^i)$ in \eqref{rho1} commute, therefore they can simultaneously be brought into an upper-triangular form. So, abelian quantum Airy structures are always, up to the action of $\mathcal{G}_{V}$, equivalent to an abelian quantum Airy structure with $C = 0$, to which Proposition~\ref{Cequal0} can be applied. 
\end{remark}

\subsection{The case \texorpdfstring{$A = 0$}{A = 0}}
\label{S35bis}
Quantum Airy structures with $A = 0$ are related in principle to quantum Airy structures with $C = 0$ by an element of $\mathcal{G}_{V}$. Indeed, the automorphism of the Weyl algebra $(x_i,\partial_{x_i}) \rightarrow (\hbar\partial_{x_i},-x_i)$ exchanges $C$-terms and $A$-terms. This transformation yields operators which are not anymore in normal form, but it can be pre- and post-composed with elements in $\exp(\mathcal{D}_{V,2}/\hbar)$ so as to be brought back to quantum Airy structures. So the partition function of a quantum Airy structures with $A = 0$ is computable in principle from Proposition~\ref{Cequal0}.

Here we take a direct route, by examining which graphs may give non-zero contributions.

\begin{lemma}
\label{Aequal0genus0} For a quantum Airy structure with $A = 0$, we have $F_{0,n} = 0$ for all $n \geq 1$.
\end{lemma}
\noindent \textbf{Proof.} The recursion for $n \geq 4$
$$
F_{0,n}(i_1,\ldots,i_{n}) = \sum_{m = 2}^{n} B^{i_1}_{i_m,a} F_{0,n - 1}(a,i_2,\ldots,\widehat{i_m},\ldots,i_n)
$$
together with the initial data $F_{0,3}(i_1,i_2,i_3) = A^{i_1}_{i_2,i_3} = 0$ has $F_{0,n} = 0$ as the unique solution. \hfill $\Box$

In higher genera, simplification occurs when furthermore $B = 0$.

\begin{definition}
Let $\mathfrak{T}_{g}$ be the set of rooted trivalent trees with $g$ leaves. The root edge is denoted $r$. If $v$ is a vertex, we denote $e_0(v)$ the edge closer to the root, and $e_{\pm}(v)$ the two other edges --- in arbitrary order. If $T \in \mathfrak{T}_{g}$, let $E(T)$ be the set of unoriented edges (including the root edge, denoted $r$), $V(T)$ the set of trivalent vertices, and $L(T)$ the set of leaves (univalent vertices distinct from the root). If $\ell$ is a leaf, we denote $e(\ell)$ its incident edge. Note that the cardinality of ${\rm Aut}\,T$ is a power of $2$, as this group consists of the permutations of $\{e_{-}(v),e_{+}(v)\}$ for each $v$, which preserve $T$.
\end{definition}

\begin{proposition}
\label{ABequal0}
The partition function of a quantum Airy structure with $A = B = 0$ is computed by
$$
F_{g,n}(i_1,\ldots,i_n) = \delta_{n,1}f_g(i_1),
$$
with $f_0 = 0$ and for $g \geq 1$
\beq
\label{fgab0} f_g(i) :=  \sum_{T \in \mathfrak{T}_{g}} \frac{1}{|{\rm Aut}\,T|} \sum_{\substack{a\,:\,E(T) \rightarrow I \\ a(r) = i}} \prod_{v \in V(T)} C^{a(e_0(v))}_{a(e_-(v)),a(e_+(v))}\,D^{a(e(\ell))}.
\eeq
\end{proposition}

\noindent \textbf{Proof.} Let us write $Z = \exp\big(\sum_{g \geq 0} \hbar^{g - 1}S_{g}\big)$. Then, $L_i \cdot Z = 0$ turns into the differential recursion
\beq
\label{Sgrel}\forall g \geq 0,\qquad \partial_{x_i} S_{g} = \tfrac{1}{2}C^i_{a,b}\big(\partial_{x_a}\partial_{x_b}S_{g - 1}(x) + \sum_{h_1 + h_2 = g} \partial_{x_a} S_{h_1} \partial_{x_b}S_{h_2}\big) + \delta_{g,1}D^i.
\eeq
We know from Lemma~\ref{Aequal0genus0} that $S_{0} = 0$. Therefore, the equation above for $g = 0$ implies $S_{1} = D^ax_a$, and by induction that $S_{g}$ must be a linear function. As $A = B = 0$, in the graphs in $\mathfrak{G}_{g,n}(1)$ with non-zero contribution the Betti number $g$ can only arise from loops receiving a $D$-weight. They should thus form $g$ leaves of the graph attached to the spanning tree. \hfill $\Box$

\section{Finite-dimensional Airy structures and representation theory}
\label{SS5}

\subsection{General properties}

\begin{lemma}
\label{Vsemi1}Assume $V$ is a finite-dimensional Lie algebra, $L\,:\,V \rightarrow \mathcal{D}_{V}$ a homomorphism of normal form and  that there exists a V-submodule $\rho_M : V \ra \End(M)$ of $ \mathbb{C}.\hbar \oplus T^*_{\hbar}V $ and an isomorphism 
\begin{equation}\label{isoM}
 \Phi\,:\,\mathbb{C}.\hbar \oplus T^*_{\hbar}V \ra \mathbb{C}.\hbar \oplus M,\qquad \Phi \tilde \rho_1 \Phi^{-1} =\rho_{\text{t}} \oplus \rho_M,
 \end{equation}
where $\rho_{\text{t}}$ is the one-dimensional trivial representation and $\Phi|_{\mathbb{C}.\hbar} = \Id$. 
Then there exists $\alpha \in M^*$ such that
$$
\forall v \in V,\qquad \varrho_{0,1}(v) \circ \Phi^{-1} = \alpha \circ \rho_M(v)\circ \pi_M,
$$
where $\pi_M:  \mathbb{C}.\hbar \oplus M \ra M$ is the projection onto $M$ with kernel $\mathbb{C}.\hbar$.
Further, 
$$   
V_{\alpha} := \{v \in V\,\,:\,\,{\rm Im}\,\rho_M(v) \subseteq {\rm Ker}\,\alpha\}
$$ 
must vanish and  
$$\pi_M \circ \Phi \circ \mathcal{I}(V) = \{ m\in M\,\,\,:\,\,\,\forall v \in V\quad \alpha(\rho_M(v)m)=0\},$$
$$\pi_M \circ \Phi \circ \mathcal{I}([V,V]) \subset {\rm Ker}\,\alpha.$$
where $\mathcal{I}$ is the Lagrangian embedding from Lemma~\ref{LLLfdsf}.
\end{lemma}

\begin{remark}
We note that if $V$ is semi-simple then the trivial submodule $\mathbb{C}.\hbar$ in $\mathbb{C}.\hbar \oplus T^*_{\hbar}V $ will have a complementary submodule $M$ as assumed in Lemma \ref{Vsemi1}, thus the conclusion of this Lemma applies in this case. We will then use the notation
$$ \tilde{\mathcal{I}} = \pi_M \circ \Phi \circ \mathcal{I}.$$

Further, since $V$ is semi-simple, it is the complexification of a compact real form $V_{\mathbb R}$ and so let us also assume $M$ is the complexification of $M_{\mathbb R}$ such that there exist $V_{\mathbb R}$-submodule $\rho_{M_{\mathbb R}} : V_{\mathbb R} \rightarrow \End(M_{\mathbb R})$ of $\mathbb{R}.\hbar \oplus T^*_{\hbar}V_{\mathbb R}$ such that $\rho_M = \rho_{M_{\mathbb R}}\otimes {\mathbb C} $ and an isomorphism
$$ \Phi_{\mathbb R}\,:\,\mathbb{R}.\hbar \oplus T^*_{\hbar}V_{\mathbb R} \ra \mathbb{R}.\hbar \oplus M_{\mathbb R}
$$  
such that $\Phi$ is the complexification of $\Phi_{\mathbb R}$. 
Then there exist $\alpha_{\mathbb R} \in M^*_{\mathbb R}$ such that $\alpha$ from Lemma \ref{Vsemi1} is the complexification of $\alpha_{\mathbb R}$. Further we then also get a real version of $\tilde{\mathcal{I}} $
$$ \tilde{\mathcal{I}}_{\mathbb R} = \pi_{M_{\mathbb R}} \circ \Phi_{\mathbb R} \circ \mathcal{I}$$
since $\mathcal{I}$ of maps $V_{\mathbb R}$ into the first summand of $V_{\mathbb R}\oplus V_{\mathbb R}$.

If we in the following Lemma \ref{Vsemi1.1} have that $\omega_M$ is the complexification of a real symplectic form $\omega_{M_{\mathbb R}}$ on $M_{\mathbb R}$, then $m_\alpha \in M_{\mathbb R}$ and
$$\alpha_{\mathbb R} = \omega_{M_{\mathbb R}}(m_\alpha, \cdot).$$
\end{remark}

\begin{lemma}\label{Vsemi1.1}
Suppose that $V$ is a finite-dimensional semi-simple Lie-algebra and $L\,:\,V \rightarrow \mathcal{D}_{V}$ a Lie algebra homomorphism of normal form. Let $\rho_M : V \ra \End(M)$ be the symplectic V-submodule of  $ \mathbb{C}.\hbar \oplus T^*_{\hbar}V $ with the properties stated in Lemma \ref{Vsemi1}. If $\omega_M$ is the $V$-invariant symplectic structure on $M$, then there exists an $m_\alpha \in \tilde{\mathcal{I}}(V) \subset M$ such that
$$\alpha = \omega_M(m_\alpha, \cdot).$$
Moreover $\tilde{\mathcal{I}}(V)$ is a Lagrangian subspace of $M$ on which $\alpha$ vanishes and
$$ \rho_M(V) m_\alpha = \tilde{\mathcal{I}}(V).$$
There exists a unique $v_\alpha \in V$ such that
$$ \rho_M(v_\alpha) m_\alpha = m_\alpha.$$
\end{lemma}

\noindent\textbf{Proof of Lemma~\ref{Vsemi1}.} We consider some $V$-module structure $\tilde{\rho}_{1}\,:\,V \rightarrow {\rm End}(\mathbb{C}.\hbar \oplus T^*_{\hbar}V)$ of the form (\ref{rho1ti}). Then $\mathbb{C}.\hbar$ is a trivial one-dimensional submodule.
By projection from $\mathbb{C}.\hbar \oplus T^*_{\hbar}V $ onto $\mathbb{C}.\hbar$ along $T^*_{\hbar}V$, we get a linear functional $\tilde \alpha \in (\mathbb{C}.\hbar \oplus T^*_{\hbar}V)^*$. If we now let $\beta$ be the projection onto $T^*_{\hbar}V$ along $\mathbb{C}.\hbar$, then $\tilde \alpha + \beta = \Id_{\mathbb{C}.\hbar \oplus T^*_{\hbar}V }$ and
$$\tilde \rho_1 = (\tilde \alpha + \beta) \circ \Phi^{-1}\circ \rho_M   \circ \Phi= 
\tilde \alpha \circ \Phi^{-1}\circ\rho_M   \circ \Phi + \beta \circ \Phi^{-1}\circ \rho_M \circ \Phi,$$
where the first factor on the right hand side gives $\rho_{0,1}$ by definition. We now let $\alpha = \tilde \alpha\circ \Phi^{-1}|_M$ and get the claimed formula for $\rho_{0,1}$.
If $(e_i)_{i \in I}$ is a basis of $V$, the linear forms $(\alpha \circ \rho_M(e_i))_{i \in I}$ on $V$ must be linearly independent. Therefore, if $v = \sum_{i \in I} v_i e_i$ is solution to  $\rho_{0,1}(v) = \sum_{i \in I} v_i\,\alpha \circ \Phi^{-1} \circ \rho_M(e_i) = 0$, we must have $v = 0$. This implies $V_{\alpha} = 0$. From the normal form of $L$, we see that 
$$ V = \{ m\in M\,\,:\,\,\,\forall v \in V\quad \rho_{0,1}(v)(\Phi^{-1}(m))=0\},$$
from which it follows that
$$ \pi_M \circ \Phi \circ \mathcal{I}(V) = \{ m\in M\,\,\,:\,\,\,\forall v \in V\quad \alpha( \rho_M(v)m)=0\}.$$
But then (\ref{commuts}) implies that
$$ \pi_M \circ \Phi \circ \mathcal{I}([V,V]) \subset {\rm Ker}\,\alpha.$$
\hfill $\Box$

\noindent\textbf{Proof of Lemma~\ref{Vsemi1.1}.}
There exists a unique $m_\alpha \in M$ such that $\alpha = \omega_{M}( m_\alpha,\cdot )$, where $\omega_{M}$ is the symplectic form, by the non-degenerateness of $\omega_{M}$. Since
$$\alpha \circ \rho_M(v) \pi_M \circ \Phi \circ \mathcal{I} (u) = 0$$
for $u,v\in V$ by the normal form assumption, we see that
$$\rho_M(V) m_\alpha \subset \pi_M \circ \Phi \circ \mathcal{I}(V).$$
As $V_\alpha = 0$, $\alpha\circ \rho_M : V \ra M^*$ is injective. Thus
$$ \rho_M(V) m_\alpha = \tilde{\mathcal{I}}(V).$$
Since $V$ is semi-simple, we get that $[V,V]=V$. Lemma~\ref{Vsemi1} then implies that $\alpha(\tilde{\mathcal{I}}(V)) =0$. Thus $m_\alpha$ must be in the Lagrangian subspace $\tilde{\mathcal{I}}(V)$, and we see that there must exist $v_\alpha\in V$ such that
$$\rho_M(v_\alpha) m_\alpha = m_\alpha.$$
If $\tilde{v}_\alpha\in V$ was another such element, then $\rho_M(v_\alpha-\tilde{v}_\alpha)m_\alpha = 0$. Thus $v_\alpha - \tilde{v}_\alpha \in V_\alpha$. By Lemma \ref{Vsemi1}, we get that $v_\alpha =\tilde{v}_\alpha$, hence the claimed uniqueness.
\hfill $\Box$

We shall now establish a kind of converse to Lemma \ref{Vsemi1.1}.

\begin{theorem}\label{SSC}
Suppose that $V$ is a finite-dimensional Lie algebra and $\rho_M : V \ra \End(M)$ is a symplectic representation of dimension twice the dimension of $V$ and further assume that there exists $m_\alpha\in M$ such that $\rho_M(V)m_\alpha \subset M$ is a Lagrangian subspace. 

 Then for any choice of a basis $(w_i)_{i \in I}$ of a Lagrangian complement $W$ to $ \rho_M(V)m_\alpha$, one can construct a quantum Airy structure $L : V \ra \mathcal{D}_V$ as follows. Let $v_i$ be the unique basis of $V$ such that $\big(w_i,(\rho_M(v_i)m_\alpha)\big)_{i \in I}$ is a symplectic basis of $M$.  Now let $(A^i,B^i,C^i)$ be defined by the matrix of $\rho_M(v_i)$ in this basis as in Equation~\ref{rho1}, and let $D^i = \frac12 \tr(B^i)$. Then $L_i$ is determined by $(A^i,B^i,C^i, D^i)$ as in Equation~\eqref{Lform}.
\end{theorem}

\noindent\textbf{Proof.}
We define the map  $\tilde{\mathcal{I}}\,:\,V \ra  \rho_M(V)m_\alpha$ by the formula
$$ \tilde{\mathcal{I}}(v) =  \rho_M(v)m_\alpha.$$
We then immediately see that
$$
\forall u,v \in V,\qquad  \rho_M(u)\tilde{\mathcal{I}}(v) - \rho_M(v) \tilde{\mathcal{I}}(v) = \tilde{\mathcal{I}}([u,v]).
$$
Further, since $\rho_M(V)m_\alpha$ is Lagrangian, we see that $\tilde{\mathcal{I}} $ is an isomorphism. For any choice of a basis $(w_i)_i$ of a Lagrangian complement $W$ to $ \rho_M(V)m_\alpha$, we pick the basis $v_i$ of $V$ as specified in the Theorem. By using the basis $\big(w_i,(\rho_M(v_i)m_\alpha)\big)_{i \in I}$ of $M$ and the basis $(v_i^*,v_i)_{i \in I}$ of $T^*_{\hbar}V$, we induce a symplectic representation $\rho_1$ of $V$ on $T^*_{\hbar}V$ and by the above property of $\tilde{\mathcal{I}}$, we see that
$$\forall u,v \in V,\qquad \rho_1(u)\mathcal{I}(v)-\rho_1(v) \mathcal{I}(u) = \mathcal{I}([u,v]).$$
But then we can simply define $L_i$ as specified in the Theorem above and because $\rho_{0,1} = \alpha \circ \rho_M$, where $\alpha = \omega_{M}(m_\alpha, \cdot)$, we see by the way the basis of $V$ is chosen that $L_i$ are of normal form. It then follows from Lemmas \ref{LLL1}-\ref{LLLfdsf} that $(A,B,C)$ satisfies the  \textbf{BB-CA}, \textbf{BC} and \textbf{BA} relations and the torsion-free condition \eqref{commuts} with respect to the Lie algebra structure on $V$. Then letting $D$ be given as in the Theorem, we see by Lemma \ref{trL} that $(A,B,C,D)$ induces a quantum Airy structure.
 \hfill $\Box$

 \subsection{The example of \texorpdfstring{$\mathfrak{sl}_{2}(\mathbb{C})$}{sl2(C)}}
 
The previous considerations allow the construction a quantum Airy structure based on $\mathfrak{sl}_{2}(\mathbb{C})$.
 
\begin{theorem}
\label{sl2th}The simple Lie algebra $\mathfrak{sl}_{2}(\mathbb{C})$ supports the following quantum Airy structure
\begin{eqnarray*}
L_1 &=& \hbar \partial_1 - \hbar(3 x_1\partial_1 + 5 x_2 \partial_2 + x_3\partial_3) - \tfrac{9\hbar}{2},\\
L_2 &=& \hbar \partial_2 - \hbar(\tfrac{8}{3} x_3\partial_1 + 3 x_1 \partial_2 ) - \tfrac{3\hbar^2}{80}\partial_3^2,\\
 L_3 &=& \hbar \partial_3 - \hbar(\tfrac{5}{3} x_2\partial_1 + 3 x_1 \partial_3 ) +60 x_3^2.
 \end{eqnarray*}
Its partition function is $Z(x) = \tfrac{\mathfrak{Z}(x)}{\mathfrak{Z}(0)}$ with
\bea
\mathfrak{Z}(x) & = & x_2^{-\frac{3}{5}}\big((1 - 3x_1)^2 -10x_2x_3\big)^{\frac{1}{2}} \exp\bigg\{\frac{8(1 - 3x_1)}{\hbar x_2}\bigg(-\frac{(1 - 3x_1)^2}{125x_2^2} + \frac{(1 - 3x_1)x_3}{5x_2} + \frac{3x_3^2}{2}\bigg)\bigg\} \nonumber \\
&& \times\widetilde{{\rm H}}^{(2)}_{\frac{1}{5}}\bigg(\frac{8\big((1 - 3x_1)^2 - 10x_2x_3\big)^{\frac{5}{2}}}{125\hbar x_2^3}\bigg), \nonumber 
\eea
where $\widetilde{{\rm H}}^{(2)}(z) := {\rm H}^{(2)}(-{\rm i}z)$ and ${\rm H}^{(2)}$ is the Hankel function of the second kind. This function satisfies the differential equation
$$
\big(z^2\partial_{z}^2 + z\partial_{z} - (\nu^2 + z^2)\big)\widetilde{{\rm H}}^{(2)}_{\nu}(z) = 0
$$
\end{theorem}

\noindent\textbf{Proof.}
$V=\mathfrak{sl}_{2}(\mathbb{C})$ has three generators $(H,E,F)$ satisfying the commutation relations
$$ [H,E] = 2E, \ [H,F] = -2F \ \ \text{ and } \ \ [E,F] = H.$$
Let $M$ be the $6$-dimensional irreducible representation of $\mathfrak{sl}_{2}(\mathbb{C})$. Recall that it has a basis $(e_0,\ldots, e_5)$ such that
\begin{eqnarray*}
\rho_M(H) e_i &=& (6-2i-1) e_i,\\
\rho_M( E) e_i &=& (6-i) e_{i-1},\\
\rho_M(F) e_i  &=& (i+1) e_{i+1},
 \end{eqnarray*}
 for $i \in \{0,\ldots,5\}$ where by convention $e_{-1}=e_6=0$.
Further $M$ is symplectic with the following symplectic structure
$$\omega_{M}(e_i,e_{5-i}) = \left\{ 
\begin{array}{ccc} 
1 && i=0\\
-5 && i=1\\
10 && i=2
\end{array} \right.
$$
and $\text{span}\{e_0,e_1,e_2\}$, $\text{span}\{e_3,e_4,e_5\}$ are Lagrangian.
We are now going to pick\footnote{By Lemma \ref{Vsemi1.1}, we know that $m_\alpha$ has to be the eigenvector of some $v_\alpha$ with eigenvalue $1$, which we can assume is a multiple of $H$, but then we see that $m_\alpha$ must be propositional to one of the $e_i$s. However $\rho_M(V)m_\alpha$ must be Lagrangian. This means that $m_\alpha$ has to be proportional to either $e_1$ or $e_4$.} $m_\alpha = e_1$.  We  get that
$$\mathcal{I}(H) = 3 e_1,\qquad \mathcal{I}(E) = 5 e_0,\qquad \mathcal{I}(F) = 2 e_2.$$
It is then easy to check that
$$ \forall u,v \in V,\qquad \rho_M(u)\mathcal{I}(v) -  \rho_M(v)\mathcal{I}(u) = \mathcal{I}([u,v])$$
as claimed in the proof of Theorem \ref{SSC}. One also finds
$$
\alpha\circ \rho_M(H)(e_4) = 15,\qquad \alpha\circ \rho_M(E)(e_5) = -5 ,\qquad  \alpha\circ \rho_M(F)(e_3) = -20
$$
and that all other evaluations on the basis $e_0,\ldots, e_5$ are zero.
Let now $(H^*,E^*,F^*)$ be the basis of $V^*$ which is dual to the basis $(H,E,F)$ of $V$.
We get a linear symplectomorphism $\phi : T^*_{\hbar} V \ra M$ by mapping
$$\phi(H^*) = \tfrac{1}{15} e_4,\qquad \phi(E^*) = -\tfrac{1}{5} e_5,\qquad \phi(F^*) = -\tfrac{1}{20} e_3,$$
and letting $\phi|_V = \mathcal{I}.$
The representation $\rho_1 = \phi^{-1}\rho_M \varphi$ reads in the basis $(e_i)_{i = 0}^5$
$$\rho_1(H) = \left(
\begin{array}{rrrrrr}
-3 & 0& 0& 0& 0& 0\\
0 & -5& 0& 0& 0& 0\\
0 & 0& -1& 0& 0& 0\\
0 & 0& 0& 3& 0& 0\\
0 & 0& 0& 0& 5& 0\\
0 & 0& 0& 0& 0& 1
\end{array}\right)
\ \ \ \
\rho_1(E) = \left(
\begin{array}{rrrrrr}
0 & -3& 0& 0& 0& 0\\
0 & 0& 0& 0& 0& 0\\
-\frac83 & 0& 0& 0& 0& 0\\
0 & 0& 0& 0& 0& \frac83\\
0 & 0& 0& 3& 0& 0\\
0 & 0& -\frac3{40}& 0& 0& 0
\end{array}\right)
$$
$$\rho_1(F) = \left(
\begin{array}{rrrrrr}
0 & 0& -3& 0& 0& 0\\
-\frac53 & 0& 0& 0& 0& 0\\
0 & 0& 0& 0& 0& -120\\
0 & 0& 0& 0& \frac53& 0\\
0 & 0& 0& 0& 0& 0\\
0 & 0& 0& 3& 0& 0
\end{array}\right)
$$
We now let $\alpha' = \alpha \circ \phi.$ Then
$$ 
\alpha' \circ \rho_1(H) = H,\qquad \alpha' \circ \rho_1(E) = E,\qquad \alpha' \circ \rho_1(F) = F,
$$
where we here think of $H,E,F$ as elements of $(T^*_{\hbar} V)^*.$ If we now define $\rho_{0,1} = \alpha' \circ \rho_1$, then it is of normal form. Comparing with \eqref{rho1}, we get the three stated operators. Alternatively, one can check directly that they satisfy the desired $\mathfrak{sl}_{2}(\mathbb{C})$ commutation relations.

The general solution of $L_1 \cdot Z = 0$ takes the form
$$
Z(x) = (1 - 3x_1)^{-\frac{3}{2}}\tilde{Z}\big((1 - 3x_1)^{-\frac{5}{3}}x_2,(1 - 3x_1)^{-\frac{1}{3}}x_3\big)
$$
In terms of the variables $y_2 = (1 - 3x_1)^{-\frac{5}{3}}$ and $y_3 = (1 - 3x_1)^{-\frac{1}{3}}x_3$, the equations $L_i\cdot Z = 0$ for $i \in \{2,3\}$ become
\bea
\Big\{\hbar \partial_{y_2} - \hbar\big(\tfrac{40}{3}y_2y_3\partial_{y_2} + \tfrac{8}{3}y_3^2\partial_{y_3} + 12y_3\big) - \tfrac{3\hbar^2}{80}\partial_{y_3}^2 \Big\}\tilde{Z}(y_2,y_3) & = & 0 \nonumber \\
\Big\{\hbar \partial_{y_3} + 60y_3^2 - \hbar\big(\tfrac{25}{3}y_2^2\partial_{y_2} + \tfrac{5}{3}y_2y_3\partial_{y_3} + \tfrac{15}{2}y_2\big)\Big\}\tilde{Z}(y_2,y_3) & = & 0 \nonumber
\eea
The two equations can be decoupled by a gauge transformation and a change of variables, and then reduce to a Bessel differential equation. Taking into account the initial conditions for the sought partition function picks up the solution ${\rm H}^{(2)}$ of this differential equation. The final outcome is
$$
\tilde{Z}(y_2,y_3) = c_{\hbar}\,(1 - 10y_2y_3)^{1/2}y_2^{-3/2}\,\widetilde{{\rm H}}^{(2)}_{\frac{1}{5}}\bigg(\frac{8}{125\hbar}\,\frac{(1 - 10y_2y_3)^{\frac{5}{2}}}{y_2^3}\bigg)\,\exp\Big\{\hbar^{-1}(-\tfrac{8}{125}y_2^{-3} + \tfrac{8}{5}y_2^{-2}y_3 - 12y_2^{-1}y_3^2)\Big\}
$$
for some $c_{\hbar}$ independent of $(y_2,y_3)$, which is the announced result.
 \hfill $\Box$

\subsection{Towards a classification for simple Lie algebras}

If $V$ is a semi-simple Lie algebra and $L\,:\,V \rightarrow \mathcal{D}_{V}$ is a homomorphism of normal form, $\rho_1$ defines a structure of symplectic $V$-module on $W = T^*V$, which must split in a direct sum of irreducible $V$-modules. We shall derive necessary condition for the choice of these modules, and initiate a classification of Airy structures supported by simple Lie algebras. Our findings are summarised in Proposition~\ref{calsssls} at the end of the section.

Let $\mathcal{R}$ be the set of equivalence class of irreducible $V$-modules which occurs in the decomposition of $W$ and $n\,:\,\mathcal{R} \rightarrow \mathbb{Z}_{> 0}$ the multiplicities. The Frobenius--Schur indicator distinguishes the following properties of an irreducible $V$-module $M$
$$
{\rm FS}(M) = \left\{\begin{array}{ccl} -1 & & {\rm if}\,\,M \cong M^*\,\,{\rm is}\,\,{\rm symplectic} \\
0 & & {\rm if}\,\,M \ncong M^* \\
1 & & {\rm if}\,\,M \cong M^*\,\,{\rm is}\,\,{\rm symmetric}
\end{array}\right.
$$
We denote $\mathcal{R}_{{\rm s}} \subset \mathcal{R}$ the subset consisting of the symplectic $V$-modules.
\begin{lemma}
\label{Ldrop} If $[M] \in \mathcal{R} \setminus \mathcal{R}_{{\rm s}}$, then $[M^*] \in \mathcal{R}\setminus \mathcal{R}_{{\rm s}}$. If ${\rm FS}(M) = 0$, then $n([M]) = n([M^*])$. If ${\rm FS}(M) = 1$, then $n([M])$ is even.
\end{lemma}
\noindent \textbf{Proof.} Suppose we have an $[M] \in \mathcal{R}\setminus \mathcal{R}_{{\rm s}}$ such that $M \ncong M^*$. Let $m = n(M)$ and $m^* = n(M^*)$. The symplectic form of $W$ restricted to $M^{m}\oplus (M^*)^{m^*}$ is then non-degenerate. Indeed, if it would be degenerate, it would induce a non-zero map from one of the $M$s or $M^*$s in this sum to some other module in the decomposition of $W$, which would then make $M$ or $M^*$ isomorphic to one of the other modules in the decomposition, thus we would have a contradiction.

Consider one copy of $M$ inside $M^{\oplus m}$ and let $$ M^\perp = \big\{ v \in M^{m}\oplus (M^*)^{m^*}\mid \omega(v, M) = 0\big\}.$$
We see that $M^{m}\subset M^\perp$. Let $M'$ be a $V$-submodule of $M^{m}\oplus (M^*)^{m^*}$, which is a complement of $M^\perp$. But then the projection from $M'$ to one of the $M^*$ factors must be an isomorphism. Then we see that $(M \oplus M')^\perp$ must be isomorphic to $M^{m- 1}\oplus (M^*)^{m^* - 1}$ by projection onto the remaining factors of $M$s and $M^*$s and we can by induction conclude that $m -1 = m^* -1$.

We now turn to the case of an $[M] \in \mathcal{R}\setminus \mathcal{R}_{{\rm s}}$ which is symmetric. Likewise, consider one copy of $M$ inside $M^{m}$ and restrict the symplectic form of $W$ to $M$. If the restriction was non-zero, it must mean that $M$ is a symplectic module and the restriction of the symplectic form of $W$ is the symplectic invariant form $M$ admits, which is unique up to non-zero rescaling. Since $M$ is assumed to be symmetric, it must then be isotropic. Consider now 
$$ M^\perp = \{ v \in M^{m}\mid \omega(v, M) = 0\}.$$
Since $M^\perp$ is a $V$-submodule, we can find a $V$-submodule $M' \subset M^{m}$ such that
$$ M^{m} = M^{\perp} \oplus M'.$$
But then the symplectic form of $W$ will induce an isomorphism between $M^*$ and $M'$ and will make $M\oplus M'$ a symplectic vector subspace of $M^{m}$, but since $M'\subset  M^{m}$, it must also be isomorphic to $M$.  To prove that $m$ is now even, we proceed inductively by considering $ (M\oplus M')^\perp \subset M^{m}$. We see that as a $V$-module $(M\oplus M')^\perp \cong M^{m-2}$. Thus, just as before, by induction, this module must be isomorphic to $M^{\oplus 2m'}$, thus $m = 2(m'+1).$ \hfill $\Box$

Let us denote $\mathcal{R}_{2}$ the quotient of $\mathcal{R}\setminus \mathcal{R}_{{\rm s}}$ by the relation identifying the objects $M$ and $M^*$, and $\mathcal{R}' = \mathcal{R}_{{\rm s}} \cup \mathcal{R}_{2}$. Lemma~\ref{Ldrop} tells us that the multiplicity drops to a function $n\,:\,\mathcal{R}' \rightarrow \mathbb{Z}_{> 0}$.
\begin{corollary}
\label{Cdim} The dimension of the $V$-module $W$ is
\beq
\label{dimf} 2\dim V = \sum_{M \in \mathcal{R}_{{\rm s}}} n(M)\,\dim M + \sum_{M \in \mathcal{R}_{2}} 2n(M)\,\dim M.
\eeq
In particular: if $M \in \mathcal{R}_{{\rm s}}$, then $\dim M \leq 2\dim V$; if $M \in \mathcal{R}_{2}$, then $\dim M \leq \dim V$. \hfill $\Box$
\end{corollary}

\begin{lemma}
\label{nohadj}$W$ cannot be isomorphic as a $V$-module to the direct sum of the adjoint module and its dual.
\end{lemma}

\noindent\textbf{Proof.} Assume that $W$ is isomorphic as a $V$-module to $V \oplus V^*$.  Since $V$ is semi-simple, it has a real compact form $V_{\mathbb R}$ such that $V$ is the complexification of $V_{\mathbb R}$. But then so does $W$ and we denote it $W_{\mathbb R}$. We use the Killing form $\langle \cdot,\cdot \rangle$ to identify $V_{\mathbb R}$ with $V^*_{\mathbb R}$ as a $V_{\mathbb R}$-module, so we obtain an isomorphism of $V_{\mathbb R}$-modules $\Psi_{\mathbb R}\,:\,W_{\mathbb R} \rightarrow V_{\mathbb R} \oplus V_{\mathbb R}$. Lemma~\ref{Vsemi1.1} gives elements $m_{\alpha} = (m_{\alpha}',m_{\alpha}'') \in V_{\mathbb R}\oplus V_{\mathbb R}$ and $v_\alpha \in V$ such that
$$ [v_\alpha, m'_\alpha] = m'_\alpha, \ \ \ \  [v_\alpha, m_\alpha''] = m_\alpha''.$$
It is now very easy to see that this implies that $m'_\alpha = 0 = m_\alpha''$, which contradicts the uniqueness of $v_\alpha$. To see the vanishing of   say $m_\alpha'$, we choose a Cartan $h \subset V$  which contains $m_\alpha'$. This is possible since any element of a compact connected Lie group is contained in a maximal torus  --- see e.g. \cite[Theorem 4.21]{Adams} and that the exponential map is a local diffeomorphisms around zero. But then it follows that $[v_\alpha, m'_\alpha]$ is contained in the sum of the root spaces corresponding to $h$, but since $h$ is not contained in the root spaces for $h$, $m'_\alpha $ must vanish. The same argument of course also applies to $m_\alpha''$. \hfill $\Box$

\begin{lemma}
\label{nozero}$\mathcal{R}$ cannot contain the trivial $V$-module.
\end{lemma}
\noindent \textbf{Proof.} Assume that the trivial representation is a $V$-submodule of $W$ which we denote $T$. According to Lemma~\ref{Ldrop}, it must have even multiplicity such that $T^{2n(T)}$ forms a symplectic $V$-submodule of $W$, and there exists a symplectic $V$-submodule $M''$ such that $W$ is isomorphic to $T^{2n(T)} \oplus M''$ as a $V$-module. Take $m_{\alpha} \in W$ as in Lemma~\ref{Vsemi1.1}, which we decompose as $(m_{\alpha},m_{\alpha}'')$ in $T^{2n(T)} \oplus M''$. Then $\rho(V)m_{\alpha} = \rho(V)m_{\alpha}'' \subset M''$. As $M''$ is a symplectic vector space, for dimensional reasons $\rho(V)m_{\alpha}$ cannot be Lagrangian, thus contradicting the result of Lemma~\ref{Vsemi1.1}. \hfill $\Box$

This dimension formula together with Lemmas~\ref{nohadj}-\ref{nozero} already puts strong constraints on $V$ and the sets of $V$-modules $\mathcal{R}_{2}$ and $\mathcal{R}_{{\rm s}}$. Since simple Lie algebras and their finite-dimensional irreducible representations are completely classified, we can use this classification to identify which $V$ and $M$ satisfy the naive dimension bounds of Corollary~\ref{Cdim} and which could belong to  $\mathcal{R}$'. For each simple $V$ there are finitely many possible $M$s. Then, one determines what are the possible $\mathcal{R}'$s made out of these $M$s and the possible multiplicity functions $n\,:\,\mathcal{R}' \rightarrow \mathbb{Z}_{> 0}$, such that the dimension formula \eqref{dimf} holds. The outcome of this process is summarised in the following proposition.

\begin{proposition} 
\label{calsssls}The following simple Lie algebras do not admit classical Airy structure (and therefore, do not admit quantum Airy structures either): $A_n$ for $n \notin \{1,5\}$, $C_n$ for $n \geq 6$, $D_n$ for any $n \geq 4$, $E_6$, $E_7$, $E_8$. For the remaining simple Lie algebras, the dimension bound \eqref{dimf} is respected for the following decompositions in irreducible modules\footnote{We indicate an irreducible module by its dimension in bold characters. If there are several non isomorphic irreducible modules of the same dimension $d$, we use the notation $\textbf{d}$, $\textbf{d}'$, $\textbf{d}''$ to distinguish them. We write $(\textbf{s})$ when the module is symplectic.}. 
\begin{itemize} 
\item[$\boxed{A_1}$] the candidates are $\textbf{6(s)}$ (realised in Theorem~\ref{sl2th}), $\textbf{4(s)} \oplus \textbf{2(s)}$ and $\textbf{2(s)}^3$. 
\item[$\boxed{A_5}$] the only candidate is $\textbf{20(s)}^2 \oplus \textbf{15}^2$. Here, if we denote $F$ the fundamental module, $\textbf{20(s)} = \Lambda^3F$ and $\textbf{15} = S^2F$.
\item[$\boxed{B_n}$] For $n \geq 3$, the only candidate is $\textbf{(2n + 1)}^{2n}$, where $\textbf{(2n + 1)}$ is the fundamental module. For $n = 2$, we have $\textbf{5}^4$, $\textbf{4(s)}^5$, $\textbf{4(s)} \oplus \textbf{16(s)}$ and $\textbf{20(s)}$. Here, $\textbf{4(s)}$ is the fundamental module for $\mathfrak{sp}(4)$ --- using the isomorphism of this Lie algebra with $\mathfrak{so}(5) = B_2$.
\item[$\boxed{C_3}$] there are three candidates: $\textbf{6(s)}^{7}$, $\textbf{14'(s)}^3$ and $\textbf{14}^2 \oplus \textbf{14'(s)}$. Here, $\textbf{6}(s)$ is the fundamental module.
\item[$\boxed{C_4}$] there are three candidates: $\textbf{8(s)}^{9}$ and $\textbf{8(s)}^{3} \oplus \textbf{48}$.
\item[$\boxed{C_5}$] the only candidate is $\textbf{110(s)}$.
\item[$\boxed{F_4}$] the only candidate is $\textbf{26}^{4}$.
\item[$\boxed{G_2}$] the only candidate is $\textbf{7}^{4}$.
\end{itemize} 
\end{proposition}
\begin{remark} \label{rem61} The full classification has been obtained since then in \cite{RHAiry}. The result is that Airy structures based on simple Lie algebras exist only for $A_1$, $C_2$ and $C_5$. For $A_1$, up to symmetries and besides the one based on $\textbf{6(s)}$, there is one Airy structure based on $\textbf{4(s)} \oplus \textbf{2(s)}$. For $B_2$, up to symmetries there are two Airy structures based on $\textbf{4(s)} \oplus \textbf{16(s)}$ (which is equal to $\textbf{4(s)} \otimes \textbf{5}$). For $C_5$, up to symmetries there are two Airy structures based on $\textbf{110(s)}$. All these Airy structures have been constructed explicitly in \cite{RHAiry}. We find remarkable and mysterious that many simple Lie algebras are excluded.
\end{remark}

\section{Low dimensional examples}
\label{S5}

\subsection{Twisted cohomology of Lie algebras in low dimensions}

Let $\Lie{g}$ be a complex Lie algebra and $M$ be a $\Lie{g}$-module. We will consider a rather special case, which however suffices for our purposes to compute the needed Lie algebra cohomologies, namely that $\Lie{g}$ has a codimension one ideal $\Lie{i}$ such that
$$ \Lie{g} = \Lie{i} \oplus \Lie{q}$$
as vector spaces. Pick $y\in \Lie{q}-\{0\}$. If $v$ is an endomorphism of a vector space $\mathcal{V}$, we denote the $v$-invariant subspace $\mathcal{V}^{\langle v \rangle} \subseteq \mathcal{V}$.

\begin{lemma}\label{q1Lie}
We have that
$$ H^q(\Lie{g},M) \cong H^q(\Lie{i}, M)^{\langle y \rangle} \oplus \frac{H^{q-1}(\Lie{i}, M)}{yH^{q-1}(\Lie{i}, M)}.$$
\end{lemma}

\noindent \textbf{Proof.} 
We consider the Hochschild--Serre spectral sequence \cite{HochSerre} for the pair $(\Lie{i},  \Lie{g})$ with $E_2$-page
$$ E^{p,q}_2 = H^p(\Lie{q}, H^q(\Lie{i},M)).$$
Since $\mathfrak{q}$ is one-dimensional, this $E_2$-page is concentrated in the columns $p=0,1$, thus the differential $\dd_2$ is trivial and the spectral sequence collapses at the $E_2$-page. Since in general this spectral sequence converges to $H^q(\Lie{g},M)$, we get that
$$ H^q(\Lie{g},M) \simeq H^0(\Lie{q},H^q(\Lie{i},M)) \oplus H^1(\Lie{q},H^{q-1}(\Lie{i},M)).$$
But
$$
H^0(\Lie{q},H^q(\Lie{i},M)) \simeq H^q(\Lie{i}, M)^{\langle y \rangle},\qquad H^1(\Lie{q},H^{q-1}(\Lie{i},M)) \simeq \frac{H^{q-1}(\Lie{i}, M)}{yH^{q-1}(\Lie{i}, M)}.$$\hfill $\Box$

We recall that $y\in \Lie{q}$ acts on $u \in Z^q(\Lie{i},M)$ by the formula
\begin{equation}\label{yaction}
 y(u)(x_1, \ldots x_q) = yu(x_1, \ldots x_q) - \sum_{i=1}^q u(x_1, \ldots, [y,x_i], \ldots, x_q).
 \end{equation}

The case where also $\Lie{i}$ is one-dimensional is particularly simple. Let $x\in \Lie{i}-\{0\}$. Then $[y,x]= ax$ for some $a\in {\mathbb C}$.  We then have that
$$
H^0(\Lie{i},M) \simeq M^{\langle x \rangle},\qquad H^1(\Lie{i},M) \simeq M/xM.
$$
Here the last isomorphism is induced by mapping $u\in Z^1(\Lie{i},M)$ to $u(x)\in M$. We then see that
$$y(u)(x) = yu(x) - u([y,x]) = (y-a \Id)u(x).$$
Thus we get that
$$H^1(\Lie{i},M)^{\langle y \rangle} = \left( M/xM\right)^{{\langle y-a \rangle}}.$$

\begin{lemma}\label{cod2}
In the case where $\Lie{g}$ is two-dimensional, with the notation as above, we have that
$$H^0(\Lie{g},M) \simeq  M^{{\langle x,y \rangle}},\qquad H^1(\Lie{g}, M) \simeq \left( M/xM\right)^{{\langle y-a \rangle}} \oplus \left(M^{\langle x \rangle}/yM^{\langle x \rangle}\right)\,,$$
and
$$ H^2(\Lie{g}, M) \simeq \frac{M}{xM + (y-a)M}.$$
\end{lemma}

\noindent \textbf{Proof.} 
Since we have already computed $H^0$ and $H^1$, we just need to compute $H^2$. From Lemma \ref{q1Lie}, we get that
$$  H^2(\Lie{g}, M)  \simeq H^1(\Lie{q}, H^1(\Lie{i},M)) \simeq \frac{M/xM}{(y-a)M/xM}$$
from which the result follows.
\hfill $\Box$

Let us now consider the case where $\Lie{i}$ has dimension two. Say $\Lie{i} = {\rm span}_{\mathbb{C}}\{x_1,x_2\}$, where  
$[x_2,x_1] = a x_1.$ We then have a description of the cohomology of $\Lie{i}$ from Proposition \ref{cod2}. However, in order to track the action of $y$, it is easier to work with the following model
$$ H^1(\Lie{i},M) \simeq K/N,$$
where 
\bea
K & = & \ker(\dd_2 : M\times M \ra M),\qquad  \dd_2(m_1,m_2) = x_1m_2 - (x_2-a)m_1 \nonumber \\
N & = & \im(\dd_1 : M \ra M\times M), \ \ \dd_1(m) = (x_1m,x_2m). \nonumber
\eea
If we now define $(a_{ij})_{i,j}$ by
\beq
 [y,x_1] = a_{11} x_1 + a_{12} x_2,\qquad   [y,x_2] = a_{21} x_1 + a_{22} x_2 \label{comut}
 \eeq
then the Jacobi identity demands that
$$ aa_{12} = 0, \qquad  a a_{22} = 0.$$
It is easy to check that the action of $y$ on $ H^1(\Lie{i},M)$ is induced from the following action of $y$ on $K$
$$ y(m_1,m_2) = \big((y-a_{11})m_1 -a_{12} m_2, (y-a_{22})m_2 - a_{21} m_1\big),$$
and using the above condition on $a$ and $ (a_{ij})_{i,j}$ one can easily verify that this action of $y$ indeed does preserve $K$ and that it maps $N$ to itself.

In order to compute the action of $y$ on $H^2(\Lie{i},M)$, we simply use (\ref{yaction}) to get that the action is induced by the following action of $y$ on $M$
$$y(m) = (y-a_{11}-a_{22})m.$$ Putting the above together, we get the following result.

\begin{lemma}\label{cod3}
In the case where $\Lie{g}$ is three-dimensional and has an ideal $\Lie{i}$ of codimension one, with the notation as above, we have that
$$H^0(\Lie{g},M) \simeq  M^{{\langle x_1,x_2,y \rangle}}, \ \ H^1(\Lie{g}, M) \simeq \left( K/N\right)^{{\langle y \rangle}} \oplus \left(M^{{\langle x_1,x_2 \rangle}}/yM^{{\langle x_1,x_2 \rangle}}\right)\,,$$
and
$$ H^2(\Lie{g}, M) \simeq  \left( \frac{M}{x_1M +x_2 M+ (y-a)M}\right)^{\langle y-a_{11}-a_{22}\rangle} \oplus \left(\frac{K}{N+ yK}\right)\,,$$
where $(a_{i,j})_{i,j}$ encode the commutation relations \eqref{comut}.
\end{lemma}

We can give an alternative description of the first cohomology group in the case where $\Lie{g}$ has dimension three with generators $x_1,x_2,y$ with structure constants as above, which is adapted to impose the extra linear constraints necessary to compute the tangent space to the moduli space of quantum (rather than quasi-) Airy structures. Let us introduce
\bea
K^{(3)}  & := &  \ker\big(\dd^{(3)}_2 \,:\,M\times M\times M \longrightarrow M\times M\times M\big),  \nonumber \\
d^{(3)}_2(m_1,m_2,m_3) & = & \big(- (x_2-a)m_1 + x_1m_2, -(y-a_{11})m_1 + a_{12}m_2 + x_1 m_3,a_{21}m_1 - (y-a_{22})m_2 + x_2m_3\big), \nonumber \\
N^{(3)}  &:= &  \im\big(\dd^{(3)}_1\,:\,M \longrightarrow M\times M \times M\big), \nonumber \\
d^{(3)}_1(m) & = & (x_1m,x_2m, ym).  \nonumber
\eea
Then we have that
$$
H^1(\Lie{g},M) \simeq K^{(3)}/N^{(3)}.
$$
In this description a cocycle $u: \Lie{g} \ra M$ is identified with $(m_1,m_2,m_3)\in K^{(3)}$ by
$$
u(x_1)=m_1,\qquad u(x_2) = m_2,\qquad u(y) = m_3.
$$

\subsection{Abelian cases}

\label{Secabelian}

From Lemma~\ref{basrem} and Section~\ref{S32} we know that when $V$ is a finite-dimensional abelian Lie algebra, the data of an abelian Lie subalgebra of $\mathfrak{sp}(T^*V)$ spanned by
$$
\varrho_1(e_i) = \left(\begin{array}{cc} -B^i & A^i \\ - C^i & (B^i)^{T}\end{array}\right)
$$
and of an arbitrary $D$ is equivalent to the data of a quantum Airy structure. To describe all abelian quantum Airy structure, we should take our matrices $\rho_1(e_i)$ belonging to a given maximal abelian Lie subalgebra of $\mathfrak{sp}(2n)$, and impose the constraint that $A$ is fully symmetric. If we wish to do so up to $\mathcal{G}_{V}$-equivalence, one should first list all isomorphism classes of maximal abelian Lie subalgebras of $\mathfrak{sp}(2n)$. As commuting matrices can always be simultaneously trigonalised, we can always achieve $C^i = 0$ for all $i$. The partition function is therefore explicitly computable by Proposition~\ref{Cequal0}.

The reference \cite{Winternitz} provides tools to classify conjugacy classes of maximal abelian subalgebras of $\mathfrak{sp}(T^*V)$. In particular it achieves a complete classification when $V$ has dimension $2$ or $3$, where there are respectively five and fourteen conjugacy classes (over $\mathbb{C}$).  For $\dim V \geq 4$, there exists continuous families of conjugacy classes of maximal abelian Lie subalgebras of $\mathfrak{sp}(T^*V)$, and the classification becomes more intricate. For this reason, the classification of abelian quantum Airy structures in finite dimensions --- and \textit{a fortiori} of quantum Airy structures --- modulo $\mathcal{G}_{V}$-action seems out of reach.

Given a maximal abelian subalgebra $\mathscr{A} \subseteq \mathfrak{sp}(2n)$, if we want to describe all quantum Airy structures in which the $\varrho_1(e_i) \in \mathscr{A}$ for all $i$, we have to impose the relations $A^i_{j,k} = A^j_{i,k}$, and $B^i_{j,k} = B^j_{i,k}$ corresponding to the vanishing of the structure constants. This step amounts to classify the isomorphisms $\mathcal{I}$ for which \eqref{Con1} is satisfied.

We give the resulting list of quantum Airy structures below for dimension two and three based on the classification of conjugacy classes of $\mathscr{A}$  respecting their labelling in \cite{Winternitz}. In all cases, we find that for each $i$, either $A^i = 0$ or $B^i = 0$. As a matter of fact, the aforementioned conditions on $A$ and $B$ often constraint the values of $\varrho_1$ within a given $\mathscr{A}$, which makes some cases in the list below special cases of others. 

In the following, $a,b,c,d$, etc. denote arbitrary scalars and $x_1 = x$, $x_2 = y$ and $x_3 = z$, and it is understood that one always has the freedom to add $-\hbar D^i$ to each $L_i$ for arbitrary constants $D^i$.

\subsubsection{Dimension two}

\begin{itemize}
\item[\boxed{1}] Special case of \boxed{5} with $b = 0$.
\item[\boxed{2}] $L_1 = \hbar(1 - ax)\partial_{x}$ and $L_2 = \hbar \partial_{y} - \frac{by^2}{2}$.
\item[\boxed{3}] $L_1 = \hbar \partial_{x} - (\frac{ax^2}{2} + bxy + \frac{cy^2}{2})$, $L_2 = \hbar \partial_{y} - (\frac{bx^2}{2} + cxy + \frac{dy^2}{2})$.
\item[\boxed{4}] Special case of $\boxed{3}$ with $b = c = d = 0$.
\item[\boxed{5}] $L_1 = \hbar(1 - ax - by)\partial_{x}$, $L_2 = \hbar(1 - bx - cy)\partial_{y}$.
\end{itemize}

\subsubsection{Dimension three}

\begin{itemize}
\item[\boxed{$1$}] $L_1 = \hbar(1 - ax)\partial_{x}$, $L_2 = \hbar(1 - by)\partial_{y}$, $L_3 = \hbar(1 - cz)\partial_{z}$. 
\item[\boxed{$2$}] $L_1 = \hbar(1 - ax)\partial_{x}$, $L_2 = \hbar(1 - by)\partial_{y}$, $L_3 = \hbar \partial_{z} - \frac{cz^2}{2}$.
\item[\boxed{$3$}] $L_1 = \hbar(1 - ex)\partial_{x}$, $L_2 = \hbar\partial_{y} - (\frac{ay^2}{2} + byz + \frac{cz^2}{2})$, $L_3 = \hbar\partial_{z} - (\frac{by^2}{2} + cyz + \frac{dz^2}{2})$.
\item[\boxed{$4$}] $L_1 = \hbar(1 - ax - by)\partial_{x}$, $L_2 = \hbar\partial_{y}$, $L_3 = \hbar\partial_{z} - \frac{cz^2}{2}$.
\item[\boxed{$5$}] $L_1 = \hbar(1 - ax)\partial_{x}$, $L_2 = \hbar(1 - by)\partial_{y} - \hbar (cy + bz)\partial_{z}$, $L_3 = \hbar(1 - by)\partial_{z}$. 
\item[\boxed{$6$}] $L_1 = \hbar \partial_{x} - \frac{ax^2}{2}$, $L_2 = \hbar(1 - by)\partial_{y} - (cy + bz)\partial_{z}$, $L_3 = \hbar(1 - bz)\partial_z$.
\item[\boxed{$7$}] $L_i = \hbar \partial_{x_i} - \frac{1}{2} \sum_{1 \leq j,k \leq 3} A^{i}_{j,k}x_jx_k$ for $i \in \{1,2,3\}$, with $A^{i}_{j,k}$ fully symmetric in its three indices.
\item[\boxed{$8$}] Special case of \boxed{7} with $A^{3} = 0$.
\item[\boxed{$9$}] Special case of \boxed{7} with $A^3 = 0$.
\item[\boxed{$10$}] Special case of \boxed{7} with $A^2 = A^3 = 0$.
\item[\boxed{$11$}]  Special case of \boxed{7} with $A^2 = A^3 = 0$.
\item[\boxed{$12$}]  $L_1 = \hbar\partial_{x} - \hbar(ax + by)\partial_{z}$, $L_2 = \hbar \partial_{y} - \hbar (bx + cy)\partial_{z}$, $L_3 = \hbar \partial_{z}$.
\item[\boxed{$13$}]  $L_1 = \hbar(1 - ax) \partial_{x} - \hbar (bx + cy)\partial_{y} - \hbar(cx + az)\partial_{z}$, $L_2 = \hbar(1 - ax)\partial_{y}$, $L_3 = \hbar(1 - ax)\partial_{z}$.
\item[\boxed{$14$}] $L_1 = \hbar(1 - ax)\partial_{x} - \hbar(bx + cy)\partial_{y} - \hbar (cx + by + az)\partial_{z}$, $L_2 = \hbar \partial_{x} - \hbar ax \partial_{y} - \hbar(bx + cy)\partial_{z}$, $L_3 = \hbar(1 - ax)\partial_{z}$.
\end{itemize}

\subsection{Non-abelian two-dimensional Lie algebra}

In dimension two, there is a unique non-abelian Lie algebra up to isomorphism. It is the Lie algebra of affine transformations of $\mathbb{R}^2$, generated by $L_0$ and $L_1$ with the commutation relation
\beq
\label{affinLie}[L_0,L_1] = -L_1\,.
\eeq
We introduce a $\mathbb{Z}_2$-grading by declaring $t_i$ and $\partial_{t_i}$ to have degree $i$ for $i \in \{0,1\}$. We remark that, using the conjugation by an operator of the form $\exp\big(\tfrac{\hbar}{2} u_{a,b}\partial_{x_a}\partial_{x_b}\big)$, one can generically bring $L_0$ to a form where $C^0 = 0$.

\begin{proposition}
\label{d2Leonid} The complete list of classical Airy structures such that
\begin{itemize}
\item[$(\mathbf{1})$] $L_i$ has degree $i$ for $i \in \{0,1\}$,
\item[$(\mathbf{2})$] $L_0$ has no differential operators of order two, \textit{i.e.} $C^0 = 0$,
\end{itemize}
reads as follows
\beq
\begin{array}{crclcrcl} \mathbf{Ia} & L_0 & = & \hbar \partial_{x_0} - \alpha x_0^2 + \tfrac{\beta\alpha}{4}x_1^2 + 2\hbar x_0\partial_{x_0} + \hbar x_1\partial_{x_1}  &&   L_1 & = & \hbar \partial_{x_1}  + \tfrac{\alpha\beta}{2} x_0x_1 - \hbar \beta x_1\partial_{x_0}  \\
\mathbf{Ib} & L_0 & = & \hbar \partial_{x_0} - \alpha x_0^2 - \beta x_1^2 - 2\hbar x_0\partial_{x_0} - \hbar  x_1 \partial_{x_1} && L_1 & = & \hbar \partial_{x_1} - 2\beta x_0x_1 - 2\hbar x_0\partial_{x_1} \\
\mathbf{Ic} & L_0 & = & \hbar \partial_{x_0} - \alpha x_0^2 - \hbar(\beta + 1)x_0\partial_{x_0} - \hbar\beta x_1\partial_{x_1} && L_1 & = & \hbar \partial_{x_1} - \hbar(\beta + 1)x_0\partial_{x_1}  \\
\mathbf{IIa} & L_0 & = & \hbar \partial_{x_0} + \tfrac{4}{\alpha}x_0^2 - \beta x_1^2 + 2\hbar x_0\partial_{x_0} - \hbar x_1\partial_{x_1} && L_1 & = &  \hbar \partial_{x_1} - 2\beta x_0x_1 - \hbar \alpha\beta x_1\partial_{x_0} - 2\hbar x_0\partial_{x_1} - \hbar^2 \alpha \partial_{x_0}\partial_{x_1}  \\
\mathbf{IIb} & L_0 & = & \hbar \partial_{x_0} + \frac{\alpha^2}{\beta}x_0^2 + \hbar \alpha x_0\partial_{x_0} + \hbar (1 - \alpha) x_1\partial_{x_1} && L_1 & = & \hbar \partial_{x_1} - \hbar \alpha x_0\partial_{x_1} - \hbar^2\gamma\partial_{x_0}\partial_{x_1} \end{array} \nonumber
\eeq
where $\alpha,\beta,\gamma \in \mathbb{C}$ (and non-zero when they appear in the denominator). Quantum Airy structures are obtained from these by adding an arbitrary constant $-\hbar D^0$ to $L_1$, while $D^1 = 0$. \hfill $\Box$
\end{proposition}
This result is obtained by inserting the form of the most general quantum Airy structure with properties (\textbf{1}) and (\textbf{2}) into the \textbf{BB-CA}, \textbf{BC} and \textbf{BA} relations, and solving the resulting equations.  The \textbf{D} relation can be analysed separately, and we indeed find that quantum Airy structures based on \eqref{affinLie} must have $D^1 = 0$, while $D^0$ is unconstrained. We omit the details of this proof. The groups \textbf{I} and \textbf{II} correspond to $C = 0$ or $C^1 \neq 0$.

Now we want to consider the existence of deformations modulo $\mathcal{G}_{V}$ of these structures. We remark that this may not preserve the homogeneity condition (\textbf{1}). In particular, with the following transformations --- which belong to $\mathcal{G}_{V}$
\beq
t_0\to e^{\xi_1}t_0,\,\,\,\partial_0\to e^{-\xi_1}\partial_0,\,\,\, t_1\to e^{\xi_2}t_1\,\,\, \partial_1\to e^{-\xi_2}\partial_1,\,\,\,\hbar\to e^{-\xi_1}\hbar,\,\,\,L_1\to e^{\xi_2-\xi_1}L_1, \qquad \xi_i \in \mathbb{C}
\label{scaling}
\eeq
and assuming the parameters $\alpha,\beta,\gamma$ non-zero, and in each case one can use them to set the parameters to $1$, except in case \textbf{Ic} where one can only set $\alpha = 1$ while $\beta$ remains a free parameter.

Now let us assume that $\alpha,\beta,\gamma$ are generic. We compute the Lie algebra cohomologies governing the deformations of quantum Airy structures modulo $\mathcal{G}_{V}$ using Lemma~\ref{cod2}, which in our case applies with $x = L_1$, $y = L_0$ and $a = -1$. The result is

\begin{center}
\begin{tabular}{|l|c|c|c|c|c|}
\hline & \textbf{Ia} & \textbf{Ib} & \textbf{Ic} & \textbf{IIa} & \textbf{IIb} \\
\hline
\hline $\dim H^0$ & $1$ & $1$ & $2$ & $1$ & $2$ \\
\hline $\dim H^1$ & $1$ & $1$ & $2$ & $1$ & $2$ \\
\hline $\dim H^2$ & $0$ & $0$ & $0$ & $0$ & $0$ \\
\hline
\end{tabular}
\end{center}

We observe that $H^2$ always vanishes, \textit{i.e.} there are no obstructions to deformations. As the conjugation by scalars acts trivially on $\mathcal{D}$, $H^0$ is at least one-dimensional. In the two cases where $H^0$ has dimension two, the extra generator in fact has a trivial action on $H^1$, so $\dim H^1$ gives the number of independent deformations of the corresponding quantum Airy structure. In all cases, the value of $D^0$ gives an extra one-parameter deformation. In \textbf{Ic}, $\beta$ gives the second deformation parameter --- one sees for instance that it cannot be gauged out by the scaling transformations \eqref{scaling}. In case \textbf{IIb}, one finds that the second one-parameter deformation (called $t$) is
\beq
\label{IIbt} \textbf{IIb}\,\,:\,\, L_i^{(t)} = L_0 + tL_i',\qquad L_0' = \hbar (1 - 2\alpha)x_0\partial_{x_1},\qquad L_1' = \hbar^2\beta \partial_{x_1}^2
\eeq

Although the cohomology groups we computed only give the number of independent deformations within quasi-Airy structures, we actually find in these five examples that they are always realised within Airy structures.

The partition functions for each of these quantum Airy structures --- with an arbitrary $D^0$ --- can be computed by hand. As the group \textbf{I} has $C = 0$, we can also use Lemma~\ref{Cequal0}.
\bea
Z_{{\rm Ia}} & = & (1 + 2x_0 + \beta x_1^2)^{\frac{a}{8\hbar} + \frac{D^0}{2}}\exp\big\{\hbar^{-1}\big(\tfrac{\alpha x_0(x_0 - 1)}{4} - \tfrac{\alpha\beta x_1^2}{8}\big)\big\} \nonumber \\
Z_{{\rm Ib}} & = & (1 - 2x_0)^{-\tfrac{\alpha}{8\hbar} -\frac{D^0}{2}}\exp\big\{\hbar^{-1}\big(-\tfrac{\alpha x_0(x_0 + 1)}{4}\tfrac{\beta x_0x_1^2}{1 - 2x_0}\big)\big\} \nonumber \\
\label{d2LeonidZ} Z_{{\rm Ic}} & = & (1 - (\beta + 1)x_0)^{-\frac{\alpha}{\hbar(\beta + 1)^3} - \frac{D^0}{\beta + 1}}\exp\big(-\tfrac{\alpha x_0(2 + (\beta + 1)x_0)}{2\hbar ( \beta + 1)^2}\big) \\
Z_{{\rm IIa}} & = & (1 + 2x_0)^{ - \frac{1}{4\hbar\alpha} + \frac{D^0}{4}}x_1^{\frac{1}{2\alpha\hbar} -\frac{D^0}{2}} \exp\big\{\hbar^{-1}\big(\tfrac{x_0(1 - x_0)}{
\alpha} -\tfrac{\beta x_1^2}{2}\big)\big\}\,J_{- \frac{1}{2\alpha\hbar} + \frac{D^0}{2}}\Big[\hbar^{-1}x_1\big(\tfrac{\beta(1 + 2x_0)}{\alpha}\big)^{\frac{1}{2}}\Big] \nonumber \\
Z_{{\rm IIb}} & = & (1 + \alpha x_0)^{-\frac{1}{\hbar\alpha\beta} + \frac{D^0}{\alpha}}\,\exp\big(\tfrac{x_0(2 - \alpha x_0)}{\beta\hbar}\big) \nonumber
\eea
The most interesting case is $Z_{{\rm IIa}}$, where we see an appearance of the Bessel function $J_{\nu}(z)$. It is characterised by
$$
\big(z^2\partial_{z}^2 + z\partial_{z} + (z^2 - \nu^2)\big)J_{\nu}(z) = 0,\qquad J_{\nu}(z) := \sum_{m \geq 0} \frac{\Gamma(\nu + 1)}{\Gamma(m + \nu + 1)}\,\frac{(-1)^{m}}{m!}\,\bigg(\frac{z}{2}\bigg)^{2\nu + m}\,.
$$
Our normalisation by a constant prefactor of $J_{\nu}(z)$ is not the conventional one, but we made it so that $J_{\nu}(z) = z^{2\nu}(1 + O(z))$ when $z \rightarrow 0$.
In the case \textbf{IIb}, the partition function for the one-parameter deformation \eqref{IIbt} is in fact independent of the parameter $t$, and independent of $x_1$.

\subsection{The case \texorpdfstring{$n=3$}.}

When considering isomorphism classes of three-dimensional non-trivial Lie algebra over $\mathbb{C}$, one finds \cite{Winternitzbook} four rigid cases and a one-parameter family.
\begin{itemize}
\item[$\bullet$] $\mathfrak{sl}_{2}$, which was treated in Theorem~\ref{sl2th}.
\item[$\bullet$] The direct sum of the non-abelian two-dimensional algebra and the abelian Lie algebra of dimension $1$. We do not study this direct sum case.
\item[$\bullet$] The Heisenberg Lie algebra $[y_1,y_2] = y_3$, $[y_1,y_3] = [y_2,y_3] = 0$, which we do not treat.
\item[$\bullet$] The Lie algebra $[y_1,y_2] = y_2$ and $[y_1,y_3] = y_2 + y_3$, which we also do not treat.
\item[$\bullet$] The Bianchi VI Lie algebra $\mathfrak{l}_{q}$  defined by $[y_1,y_2] = y_2$, $[y_1,y_3] = qy_3$, $[y_2,y_3] = 0$, for $q \in \mathbb{C}^*$. We have $\mathfrak{l}_{q} \simeq \mathfrak{l}_{r}$ if and only if $q = r$ or $qr = 1$. For $q = -1$, this is the Lie algebra of infinitesimal isometries of euclidean $\mathbb{R}^2$.
\end{itemize}

We will not attempt here at the classification of Airy structures supported by each of these Lie algebras. We rather look for a non-trivial example of quantum Airy structure based on $\mathfrak{l}_{q}$, keeping as many non-zero elements as possible, which illustrates the non-triviality of the problem of exhibiting finite-dimensional quantum Airy structures. We rename the variables $(x_0,x_0',x_1)$, and assign a $\mathbb{Z}_{2}$-degree $0$ to $x_0,x_0'$, and $1$ to $x_1$. We postulate the commutations relation of $\mathfrak{l}_{q}$ in the following form
$$
[L_0,L_1] = L_1,\qquad [L_0,L_0'] = qL_0'\qquad [L_0',L_1] = 0,\qquad \qquad {\rm that}\,\,{\rm is}\,\,L\,:\,\,(y_1,y_2,y_3) \rightarrow (L_0,L_1,L_0')
$$
and look for a quantum Airy structure such that $L_0,L_0'$ have degree $0$ and $L_1$ has degree $1$. For the same Lie algebra, our computations have shown that the other choice of grading $(y_1,y_2,y_3) \mapsto (L_0,L_1,L_1')$ does not support non-trivial Airy structures.

\begin{proposition}
\label{Ldim3} Let $\zeta$ be a root of $P(\zeta) := 2\zeta^3 - 2\zeta^2 + 3\zeta - 1$ and $\alpha$ a complex parameter. The matrices, which we write with respect to the basis $(e_0,e_0',e_1)$, $C^0 = 0$ and
\bea
A^0 = \left(\begin{array}{ccc} 1 - \tfrac{5}{3}\zeta & \tfrac{4}{3}(-2\zeta^2 + \zeta - 1) & 0 \\ \tfrac{4}{3}(-2\zeta^2 + \zeta - 1) & \tfrac{2}{3}(-\zeta + 1) & 0 \\ 0 & 0 & 2\alpha \end{array}\right) && B^0 = \left(\begin{array}{ccc} 1 -3\zeta & 3(-2\zeta^2 + \zeta - 1) & 0 \\ 3(-2\zeta^2 + \zeta - 1) & 3\zeta + 1 & 0 \\ 0 & 0 & 1 \end{array}\right) \nonumber \\
A'^0 = \left(\begin{array}{ccc} \tfrac{4}{3}(-2\zeta^2 + \zeta - 1) & \tfrac{2}{3}(-\zeta + 1) & 0 \\ \tfrac{2}{3}(-\zeta + 1) & 0 & 0 \\ 0 & 0 & -\alpha(2\zeta^2 + 1)  \end{array}\right) && 
B'^0 = \left(\begin{array}{ccc} 3(-2\zeta^2 + \zeta - 1) & 3\zeta - 1 & 0 \\ -\zeta + 1 & -2\zeta^2 + \zeta - 1 & 0 \\ 0 & 0 & -2\zeta^2 - 1 \end{array}\right) \nonumber \\
A^1 = \left(\begin{array}{ccc} 0 & 0 & 2\alpha \\ 0 & 0 & -\alpha(2\zeta^2 + 1) \\ 2\alpha & -\alpha(2\zeta^2 +1) & 0  \end{array}\right) && B^1 = \left(\begin{array}{ccc} 0 & 0 & 2 \\ 0 & 0 & -2\zeta^2 - 1 \\ 6\alpha & 3\alpha(2\zeta^2 + 1) & 0 \end{array}\right) \nonumber \\
C'^0 = \left(\begin{array}{ccc} 6(-2\zeta^2 + \zeta - 1) & 6\zeta & 0 \\ 6(2\zeta^2 - \zeta + 1) & 0 & 0 \\ 0 & 0 & -\alpha^{-1}(2\zeta^2 + 1) \end{array}\right) && C^1 = \left(\begin{array}{ccc} 0 & 0 & 6 \\ 0 & 0 & 3(2\zeta^2 + 1) \\ 6 & 3(2\zeta^2 + 1) & 0 \end{array}\right) \nonumber
\eea
together with
$$
D^0\,\,{\rm arbitrary},\qquad D'^{0} = \tfrac{1}{2}\,{\rm Tr}\,B'^{0} = \tfrac{1}{2}(-5\zeta^2 + 4\zeta - 5),\qquad D^1 = 0
$$
define a quantum Airy structure based on the Lie algebra $\mathfrak{l}_{-2}$. It has
$$
\dim H^i_{L}(V,\mathcal{D}_{V}) =  \left\{\begin{array}{rcl} 1 & & i = 0 \\ 2  & & i = 1 \\ 1 & & i = 2 \end{array}\right. \,.
$$
The stabiliser $\mathcal{G}_{V}(L)$ is trivial. $D^0$ is the only deformation of this quantum Airy structure.
\end{proposition}

\noindent \textbf{Proof.} One can check by direct computation that the assignments in this Lemma indeed satisfies the desired commutation relations.  We have however opted to provide some details explaining how we were led to this quantum Airy structure, in the hope that it may lead to the construction of other quantum Airy structures by similar techniques.

The main task is to find $(A,B,C)$ defining a classical Airy structure, \textit{i.e.} such that
$$
[\tilde{\varrho}_{1}(e_0),\tilde{\varrho}_{1}(e_1)] = \tilde{\varrho}_{1}(e_1),\qquad [\tilde{\varrho}_{1}(e_0),\tilde{\varrho}_1(e_0')] = q\tilde{\varrho}_{1}(e_0'),\qquad [\tilde{\varrho}_{1}(e_0'),\tilde{\varrho}_{1}(e_1)] = 0\,,
$$
with $\tilde{\rho}_{1}$ of the form \eqref{rho1ti}. As our goal is not to find all possible solutions, we are going to postulate certain properties, which eventually lead to a solution.

\vspace{0.1cm}

\noindent \textbf{1st postulate.} $L_0,L_0'$ have degree $0$, $L_1$ has degree $1$. \\
\noindent \textbf{2nd postulate.} $C^0 = 0$ --- for generic operators this can be achieved by a $\mathcal{G}_{V}$ transformation.

\vspace{0.1cm}

The matrices then take the form
\begin{align}\label{block}
  &B^0=\left(\begin{array}{c|c}
              \tilde{B} & 0 \\
              \hline
              0 & b_{33}
            \end{array} \right), \quad B'^0=\left(\begin{array}{c|c}
              \tilde{B}' & 0 \\
              \hline
              0 & b'_{33}
            \end{array} \right), \quad C'^0=\left(\begin{array}{c|c}
              2\tilde{C}' & 0 \\
              \hline
              0 & 2c'_{33}
            \end{array} \right), \quad  A^0=\left(\begin{array}{c|c}
              2\tilde{A} & 0 \\
              \hline
              0 & 2a_{33}
            \end{array} \right), \nonumber\\
   & A'^0=\left(\begin{array}{c|c}
              2\tilde{A}' & 0 \\
              \hline
              0 & 2a'_{33}
            \end{array} \right), \quad B_1=\left(\begin{array}{c|c}
              0 & {\mathbf v}^{\text{T}} \\
              \hline
              {\mathbf u} & 0
            \end{array} \right), \quad  C^1=\left(\begin{array}{c|c}
              0 & 2{\mathbf c}^{\text{T}} \\
              \hline
              2{\mathbf c} & 0
            \end{array} \right), \quad A^1=\left(\begin{array}{c|c}
              0 & 2{\mathbf a}^{\text{T}} \\
              \hline
              2{\mathbf a} & 0
            \end{array} \right), \\ 
  & \tilde{A}^{\text{T}}=\tilde{A},\ (\tilde{A}')^{\text{T}}=\tilde{A}', \ (\tilde{C}')^{\text{T}}=\tilde{C}'\quad{\hbox{and}}\quad a_{12}=a'_{11},\ a_{22}=a'_{12},\ a_{33}={\mathbf a}_1,\
  a'_{33}={\mathbf a}_2.\nonumber 
\end{align}
where $\tilde{A},\tilde{A}',\tilde{C},\tilde{C}'$ are $2 \times 2$ matrices, and $\mathbf{a},\mathbf{c},\mathbf{u},\mathbf{v}$ are row vectors of size two. The relation \eqref{f2B} between the structure constants of the Lie algebra and antisymmetric part of $B$ results in
\beq
b'_{1,i}-b_{2,i}=q\delta_{i,2} \quad {\rm for}\,\, i=1,2;\qquad {\mathbf v}_2=b'_{33},\qquad  {\mathbf v}_1=b_{33}+1.
\label{n3.1}
\eeq
We first address the relation $[\tilde{\varrho}_{1}(0),\tilde{\varrho}_1(e'_0)]=q\tilde{\varrho}(e_0')$. It results in the following relations
\begin{align}
   &-\tilde{C}'\tilde{B}-\tilde{B}^{\text{T}}\tilde{C}'=q\tilde{C}',\qquad -2c'_{33}b_{33}=qc'_{33}\,,\label{L001}\\
   & [\tilde{B},\tilde{B}']-4\tilde{A}\tilde{C}'=q\tilde{B}',\qquad -4c'_{33}a_{33}=q b'_{33}\,,\label{L002}\\
   & \tilde{B}\tilde{A}'+\tilde{A}'\tilde{B}^{\text{T}}-\tilde{B}'\tilde{A}-\tilde{A}(\tilde{B}')^{\text{T}}=q\tilde{A}',\qquad 2b_{33}a'_{33}-2b'_{33}a_{33}=q a'_{33}\,.\label{L003}
\end{align}

\vspace{0.1cm}

\noindent \textbf{3rd postulate.} We assume $C'^0$ is non-degenerate, \textit{i.e.} $\det \tilde{C}'\ne 0$ and $c'_{33}\ne 0$.

\vspace{0.1cm}

Then, from the second equation in (\ref{L001}), we have that $b_{33}=-\tfrac{q}{2}$. From (\ref{L003}) we then obtain
$$
b'_{33}a_{33}=-qa'_{33}.
$$
The first equation in (\ref{L001}) implies that $\tilde{B} = S-\tfrac{q}{2}\mathbb {\rm Id}$, where the matrix $S$ satisfies the equation
\beq
\label{STCeq} \tilde{C}'S=-S^{\text{T}}\tilde{C}',
\eeq
which admits a one-parameter family of solutions: we can express $\tilde{B}$ in terms of $\tilde{C}'$ and the parameter $t$
\beq
    \label{B} \tilde{B}=-\tfrac{q}{2} {\rm Id} + t\left(\begin{array}{cc}
                               -c'_{12} & c'_{11} \\
                               -c'_{22} & c'_{12}
                             \end{array}\right)\,.
\eeq
From (\ref{n3.1}) we then have $b'_{11}=-tc'_{22}$, $b'_{12}=tc'_{12}+\tfrac{q}{2}$, $\mathbf v_1=1-\tfrac{q}{2}$, and $\mathbf v_1=b'_{33}$. Due to general scaling transformations in $\mathcal{G}_{V}$ for even and odd variables, we have two \emph{free parameters}: $t$ and $a_{33}$.

\vspace{0.1cm}

\noindent \textbf{4th postulate} $t = 1$ -- we however keep $a_{33} = 2\alpha$ arbitrary.

We now solve (\ref{L002}) with respect to $\tilde{A}$. We first rewrite it in terms of $S$
$$
[S,\tilde{B}']-4\tilde{A}\tilde{C}'=q\tilde{B}'.
$$
Because $-S(\tilde{C}')^{-1}=(\tilde{C}')^{-1}\tilde{S}^{\text{T}}$ from \eqref{STCeq}, multiplying by $(\tilde{C}')^{-1}$ from the right, we obtain that
\beq
4\tilde{A}=S\tilde{B}'(\tilde{C}')^{-1}+\tilde{B}'(\tilde{C}')^{-1}S^{\text{T}}-q\tilde{B}'(\tilde{C}')^{-1}.
\label{A->SBC}
\eeq
Introducing $R:=\tilde{B}'(\tilde{C}')^{-1}$, we obtain that $4\tilde{A}=SR+RS^{\text{T}}-qR$. Another useful information can be extracted from the fact that
$\tilde{A}$ is symmetric: splitting $R$ into symmetric and skew-symmetric parts, $R=R_{\text{sym}}+R_{\text{asym}}$, we observe that
$$
SR_{\text{asym}}+R_{\text{asym}}S^{\text{T}}-qR_{\text{asym}}=0,
$$
wherefrom, accounting for an explicit form of $S$, the first two terms are mutually cancelled, and we obtain that for $q\ne 0$,
$$
R_{\text{asym}}=0,
$$
that is,
\beq
\tilde{B}'(\tilde{C}')^{-1}=(\tilde{C}')^{-1}(\tilde{B}')^{\text{T}},\ \ \hbox{or}\ \ \tilde{C}'\tilde{B}'=(\tilde{B}')^{\text{T}}\tilde{C}',
\label{L00BC}
\eeq
which in the component form results in the condition
\beq
0=c'_{12}(c'_{11}+c'_{22})-c'_{11}+c'_{12}b'_{22}-c'_{22}b'_{21}\,.
\label{eq-2}
\eeq

Because $\tilde{B}=S-\tfrac{q}{2}{\rm Id}$, equation (\ref{L003}) in terms of the matrices $S$, $\tilde{A}$, $\tilde{A}'$, and $\tilde{B}'$ reads
$$
S\tilde{A}'+\tilde{A}'S^{\text{T}}-\tilde{B}'\tilde{A}-\tilde{A}(\tilde{B}')^{\text{T}}=2q\tilde{A}'.
$$
We solve this equation as follows. It follows from this equation that
\beq
\tilde{A}'[S^{\text{T}}-q{\rm Id}]=\tilde{B}'\tilde{A}+S_A,\qquad S_A=\left(\begin{array}{cc}
                   0 & z \\
                   -z & 0
                 \end{array}\right)
\label{A-A}
\eeq
for some $z \in \mathbb{C}$.

Consider now the equation $[\tilde{\varrho}_{1}(e_0),\tilde{\varrho}_{1}(e_1)]=\tilde{\varrho}_{1}(e_1)$. Substituting the form (\ref{block}) for the respective matrices, we obtain
\begin{align}
  &(\tilde{B}-(b_{33} + 1){\rm Id}){\mathbf v}^{\text{T}}-4\tilde{A}{\mathbf c}^{\text{T}}=0, \label{L011}\\
  &{\mathbf u}(\tilde{B}-(b_{33} + 1){\rm Id})-4a_{33}{\mathbf c}=0, \label{L012}\\
  & {\mathbf c}\bigl[ \tilde{B}+(b_{33} + 1){\rm Id}\bigr]=0, \label{L013}\\
  &\bigl[\tilde{B}+(b_{33} - 1){\rm Id}\bigr]{\mathbf a}^{\text{T}}-\tilde{A}{\mathbf u}^{\text{T}}-a_{33}{\mathbf v}^{\text{T}}=0. \label{L014}
\end{align}

\vspace{0.1cm}

\noindent\textbf{5th postulate.} $\mathbf c\ne 0$.

\vspace{0.1cm}

Then, $\mathbf{c}$  has to be in the kernel of the matrix $\bigl[ S+ (1-q){\rm Id}\bigr]$.
From the form of the matrix $S$ we have that
\beq
\det(S+\lambda{\rm Id})=\lambda^2-[c'_{12}]^2+c'_{11}c'_{22}.
\label{DET}
\eeq

\vspace{0.1cm}
\noindent \textbf{6th postulate.} $A_{33}\ne 0$, and $\pm 1$ are not eigenvalues of $S$. From \eqref{DET}, this implies $q \notin \{0,2\}$.

\vspace{0.1cm}

The above system of equations become
\begin{align}
  &[S-{\rm Id}]{\mathbf v}^{\text{T}}=4\tilde{A}{\mathbf c}^{\text{T}}, \label{L011a}\\
  &-{\mathbf u}[S+ {\rm Id}]=4a_{33}{\mathbf c}, \label{L012a}\\
  & {\mathbf c}\bigl[S+(1-q){\rm Id}\bigr]=0, \label{L013a}\\
  &[S-(1+q){\rm Id}]{\mathbf a}^{\text{T}}=\tilde{A}{\mathbf u}^{\text{T}}+a_{33}{\mathbf v}^{\text{T}}. \label{L014a}
\end{align}
Acting by $[S+(1-q){\rm Id}]$ on (\ref{L012a}) from the left, we obtain zero on the both sides; because $S+ {\rm Id}$ is non-degenerate by our 6th postulate, and commutes with $[S+(1-q){\rm Id}]$, the vector $\mathbf u$ is also a zero vector of $[S+(1-q){\rm Id}]$ and it is therefore non-zero itself and proportional to $\mathbf c$, so
\beq
\mathbf u=-\frac{4a_{33}}{q}\mathbf c.
\label{L01b}
\eeq
Next, we transform (\ref{L014a}) using (\ref{L01b}) and (\ref{L011a}) we have that
$$
\tilde{A}{\mathbf u}^{\text{T}}=-\frac{4a_{33}}{q}\tilde{A}{\mathbf c}^{\text{T}}=-\frac {a_{33}}{q}[S-{\rm Id}]{\mathbf v}^{\text{T}}
$$
and (\ref{L014a}) becomes
\beq
[S-(q+1){\rm Id}]\Bigl[{\mathbf a}^{\text{T}}+\frac{a_{33}}{q}{\mathbf v}^{\text{T}}\Bigr]=0.
\eeq

\vspace{0.1cm}
\noindent \textbf{7th postulate.} $\det(S-(q+1){\rm Id})\ne 0$.

\vspace{0.1cm}

Then
\beq
\label{LLL}\mathbf a=-\tfrac{a_{33}}{q}\mathbf v=
\left(-\tfrac{a_{33}}{q}\mathbf v_1,\ -\tfrac{a_{33}}{q}\mathbf v_2 \right)= \left(-\tfrac{a_{33}}{q}\Bigl(1-\tfrac {q}{2}\Bigr),\ +a'_{33} \right).
\eeq
The condition $a'_{33}=\mathbf a_2$ is satisfied automatically, whereas the condition $a_{33}=\mathbf a_1$ implies
\beq
\boxed{q=-2.}
\label{b-2}
\eeq
From now on, we set $q=-2$ in all further calculations. Note that thus the chosen $q$ implies the non-degeneracy of the above determinants $\det(S+\lambda {\rm Id})$ with $\lambda =-1,1$, and with $\lambda=1+q=-1$.

Because $S+3{\rm Id}$ has a non-zero kernel, we have that
\beq
\det (S+3{\rm Id})=9-[c'_{12}]^2+c'_{11}c'_{22}=0.
\label{eq-1}
\eeq
We now express $a_{ij}$ from the formula (\ref{A->SBC}). After a few computations we obtain
\begin{align}
  4a_{11}&= -\tfrac{1}{9} \Bigl[2(1-c'_{12})\bigl( -[c'_{22}]^2 -[c'_{12}]^2 +c'_{12}\bigr)+2c'_{11}\bigl( c'_{12}(c'_{11}+c'_{22})-c'_{11}\bigr) \Bigr], \label{eq-3}\\
  4a_{12}&=-\tfrac{1}{9} \Bigl[c'_{11}(-c'_{12}b'_{21}+c'_{11}b'_{22}) -c'_{22}\bigl( -[c'_{22}]^2 -[c'_{12}]^2 +c'_{12}\bigr)+2\bigl( c'_{12}(c'_{11}+c'_{22})-c'_{11}\bigr) \Bigr],  \label{eq-4}\\
  4a_{22}&=-\tfrac{1}{9} \Bigl[2(c'_{12}+1)(-c'_{12}b'_{21}+c'_{11}b'_{22}) -2c'_{22}\bigl( c'_{12}(c'_{11}+c'_{22})-c'_{11}\bigr) \Bigr]. \label{eq-5}
\end{align}

Finally, let us consider the last remaining condition $[\tilde{\varrho}_1(e_0'),\tilde{\varrho}_{1}(e_1)]=0$. We obtain the system
\begin{align*}
  &[\tilde{B}'-b'_{33}{\rm Id}]{\mathbf v}^{\text{T}}=4\tilde{A}'{\mathbf c}^{\text{T}}-4c'_{33}{\mathbf a}^{\text{T}},\\
  &{\mathbf u}[\tilde{B}'-b'_{33}{\rm Id}]=4\mathbf a \tilde{C}'-4a'_{33}{\mathbf c},\\
  &{\mathbf c}\bigl[\tilde{B}'+b'_{33}{\rm Id}\bigr]=\mathbf v \tilde{C}'+\mathbf u c'_{33},\\
  &[\tilde{B}'+b'_{33}{\rm Id}]{\mathbf a}^{\text{T}}=\tilde{A}'{\mathbf u}^{\text{T}}+a'_{33}{\mathbf v}^{\text{T}}.
\end{align*}
Because $\mathbf a+\tfrac{a_{33}}{q}\mathbf v=0$, $2\mathbf u=4a_{33}\mathbf c$, $q=-2$, and $-4a_{33}c'_{33}=qb'_{33}$, the above
system simplifies dramatically. Two out of four equations become redundant, and the only nontrivial equations that survive are
\begin{align}
  & \mathbf u \tilde{B}'=4\mathbf a \tilde{C}', \label{L11a}\\
  & \tilde{B}' {\mathbf a}^{\text{T}}=\tilde{A}' {\mathbf u}^{\text{T}}. \label{L11b}
\end{align}
We add to this system equation (\ref{L014a}), which, upon implementing (\ref{LLL}) becomes
\beq
\mathbf a[S^{\text{T}}-{\rm Id}]=\mathbf u \tilde{A}.
\label{MMM}
\eeq

\vspace{0.1cm}

\noindent \textbf{8th postulate.} $\det B'\ne 0$.

\vspace{0.1cm}

Recall that $\mathbf u[S+3{\rm Id}]=\mathbf c[S+3{\rm Id}]=0$. From equations (\ref{L11a}) and
(\ref{MMM}), we obtain that
$$
\mathbf a\bigl[ S^{\text{T}}-{\rm Id} -4\tilde{C}'(\tilde{B}')^{-1}\tilde{A}\bigr]=0.
$$
Substituting (\ref{A->SBC}), we obtain
\begin{align*}
  &\mathbf a\Big( S^{\text{T}}- {\rm Id} -\tilde{C}'(\tilde{B}')^{-1}\bigl[S\tilde{B}'(\tilde{C}')^{-1}+\tilde{B}'(\tilde{C}')^{-1}S^{\text{T}}+2\tilde{B}'(\tilde{C}')^{-1}\bigr]\Big)  \\
  =&\mathbf a \bigl[-3{\rm Id}-\tilde{C}'(\tilde{B}')^{-1}S\tilde{B}'(\tilde{C}')^{-1}\bigr]=\mathbf a \tilde{C}'(\tilde{B}')^{-1}[-3{\rm Id}-S]\tilde{B}'(\tilde{C}')^{-1}\\
  =&-\tfrac{1}{4} \mathbf u[S+3{\rm Id}]\tilde{B}'(\tilde{C}')^{-1}=0,
\end{align*}
by (\ref{L013}), so the condition (\ref{L11a}) is satisfied automatically.

From (\ref{L11a}) and (\ref{L11b}), we obtain that
\beq
\mathbf a\bigl[ (\tilde{B}')^{\text{T}} -4\tilde{C}'(\tilde{B}')^{-1}\tilde{A}'\bigr]=0.
\label{QQQ}
\eeq 
Because $\tilde{A}'[2{\rm Id}+S^{\text{T}}]=\tilde{B}'\tilde{A}+S_{A}$, see (\ref{A-A}), multiplying from the right by $[2{\rm Id}+S^{\text{T}}]$,
we can perform the chain of transformations --- where we have that $\tilde{B}'(\tilde{C}')^{-1}$ is symmetric ---
\begin{align*}
  0&=\mathbf a\Big((\tilde{B}')^{\text{T}}[2{\rm Id}+S^{\text{T}}] -4\tilde{C}'\tilde{A} -4\tilde{C}'(\tilde{B}')^{-1}S_A\Big)\\
  &=\mathbf a \tilde{C}'(\tilde{B}')^{-1}\Big(\tilde{B}'\bigl[(\tilde{C}')^{-1}(\tilde{B}')^{\text{T}}\bigr][2{\rm Id}+S^{\text{T}}] -4\tilde{B}'\tilde{A} -4S_A\Big)
  \\
  &=\mathbf a \tilde{C}'(\tilde{B}')^{-1}\Big((\tilde{B}')^2(\tilde{C}')^{-1}[2{\rm Id}+S^{\text{T}}] -\tilde{B}'\bigl[S\tilde{B}'(\tilde{C}')^{-1}+\tilde{B}'(\tilde{C}')^{-1}S^{\text{T}}+2\tilde{B}'(\tilde{C}')^{-1}\bigr] -4S_A\Big)\\
  &=\tfrac{1}{4} \mathbf u \bigl[-\tilde{B}'S\tilde{B}'(\tilde{C}')^{-1}-4S_A\bigr] \\ 
  &=\tfrac{1}{4} \mathbf u \bigl[-\tilde{B}'S-4S_A\tilde{C}'(\tilde{B}')^{-1}\Big)\tilde{B}'(\tilde{C}')^{-1},
\end{align*}
that is, both $S+3{\rm Id}$ and $\tilde{B}'S+4S_A\tilde{C}'(\tilde{B}')^{-1}$ share the same null vector $\mathbf u$. This imposes two more constraints on
the matrix elements of $\tilde{B}'S\tilde{B}'+4S_A\tilde{C}'$. Note that $S_A\tilde{C}'=-z S^{\text{T}}$, so both columns of the matrix $\tilde{B}'S\tilde{B}'-4zS^{\text{T}}$
must be proportional
to $\big(3-c'_{12},-c'_{22}\big)^{{\rm T}}$.
    
\vspace{0.1cm}
\noindent \textbf{9th postulate.} The row vector $(3-c'_{12},c'_{22})$ is non-zero.
\vspace{0.1cm}

Hence we obtain
\beq
-c_{22}'\bigl(-(c'_{22})^2+(9-c'_{12})b'_{21}+4zc'_{12}\bigr) =\bigl(b'_{21}c'_{12}c'_{22}+b'_{22}(c'_{22})^2+[b'_{21}]^2 c'_{11}+b'_{21}b'_{22}c'_{12}-4zc'_{11}\bigr)(3-c'_{12})
\label{eq-6}
\eeq
and
\begin{align}
&-c_{22}'\bigl(c'_{22}(c'_{12}-1)+(9-c'_{12})b'_{22}+4zc'_{22}\bigr) \nonumber\\
&\qquad\qquad=\bigl(-b'_{21}c'_{12}(c'_{12}-1)-b'_{22}c'_{22}(c'_{12}-1)+b'_{21}b'_{22} c'_{11}+[b'_{22}]^2c'_{12}-4zc'_{12}\bigr)(3-c'_{12}).
\label{eq-7}
\end{align}

The last set of relations comes from (\ref{A-A}). Here, we use that the matrix elements of $\tilde{A}$ and $\tilde{A}'$ are not independent. We obtain four equations:
\begin{align}
  & (3-2c'_{12})a_{12}+c'_{11}a_{22}+c'_{22}a_{11}=0, \label{eq-8}\\
  & (-c'_{12}+5)a_{22}+c'_{11}a'_{22}-b'_{21}a_{11}-b'_{22}a_{12}=0, \label{eq-9}\\
  & (c'_{12}+2)a'_{22}-c'_{22}a_{22}-b'_{21}a_{12}-b'_{22}a_{22}=0, \label{eq-10}\\
  & z=(2c'_{12}-3)a_{22}+c'_{11}a'_{22}-c'_{12}a_{12}. \label{eq-11}
\end{align}
We now have eleven (inhomogeneous and, in principle, nonlinear) equations on the ten variables $a_{11}$, $a_{12}$, $a_{22}$, $a'_{22}$, $b'_{21}$, $b'_{22}$, $z$, $c'_{11}$, $c'_{22}$, and $c'_{12}$. Equations (\ref{eq-3})--(\ref{eq-5}) and (\ref{eq-11}) are substitutions for $a_{11}$, $a_{12}$, $a_{22}$, and $z$. Equations (\ref{eq-2}) and (\ref{eq-8}) constitute a linear system determining $b'_{21}$ and $b'_{22}$. We then determine $a'_{22}$ from (\ref{eq-9}) and $c'_{22}$ from the determinant equation (\ref{eq-1}) and substitute all the obtained expressions into the remaining three nonlinear equations (\ref{eq-10}), (\ref{eq-6}) and (\ref{eq-7}) on two variables $c'_{11}$ and $c'_{12}$ using \textsc{Maple}\footnote{We acknowledge a valuable help of Misha Shapiro at this part of the work.}. It turns out that \eqref{eq-10} then factorises into two factors, one rather large factor and the another is $9-(c'_{12})^2-(c'_{11})^2$. For the huge factor, its resultant with equations (\ref{eq-6}) and (\ref{eq-7}) is empty, whereas for the second factor we obtain a nonempty set of solutions. We now describe these solutions.

Let $9-(c'_{12})^2-(c'_{11})^2=0$. Then, using (\ref{eq-1}), either $c'_{11}=0$, which results in inconsistencies in all the above calculations, or
$$
c'_{11}+c'_{22}=0,
$$
which we assume in what follows. The substitutions then give
\beq
a_{11}=-\tfrac{1}{9} (5c'_{12}-9),\quad a_{12}=\tfrac{2}{9} c'_{11}, \quad a_{22}=-\tfrac{1}{9} (c'_{12}-3),
\eeq
and
\beq
b'_{21}=1-\tfrac{c'_{12}}{3},\quad b'_{22}=\tfrac{c'_{11}}{3},\quad a'_{22}=0.
\eeq
Equation (\ref{eq-10}) is satisfied identically, expressing $z$ out of (\ref{eq-11}), we obtain
$$
z=-\tfrac{1}{9} (2c'_{12}-3)(c'_{12}-3)-\tfrac{2}{9} c'_{12}c'_{11},
$$
and substituting all the above quantities into the matrix $\tilde{B}'S\tilde{B}'-4zS^{\text{T}}$, we obtain that all matrix elements are multiplied by the same factor:
$$
\tilde{B}'S\tilde{B}'-4zS^{\text{T}}=\beta \left(\begin{array}{rr} -c'_{12} & c'_{11}\\ c'_{11} & c'_{12}\end{array}\right),\qquad
\beta = 2(c'_{12}-3)^2+2c'_{12}c'_{11}.
$$
So, the only possibility for this matrix to be degenerate is to set $\beta=0$. Then this matrix vanishes and all vectors are its null vectors. We therefore have that
\beq
c'_{11}=-\frac{(c'_{12}-1)^2}{c'_{12}},
\label{c11}
\eeq
and since we require $c'_{12}-3\ne 0$, using (\ref{eq-1}) we obtain a cubic equation determining $c'_{12}$
\beq
(3-c'_{12})^3=(c'_{12}+3)[c'_{12}]^2\ \ \hbox{or}\ \ 2[c'_{12}]^3-6[c'_{12}]^2+27 c'_{12}-27=0.
\label{c12}
\eeq
This equation admits one real root $c'_{12}=1+\Bigl[1+\frac34 \sqrt{78} \Bigr]^{1/3}+\Bigl[1-\frac34 \sqrt{78}\Bigr]^{1/3}\approx 1.1898$ and two complex conjugate roots $c'_{12}\approx 0.9051 \pm 3.2445{\rm i}$. If we set $c'_{12} = 3\zeta$, we find that $\zeta$ are the roots of
\beq
\label{Proot} P(\zeta) = 2\zeta^3 - 2\zeta^2 + 3\zeta - 1.
\eeq
Collecting all previous results and calling $\alpha = a_{33}$, we find that all coefficients of the matrices $(A,B,C)$ as explicit elements in $\mathbb{Q}(\zeta) + \alpha.\mathbb{Q}(\zeta)$. We can use \eqref{Proot} to express elements of $\mathbb{Q}(\zeta)$ as degree two polynomials in $\zeta$. The result is the one announced in Proposition~\ref{Ldim3}. Note that since the matrix $\tilde{B}'S\tilde{B}'-4zS^{\text{T}}$ is null, from (\ref{QQQ}) we have another nice relation
$$
\tilde{A}'=\tfrac{1}{4} \tilde{B}'(\tilde{C}')^{-1}(\tilde{B}')^{\text{T}}.
$$

We have thus found a classical Airy structure -- as can be directly checked. Lemma~\ref{basrem} tells us the quantum Airy structures which projects to this classical Airy structure are
$$
D^i = \tfrac{1}{2}{\rm Tr}\,B^i + \Delta^i,
$$
where $\Delta$ is an arbitrary elements of $V^*$ such that $\Delta$ is zero on the space of commutators in the Lie algebra. As the latter is spanned by $L_0'$ and $L_1$, we deduce that $\Delta'^0 = \Delta^1 = 0$ while $\Delta^0$ remains arbitrary.

These quantum Airy structures determine an (adjoint) module structure of $V$ on $\mathcal{D}_{V}$, and the cohomology spaces $H^p(V,\mathcal{D}_{V})$ for $p \in \{0,1,2\}$ can be computed with Lemma~\ref{cod3}. We are indeed in a three-dimensional situation with a codimension $1$ ideal $\mathfrak{i}$ spanned by $(x_1,x_2) = (L_1,L_0')$, the extra generator being $y = L_0 \in \mathfrak{q}$, and commutation relations \eqref{comut} determined by
$$
a = 0,\qquad \left(\begin{array}{cc} a_{11} & a_{12} \\ a_{21} & a_{22} \end{array}\right) = {\rm diag}(-1,2).
$$
The result is that $\dim H^0_{L}(V,\mathcal{D}_{V}) = 1$ corresponding to the constant in $\mathcal{D}_{V}$, $\dim H^1 = 2$ with one generator corresponding to deformation by the constant $\Delta^0$,  and $\dim H^2 = 1$. \hfill $\Box$

\section{Four classes of examples from geometry}

In the remaining of the paper, we consider examples of quantum Airy structures which have a stronger geometric flavour --- in relation with topological quantum field theories and enumerative geometry.

\label{S4}

\subsection{From Frobenius algebras (2d TQFTs)}
\label{S4Frob}

\subsubsection{Quantum Airy structures}

Let $\mathbb{A}$ be a Frobenius algebra (not necessarily unital), \textit{i.e.} a finite-dimensional vector space together with a commutative, associative product $\mathbb{A} \otimes \mathbb{A} \rightarrow \mathbb{A}$ and a linear form $\varphi\,:\,\mathbb{A} \rightarrow \mathbb{C}$ such that the pairing $\mathbb{A} \otimes \mathbb{A} \rightarrow \mathbb{C}$ defined by $\langle v_1,v_2 \rangle = \varphi(v_1v_2)$ is non-degenerate. We recall that
\begin{lemma} \cite{Abrams}
\label{defFrob} If the product is given, all the other Frobenius algebra structures on $\mathbb{A}$ are of the form $\tilde{\varphi}(v) := \varphi(vu)$ for some invertible element $u \in \mathbb{A}$.
\end{lemma}

Let $(e_{i})_{i \in I}$ be a basis, and $(e_{i}^*)_{i \in I}$ the basis such that
$$
\langle e_{i},e_{j}^* \rangle = \delta_{i,j}.
$$

\begin{proposition}
\label{FrobABCD} For any $\theta_{A},\theta_{B},\theta_{C} \in \mathbb{A}$
\begin{eqnarray}
A_{j,k}^{i} & = & \varphi(e^*_{i}e^*_{j}e^*_{k}\theta_{A}), \nonumber \\
B_{j,k}^{i} & = & \varphi(e^*_{i}e_{j}^*e_{k}\theta_{B}), \nonumber \\
C_{j,k}^{i} & = & \varphi(e^*_{i}e_{j}e_{k}\theta_{C}), \nonumber
\end{eqnarray}
and arbitrary $D^{i}$ defines a quantum Airy structure on $V = \mathbb{A}$. It has $f^{i}_{j,k} = 0$, \textit{i.e.} the underlying Lie algebra is abelian.
\end{proposition}
\noindent \textbf{Proof.} Commutativity (resp. invariance) of the product shows that $B^{i}_{j,k} = B^{j}_{i,k}$ (resp. $A$ is fully symmetric). Therefore, we just need to check the \textbf{BB-CA}, \textbf{BC} and \textbf{BA} relations of Section~\ref{S22}. By definition of the dual basis and the pairing we remark that
\beq
\label{propair} \forall v_1,v_2 \in \mathbb{A},\qquad v_1v_2 = \varphi(v_1v_2e_{a})\,e_{a}^*.
\eeq
We recall the matrix notations $X^{i} = (X^{i}_{j,k})_{j,k}$. Using invariance, associativity and commutativity of the product, we find
\begin{equation}
(B^{i}B^{j})_{k,\ell} = \varphi(e_{i}^*e_{k}^*e_{a}\theta_{B})\varphi(e_{j}^*e_{a}^*e_{\ell}\theta_{B}) = \varphi(e_{i}^*e_{j}^*e_{k}^*e_{\ell}\theta^2_{B}),
\end{equation}
and likewise
\beq
(B^{i}A^{j})_{k,\ell} = \varphi(e_{i}^*e_{k}^*e_{j}^*e_{\ell}\theta_{B}\theta_{A}),\qquad
(C^{i}B^{j})_{k,\ell} = \varphi(e_{i}^*e_{k}e_{j}^*e_{\ell}\theta_{C}\theta_{B}). \nonumber
\eeq
These expressions are completely symmetric under permutation of $i$ and $j$. So, pairs of terms in the \textbf{BB-CA}, \textbf{BC} and \textbf{BA} relations cancel each other. \hfill $\Box$

As we see from the proof, this example provides a rather trivial solution of the three relations in the sense that the three terms in each of them are already symmetric in $i$ and $j$.

Up to a change of basis in $\mathbb{A}$, we can and will assume in the remaining of this section that $(e_i)_{i \in I}$ is an orthonormal basis. We will sometimes need the element
$$
H := e_{a}^2.
$$

Let us make a last observation, in case $\theta_{A}$ is invertible. Using the symmetry \eqref{SymU} with $u_{i,j} = \varphi(e_{i}e_{j}\nu)$ and choosing $\nu = -\theta_B/\theta_{A}$, we find an equivalent quantum Airy structure with $\tilde{B} = 0$ and
\bea
\tilde{A}^i_{j,k} & = & \varphi(e_ie_je_k\theta_{A}), \nonumber \\
\tilde{C}^i_{j,k} & = & \varphi\big(e_ie_je_k(\theta_{C} - \theta_{A}^{-1}\theta_{B}^2)\big), \nonumber \\
\tilde{D}^i & = & D^i - \tfrac{1}{2}\varphi(e_iH\theta_{B}). \nonumber 
\eea
In the special case $\theta_{B}^2 = \theta_{A}\theta_{C}$, we have $\tilde{C} = 0$. On the contrary, if $\theta_{B}^2 \neq \theta_{A}\theta_{C}$, we cannot \textit{a priori} get rid of $\tilde{C}$ with the group action while keeping $\tilde{B} = 0$.

We can illustrate this phenomenon in a simple way when $\mathbb{A}$ has dimension $1$. In this case, we can perform a rescaling of $x$ (this transformation is in $\mathcal{G}_{V}$) to achieve $\theta_{A} = 1$, and conjugation by an exponential of the Laplacian as above to achieve $\theta_{B} = 0$. However, when $\theta_{B}^2 - \theta_{A}\theta_{C} \neq 0$, we do not have further freedom to modify $\theta_{C}$. Thus, $\theta_{B}^2 - \theta_{A}\theta_{C}$ is a continuous parameter of deformation for $\mathcal{G}_{V}$-orbits. If we allowed $\hbar$-rescalings --- this is not in the group $\mathcal{G}_{V}$ --- we could in fact achieve $\theta_{C} = 0$ or $1$, so this deformation parameter coincides with $\hbar$-rescalings.

The action by the group of invertible elements of $\mathbb{A}$ deforming the Frobenius structure according to Lemma~\ref{defFrob} also provides a family of commuting flows on the moduli space of quantum Airy structures on the abelian Lie algebra $\mathbb{A}$, although we do not make statements about the independence of these flows.

The classical Airy structure is in this case simply given by  the infinitesimal symplectomorphisms $({\cal L}_i)_{i \in I}$, where
\bea
{\cal L}_i e_j^* & = & - \varphi(\theta_{B}e_i^*e_je_a^*)e_a^* - \varphi(\theta_{C}e_i^*e_je_a) e_a \nonumber  \\
\mathcal{L}_{i} e_j & = & -\varphi(\theta_{A}e_i^*e_j^*e_a^*)e_a^* + \varphi(\theta_{B}e_i^*e_j^*e_a\big)e_a  \nonumber
\eea
The commutativity of the product directly implies the commutation of the ${\cal L}_i$s and the full symmetry of $A$ and symmetry of $B$ with respect to the two first indices.

\subsubsection{Partition function for $\dim\,\mathbb{A} = 1$}
\label{1dZ}

Quantum Airy structures on a vector space of dimension $1$ are just differential operators of the form
$$
L = \hbar \partial_{x} - \tfrac{\theta_{A} x^2}{2} - \hbar \theta_{B} x\partial_{x} - \tfrac{\theta_C}{2}\partial_{x}^2 - \hbar D,
$$
where $\theta_{A},\theta_{B},\theta_{C},D$ are scalars. The differential equation $L\cdot Z = 0$ can be solved by elementary means to obtain the partition function. We have to distinguish several cases, all related by limiting procedures, which we will not discuss. 

\vspace{0.2cm}

\noindent\boxed{\textit{$\theta_{C} = 0$}}

We have $Z(x) = \exp\big(\tfrac{S_0(x)}{\hbar} + S_1(x)\big)$ with
$$ 
S_0(x) = \tfrac{\theta_{A}}{2\theta_{B}^3}\Big(-\ln(1 - \theta_{B}x) - \theta_{B}x - \tfrac{\theta_{B}^2x^2}{2}\big),\qquad S_1(x) = -\tfrac{D}{\theta_{B}}\ln(1 - \theta_{B}x).
$$
The answer can be obtained combinatorially, as a corollary to Proposition~\ref{Cequal0}. This formula still makes sense when $\theta_{B} = 0$, it becomes  $Z(x) = \exp\big(\tfrac{\theta_{A}x^3}{6\hbar} + Dx\big)$.

\vspace{0.2cm}

\noindent \boxed{\textit{$\theta_{C} \neq 0$ \textit{and} $\theta_{B}^2 = \theta_{A}\theta_{C}$}}

We find $Z(x) = \mathfrak{Z}(x)/\mathfrak{Z}(0)$ with
\bea 
\mathfrak{Z}(x) & = & \exp\Big\{\tfrac{1}{\hbar}\big(\tfrac{x}{\theta_{C}} - \tfrac{\theta_{B}x^2}{2\theta_{C}}\big)\Big\}\,{\rm Bi}\Big(\tfrac{1 - 2\theta_{B}x + \hbar\theta_{C}(\theta_{B} - 2D)}{(2\theta_{B}\theta_{C}\hbar)^{2/3}}\Big) \nonumber \\
& = & c_{\hbar}\,\exp\Big\{\tfrac{1}{\hbar}\big(\tfrac{x}{\theta_{C}}- \tfrac{\theta_{B}x^2}{2\theta_{C}}\big)\Big\} \int \dd t\,\exp\Big(\tfrac{1}{\hbar}\big(\tfrac{1 - 2\theta_{B}x + \hbar\theta_{C}(\theta_{B} - 2D)}{(2\theta_{B}\theta_{C})^{2/3}}\,t - \tfrac{t^3}{3}\big)\Big). \nonumber
\eea
where $c_{\hbar}$ only depends on $\hbar$. Here ${\rm Bi}(z)$ is the Bairy function, which is the solution of the differential equation ${\rm Bi}''(z) = z{\rm Bi}(z)$, whose asymptotics at $z \rightarrow +\infty$  is
$$
{\rm Bi}(z) = \frac{e^{\frac{2}{3}z^{\frac{3}{2}}}}{\pi^{\frac{1}{2}}z^{\frac{1}{4}}}\big(1 + o(1)\big).
$$
In particular, $Z(x) = \exp\big(\tfrac{S_0(x)}{\hbar} + S_1(x) + O(\hbar)\big)$ with
\bea
S_0(x) & = & \frac{(1 - 2\theta_{B}x)^{\frac{3}{2}} - 1 + \tfrac{x}{\theta_{C}} - \tfrac{\theta_{B}x^2}{2\theta_{C}}}{3\theta_{B}\theta_{C}} \nonumber \\
S_1(x) & = & \tfrac{\theta_{B} - 2D}{2\theta_{B}}\big((1 - 2\theta_{B}x)^{\frac{1}{2}}- 1\big) - \tfrac{1}{4}\ln(1 - 2\theta_{B}x).
\eea
The integral above is a formal integral, \textit{i.e.} it is to be evaluated by expansion around the (unique) saddle point which realises $F_{0,3} = \theta_{A}$. However, choosing the corresponding steepest descent contour offers the possibility to define $Z(x)$ here as an entire function of $x$. This is achieved by the Bairy function does. We retrieve the Taylor coefficients $F_{g,n}$ by expanding ${\rm Bi}(z)$ at $z \rightarrow +\infty$. Elementary properties of the full asymptotic expansion of the Bairy function are collected in Appendix~\ref{Bairyapp}.

\vspace{0.2cm}

\noindent \boxed{\textit{$\theta_{C} \neq 0$ and $\sigma^2 := \theta_B^2 - \theta_{A}\theta_{C} \neq 0$}}

We find that $Z(x) = \mathfrak{Z}(x)/\mathfrak{Z}(0)$ with
\bea
\label{Whity} \mathfrak{Z}(x) & = & c_{\hbar} \exp\Big\{\tfrac{1}{\hbar}\big(\tfrac{x}{\theta_{C}} - \tfrac{\theta_{B}x^2}{2\theta_{C}}\big) + \ln\big(1 - \tfrac{\sigma^2x}{\theta_{B}}\big)\Big\}\,M\Big(\tfrac{1}{4}\big(\tfrac{\theta_{A}}{\hbar\sigma^3} + \tfrac{2D - \theta_{B}}{\sigma}\big)\,;\,\tfrac{1}{4}\,;\,\tfrac{\theta_{B}^2}{\hbar\sigma^3\theta_{C}}\big(1 - \tfrac{\sigma^2x}{\theta_{B}}\big)^2\Big)  \\
& = & \exp\Big\{\tfrac{1}{\hbar}\big(\tfrac{x}{\theta_{C}} - \tfrac{\theta_{B}x^2}{2\theta_{C}}\big) + \ln\big(1 - \tfrac{\sigma^2x}{\theta_{B}}\big)\Big\} \\
&& \quad \times  \int_{-\frac{1}{2}}^{\frac{1}{2}} \dd t\,\exp\Big\{\tfrac{\theta_{B}^2}{\hbar\sigma^3\theta_{C}}\big(1 - \tfrac{\sigma^2x}{\theta_{B}}\big)^2\,t\Big\}(\tfrac{1}{4} - t^2\big)^{-\frac{1}{4}}\bigg(\frac{\frac{1}{2} - t}{\frac{1}{2} + t}\bigg)^{\frac{1}{4}\big(\frac{\theta_{A}}{\hbar \sigma^3} + \frac{2D - \theta_{B}}{\sigma}\big)}. \nonumber
\eea  
Here $\mathcal{W}_{M}(\mu;\nu;z)$ is the Whittaker-M function, solving the differential equation
$$
\Big\{\partial_{z}^2 + \big(-\tfrac{1}{4} + \tfrac{\mu}{z} + \tfrac{\frac{1}{4} - \nu^2}{z^2}\big)\Big\}\mathcal{W}_{M}(\mu;\nu;z) = 0.
$$
It is an entire function of $z$, but only its asymptotics near $z \rightarrow +\infty$ matter to obtain the $F_{g,n}$s. These asymptotics can be found by saddle point analysis in the integral formula, which leads to $Z(x) = \exp\big(\tfrac{S_0(x)}{\hbar} + S_1(x) + O(\hbar)\big)$ and
\bea
S_0(x) & = & \frac{1}{2}\bigg(\frac{\theta_{B}}{\sigma^2\theta_{C}} - \frac{x}{\theta_C}\bigg)\sqrt{1 - 2\theta_{B}x + \sigma^2x^2} - \frac{\theta_B}{2\sigma^2\theta_C} \\
& & \quad + \frac{\theta_{A}}{2\sigma^3}\,\bigg\{{\rm argtanh}\bigg(\frac{\sigma}{\theta_{B}}\,\frac{\sqrt{1 - 2\theta_{B}x + \sigma^2x^2}}{1 - \frac{\sigma^2 x}{\theta_{B}}}\bigg) - {\rm argtanh}\bigg(\frac{\sigma}{\theta_B}\bigg)\bigg\}  + \frac{x}{\theta_{C}} - \frac{\theta_{B}x^2}{2\theta_{C}}, \nonumber \\
S_1(x) & = & -\frac{1}{4}\ln(1 - 2\theta_{B}x + \sigma^2x^2) + \frac{2D - \theta_{B}}{2\sigma}\bigg\{{\rm argtanh}\bigg(\frac{\sigma}{\theta_{B}}\,\frac{\sqrt{1 - 2\theta_{B}x + \sigma^2x^2}}{1 - \frac{\sigma^2x}{\theta_{B}}}\bigg) - {\rm argtanh}\bigg(\frac{\sigma}{\theta_B}\bigg)\bigg\}. \nonumber
\eea
This formula still makes sense for $\theta_{B} = 0$, the main simplification being that the variable of the Whittaker function should be specialised to $z = \frac{\sigma x^2}{\hbar\theta_{C}}$. In fact, as a result of Proposition~\ref{Propmain}, $F_{g,n}$ is a polynomial in $\theta_{A},\theta_{B},\theta_{C}$ and $D$. The first values are given in Appendix~\ref{Whitapp}.

\subsubsection{Partition function in general}

In this paragraph, $\mathbb{A}$ is assumed unital. When $\dim \mathbb{A} > 1$, we are going to show that the partition function for the quantum Airy structure of Proposition~\ref{FrobABCD} is computed by a higher-dimensional version of the integral formulas of the previous paragraph. We heavily rely on the commutativity of the product in $\mathbb{A}$. We denote
$$
D := D^ae_{a} \in \mathbb{A}.
$$
An easy rewriting of Proposition~\ref{Cequal0} in terms of the linear form $\varphi$ leads to
\begin{proposition}
\label{Cequal0Frob} If $\theta_{C} = 0$, then we have that $Z = \exp\big(\tfrac{S_0(x)}{\hbar} + S_1(x)\big)$ with
\bea
S_0(x) & = & \varphi\Big(\tfrac{\theta_A}{2\theta_{B}^3}\big(-\ln(1 - \theta_{B}x) - \theta_Bx - \tfrac{\theta_B^2x^2}{2}\big)\Big). \nonumber \\
S_1(x) & = & \varphi\Big(-\tfrac{D}{\theta_{B}}\ln(1 - \theta_Bx)\Big). \nonumber
\eea
This formula also makes sense when $\theta_{B}$ is not invertible.\hfill $\Box$
\end{proposition}
When $\theta_{C} \neq 0$, we will check the formulas below by direct computation with help of Dyson--Schwinger equations. Other proofs could be obtained exploiting the action of $\mathcal{G}_{V}$ on the partition function of Proposition~\ref{Cequal0Frob}.
\begin{proposition}
\label{AiryZFrob} If $\theta_{C}$ is invertible, and $\theta_{B}^2 = \theta_{A}\theta_{C}$, we have $Z(x) = \mathfrak{Z}(x)/\mathfrak{Z}(0)$ with
\beq
\mathfrak{Z}(x) = c_{\hbar}\,\exp\Big\{\tfrac{1}{\hbar}\varphi\big(\tfrac{x}{\theta_{C}}- \tfrac{\theta_{B}x^2}{2\theta_{C}}\big)\Big\} \int \,\exp\Big\{\tfrac{1}{\hbar}\varphi\big(\tfrac{1 - 2\theta_{B}x + \hbar\theta_{C}(\theta_{B} - 2D)}{(2\theta_{B}\theta_{C})^{2/3}}\,t - \tfrac{t^3}{3}\big)\Big\}\dd t,
\eeq
where $\dd t$ is the Lebesgue measure on the linear coordinates on $\mathbb{A}$ with respect to an orthonormal basis.
\end{proposition}

\begin{proposition}
\label{WhitZFrob} If $\theta_{C} \neq 0$ and $\sigma^2 = \theta_B^2 - \theta_{A}\theta_{C}$ is invertible, we have $Z(x) = \mathfrak{Z}(x)/\mathfrak{Z}(0)$ with
\bea
\mathfrak{Z}(x) & = & \exp\Big\{\tfrac{1}{\hbar}\varphi\big(\tfrac{x}{\theta_{C}} - \tfrac{\theta_{B}x^2}{2\theta_{C}}\big)x + H\,\ln\big(1 - \tfrac{\sigma^2x}{\theta_{B}}\big)\Big\} \int  \exp\Big\{\varphi\Big(\tfrac{\theta_{B}^2}{\hbar\sigma^3\theta_{C}}\big(1 - \tfrac{\sigma^2x}{\theta_{B}}\big)^2 t \nonumber \\
&& + \tfrac{1}{4}\big(\tfrac{\theta_{A}}{\hbar \sigma^3} + \tfrac{2D - H\theta_{B}}{\sigma} - H\big)\ln\big(\tfrac{1}{2} - t\big) + \tfrac{1}{4}\big(-\tfrac{\theta_{A}}{\hbar\sigma^3} - \tfrac{2D - H\theta_{B}}{\sigma} - H\big)\ln\big(\tfrac{1}{2} + t\big)\Big)\Big\}\dd t, \nonumber 
\eea
and we recall that  $H = \sum_{i \in I} e_{i}^2$. This formula also makes sense when $\theta_{B}$ is not invertible.
\end{proposition}

In the case $\dim \mathbb{A} = 1$ with the canonical scalar product ($H = 1$), we recover formula \eqref{Whity}.  The intermediate cases in Proposition~\ref{AiryZFrob}, when $\theta_{C}$ is non-zero but non-invertible, or in Proposition~\ref{WhitZFrob} when $\sigma$ is non-zero but non-invertible, will not be discussed, since they can be obtained by limiting procedures from \ref{WhitZFrob}. For this reason, we will only write a proof of Proposition~\ref{WhitZFrob}. It is in fact a good exercise for the reader to repeat the scheme of this proof in the specific (and simpler) case of Proposition~\ref{AiryZFrob}.

\vspace{0.2cm}

\noindent \textbf{Proof.}  We consider a formal integral of the form
$$
\mathbb{I}(x) = \exp\Big\{\varphi\big(\alpha_1x + \tfrac{\alpha_2x^2}{2} + \nu\ln(1 - \beta x)\big)\Big\} \int \exp\Big\{\varphi\big(\gamma(1 - \beta x)^2t + \tau_1\ln(\tfrac{1}{2} - t) + \tau_2\ln(\tfrac{1}{2} + t)\big)\Big\}\dd t,
$$
where $\alpha_1,\alpha_2,\nu,\gamma,\beta,\tau_1,\tau_2 \in \mathbb{A}$ are considered as parameters. If $f\,:\,\mathbb{A} \rightarrow \mathbb{K}$, we denote
$$
\langle f(t) \rangle := \frac{\int \,f(t)\,\exp\Big\{\varphi\big(\gamma(1 - \beta x)^2 t + \tau_1\ln(\tfrac{1}{2} - t) + \tau_2\ln(\tfrac{1}{2} + t)\big)\Big\}\dd t}{\int \,\exp\Big\{\varphi\big(\gamma(1 - \beta x)^2 t + \tau_1\ln(\tfrac{1}{2} - t) + \tau_2\ln(\tfrac{1}{2} + t)\big)\Big\}\dd t},
$$
which implicitly depends on $x$. This definition can be extended by linearity to a function $f\,:\,\mathbb{A} \rightarrow \mathbb{A}$. We also introduce $\varphi_i(v) := \varphi(e_iv)$ for $v \in \mathbb{A}$.

Let us compute the action of the differential operators of the quantum Airy structure on $\mathbb{I}(x)$.
\bea
-\mathbb{I}^{-1}\,\hbar \partial_{x_i} \mathbb{I} & = & \hbar \varphi_i\big( - \alpha_1 - \alpha_2x + \tfrac{\nu\beta}{1 - \beta x} + 2\gamma\beta(1 - \beta x) \langle t \rangle\big). \nonumber \\
\mathbb{I}^{-1}\tfrac{1}{2}\varphi(e_ie_ae_b\theta_{A})x_ax_b\mathbb{I} & = & \varphi_i(\tfrac{x^2}{2}\theta_{A}). \nonumber \\
\mathbb{I}^{-1} \hbar \varphi(e_ie_ae_b\theta_{B}) x_a \partial_{x_b} \mathbb{I} & = & \hbar\varphi(e_ixe_b\theta_{B})\varphi\Big(e_b\big(\alpha_1 + \alpha_2x - \tfrac{\nu\beta}{1 - \beta x} - 2\gamma\beta(1 - \beta x) \langle t \rangle\big)\Big) \nonumber \\
& = & \hbar \varphi_i\Big(\theta_{B}x\big(\alpha_1 + \alpha_2x - \tfrac{\nu\beta}{1 - \beta x} - 2\gamma\beta(1 - \beta x) \langle t \rangle\big)\Big). \nonumber
\eea
In the last line, we exploited the property \eqref{propair} of the pairing in an orthonormal basis, \textit{i.e.} such that $e_i^* = e_i$. In a similar fashion
\bea
& & \mathbb{I}^{-1}\,\tfrac{\hbar^2}{2}\varphi(e_ie_ae_b\theta_{C})\partial_{x_a}\partial_{x_b}\mathbb{I} \\
& = & \tfrac{\hbar^2}{2}\Big\{\big(\alpha_1 + \alpha_2x - \tfrac{\nu \beta}{1 - \beta x}\big)^2  - 4\gamma\beta(1 - \beta x)\big(\alpha_1 + \alpha_2 - \tfrac{\nu\beta}{1 - \beta x}\big)\langle t \rangle + 4\gamma^2\beta^2(1 - \beta x)^2 \langle t^2 \rangle \\
& & \quad\,\,  + \big(\alpha_2 - \tfrac{\nu\beta^2}{(1 - \beta x)^2} + 2\gamma\beta^2 \langle t \rangle \big)H \Big\},
\nonumber
\eea
where we have used, for fixed $i \in I$ and $X,Y,Z \in \mathbb{A}$, the identities $\varphi(e_ie_ae_bX) \varphi(e_aY)\varphi(e_bZ) = \varphi(e_iXYZ)$ and $\varphi(e_ie_ae_bX) \varphi(e_ae_bY) = \varphi(e_iHXY)$. Finally, we observe that $\mathbb{I}^{-1} \hbar D^i \mathbb{I} = \varphi_i(\hbar D)$. Summing all the terms yields an expression for $\mathbb{I}^{-1} L_i \mathbb{I}$, which involves in particular $\langle t^2 \rangle$.

We now write the Dyson--Schwinger relations satisfied by the averages $\langle \cdot \rangle$. They express the fact that (formal) integrals are invariant under change of variables. Here we use the infinitesimal change of variable $t \rightarrow t + \epsilon \big(\tfrac{1}{4} - t^2)e_j$ for a given $j$. We remark that $\dd t \rightarrow \dd t\big(1 - \epsilon \varphi(2tHe_j) + O(\epsilon^2)\big)$, and invariance of the integral at order $\epsilon$ gives the exact formula
$$
\Big\langle \varphi\Big(e_j\Big\{-2tH + \gamma(1 - \beta x)^2\big(\tfrac{1}{4} -t^2\big) - \tau_1\big(\tfrac{1}{2} +  t) + \tau_2\big(\tfrac{1}{2} - t\big)\Big\}\Big)\Big\rangle  = 0.
$$
This formula can equivalently be obtained by integration by parts. As it holds for any $j$, we also have the identity in $\mathbb{A}$
$$
-2H\langle t \rangle + \gamma(1 - \beta x)^2\big(\tfrac{1}{4} -\langle t^2 \rangle\big) - \tau_1\big(\tfrac{1}{2} +  \langle t \rangle) + \tau_2\big(\tfrac{1}{2} - \langle t \rangle \big) = 0.
$$
We use this identity to eliminate $\langle t^2 \rangle$ from $-\mathbb{I}^{-1} L_i \mathbb{I}$. After tedious but straightforward algebra, the terms can be collected as follows
\bea
-\mathbb{I}^{-1}L_i\mathbb{I} & = & \varphi_i\bigg( \tfrac{\hbar^2\theta_{C}}{2}(\alpha_1^2 + \alpha_2H + 2\nu\alpha_2) - \hbar\alpha_1 + \hbar\nu\theta_{B} + \hbar D + \tfrac{\hbar^2\theta_{C}}{2}\big(\gamma^2\beta^2 + 2\gamma\beta^2(\tau_2 - \tau_1)\big) \nonumber \\
&& + x\big\{- \hbar\alpha_2 + \hbar \alpha_1\theta_{B} + \hbar^2\theta_{C}\alpha_1\alpha_2 - \hbar^2\theta_{C}\gamma^2\beta^3\big\} + x^2\big\{\tfrac{\theta_{A}}{2} + \hbar\theta_{B}\alpha_2 + \tfrac{\hbar^2\theta_{C}}{2}(\alpha_2^2 + \gamma^2\beta^4)\big\} \nonumber \\
&& + \tfrac{\nu\beta}{1 - \beta x}\big\{\hbar\beta - \hbar\theta_{B} - \hbar^2\theta_{C}(\alpha_1\beta + \alpha_2)\big\} + \tfrac{\hbar^2\theta_{C}\nu\beta^2}{2(1 - \beta x)^2}(\nu - H) \nonumber \\
&& + 2\langle t \rangle \Big\{ \hbar\gamma\beta + \hbar^2\theta_{C}\gamma\beta\big(-\alpha_1 + \nu\beta - \beta(\tau_1 + \tau_2 + \tfrac{3H}{2})\big)\Big\} \nonumber \\
&& + x\,\gamma\beta\big\{-\beta - \hbar\theta_{B} + \hbar^2\theta_{C}(\beta\alpha_1 - \alpha_2)\big\} + x^2\,\hbar\gamma\beta^2\big\{\hbar\theta_{C}\alpha_2 + \theta_{B}\big\}\Big)\bigg). \nonumber 
\eea
Imposing that the coefficients --- inside $\varphi_i$ --- of $1$, $x$, $x^2$, $(1 - \beta x)^{-1}$, $(1 - \beta x)^{-2}$, $\langle t \rangle$, $\langle t \rangle x$ and $\langle t \rangle x^2$ separately vanish uniquely fixes
$$
\alpha_1 = \frac{1}{\hbar\theta_{C}},\qquad \alpha_2 = -\frac{\theta_{B}}{\hbar\theta_{C}},\qquad \beta = \frac{\sigma^2}{\theta_{B}},\qquad \gamma = \frac{\theta_{B}^2}{\hbar \sigma^3 \theta_{C}},
$$
and
$$
\nu = H,\qquad \tau_1 = \tfrac{1}{4}\big(\tfrac{\theta_{A}}{\hbar\sigma^3} + \tfrac{2D - H\theta_{B}}{\sigma} - H\big),\qquad \tau_2 = -\tfrac{1}{4}\big(\tfrac{\theta_{A}}{\hbar\sigma^3} + \tfrac{2D - H\theta_{B}}{\sigma} + H\big).
$$
where we used the notation $\sigma^2 = \theta_B^2 - \theta_A \theta_C$.
We then have found a common solution of $L_i Z = 0$ for all $i \in I$, and one can check by saddle point analysis that, for the above choice of parameters, it is indeed of the form
$$
\mathbb{I}(x) = \exp\bigg(\sum_{\substack{g \geq 0 \\ n \geq 1}} \sum_{i_1,\ldots,i_n \in I} \tfrac{\hbar^{g - 1}}{n!}\,F_{g,n}(i_1,\ldots,i_n)\,x_{i_1}\cdots x_{i_n}\bigg),\qquad F_{0,1}(i) = F_{0,2}(i,j) = 0.
$$
Invoking the uniqueness of such a solution (Proposition~\ref{Propmain}) concludes the proof. 
\hfill $\Box$

\subsubsection{Example: TQFT partition function}

\label{TQFTS4}
It is well-known that unital Frobenius algebras are in one-to-one correspondence with two-dimensional topological quantum field theories \cite{Abrams}. In this context, the product on $\mathbb{A}$ comes from the map $\mathbb{A}^{\otimes 2} \rightarrow \mathbb{A}$ that the TQFT attaches to a pair of pants. Let us choose $\theta_{A} = \theta_{B} = \theta_{C} = 1$ in Proposition~\ref{FrobABCD}. We notice that, up to dualisation $F_{0,3}(i,j,k) = A^{i}_{j,k}$ represents the product. This is the manifestation of a more general, easy fact. Let $\Sigma_{g,n}$ is a topological surface with $n$ boundaries oriented inward, and denote $\mathcal{F}(\Sigma_{g,n}) \in {\rm Hom}(\mathbb{A}^{\otimes},\mathbb{C})$ the amplitude assigned to $\Sigma_{g,n}$ by the TQFT $\mathcal{F}$. Using the TQFT glueing rules, it can be computed (or defined) as follows. Take a pair of pants decomposition of $\Sigma_{g,n}$. Take the tensor product over all pairs of pants of the maps $\mathcal{F}(\Sigma_{0,3}): \mathbb{A}^{\otimes 3} \rightarrow \mathbb{C}$, and apply the pairing $\mathbb{A} \otimes \mathbb{A} \rightarrow \mathbb{C}$ on the factors of $\mathbb{A}$ corresponding to coinciding boundary components of the pair of pants. The result is a multilinear map $\mathbb{A}^{\otimes n} \rightarrow \mathbb{C}$. Owing to the associativity, commutativity and invariance of the product, it does not depend on the choice of pair of pants decomposition.

\begin{lemma}
\label{TQFTpart} Assume $\mathbb{A}$ unital. Then the partition function corresponding to the quantum Airy structure of Proposition~\ref{FrobABCD} with the choice of $D = \tfrac{H}{2}$ and $\theta_{A} = \theta_{B} = \theta_{C} = 1$,  gives the amplitudes of the two-dimensional TQFT corresponding to $\mathbb{A}$, e.g. for $2g - 2 + n > 0$
\beq 
\label{TQFTAiry} F_{g,n} = |\mathfrak{G}_{g,n}|\,\mathcal{F}(\Sigma_{g,n}).
\eeq
Here $\mathfrak{G}_{g,n}$ is the set of graphs described in the caption of Figure~\ref{TRgraph}, and its weighted cardinality
$$
|\mathfrak{G}_{g,n}(1)| := \sum_{G \in \mathfrak{G}_{g,n}(1)} \frac{1}{|{\rm Aut}\,G|}
$$
is computed by the generating series
$$
\exp\bigg(\sum_{g \geq 0} \sum_{n \geq 1} \frac{\hbar^{g - 1}}{n!}\,|\mathfrak{G}_{g,n}|\bigg) = \exp\bigg(\frac{1}{\hbar}\Big(x - \frac{x^2}{2}\Big)\bigg)\,{\rm Bi}\bigg(\frac{1 - 2x - \hbar}{(2\hbar)^{2/3}}\bigg).
$$
\end{lemma}
\textbf{Proof.} We obtain $\Sigma_{1,1}$ by glueing two boundaries of the same pair of pants. The TQFT assigns to this $\mathcal{F}(\Sigma_{1,1})(e_{i}) = \varphi(e_{i}e_{a}e_{a}) = \varphi(e_iH) = 2D^i = 2F_{1,1}(i)$. This coincides with \eqref{TQFTAiry} as the unique graph in $\mathfrak{G}_{g,n}(1)$ has a symmetry factor $\tfrac{1}{2}$. Unfolding \eqref{TRForm} using the fact that $B$ and $C$ here represent the product (up to identification $\mathbb{A} \simeq \mathbb{A}^*$ coming from the pairing), we deduce that $F_{g,n}(i_1,\ldots,i_n)$ is the sum over $\mathfrak{G}_{g,n}(1)$ of the TQFT amplitudes computed with a pair of pants decomposition canonically defined by the graph, and evaluated on $e_{i_1} \otimes \cdots \otimes e_{i_n}$. The fact that this amplitude is independent of the pair of pants decomposition leads to the factorisation of the number  (weighted by the order of the automorphism group) of graphs in $\mathfrak{G}_{g,n}(1)$. \hfill $\Box$

It is clear from the proof, that if one chooses for $\theta_{A},\theta_{B},\theta_{C}$ arbitrary scalars, one will rather obtain $F_{g,n} = G_{g,n}(\theta_{A},\theta_{B},\theta_{C}) \mathcal{F}(\Sigma_{g,n})$ where $G_{g,n}(\theta_{A},\theta_{B},\theta_{C})$ are the Taylor coefficients of the partition function of the 1-dimensional quantum Airy structure $\hbar\partial_{x} - \tfrac{\theta_{A}x^2}{2} - \hbar\theta_{B}x\partial_{x} - \tfrac{\hbar^2\theta_{C}}{2}\partial_{x}^2 - \tfrac{\hbar}{2}$, see Section~\ref{1dZ}.

\subsection{From non-commutative Frobenius algebras}
\label{S4NCFrob}

Consider now a non-commutative Frobenius algebra $\mathbb{A}$. It is defined as a Frobenius algebra except that we drop the commutativity axiom, and impose the tracial condition $\varphi([v_1,v_2]) = 0$ for any $v_1,v_2 \in \mathbb{A}$. Here we denote $[\cdot,\cdot]$ the commutator, and $\{\cdot,\cdot\}$ the anticommutator. We choose to work directly with an orthonormal basis $(e_i)_{i \in I}$.
\begin{proposition}
\label{NCFrob} Let $\lambda_{A},\lambda_{B},\lambda_{C}$ be central elements in $\mathbb{A}$, such that $\lambda_{B}^2 + \lambda_{A}\lambda_{C} = 0$. The assignments
\begin{eqnarray}
A^{i}_{j,k} & = & \varphi\big(\lambda_{A}\{e_j,e_k\}e_i\big), \nonumber \\
B^i_{j,k} & = & \varphi\big(\lambda_{B}[e_i,e_j]e_k\big), \nonumber \\
C^i_{j,k} & = & \varphi\big(\lambda_{C}\{e_i,e_j\}e_k\big), \nonumber
\end{eqnarray}
and $D = D^ae_{a} \in \mathbb{A}$ such that $\lambda_{B}D$ lies in the orthogonal complement of $[\mathbb{A},\mathbb{A}]$, define a quantum Airy structure.
\end{proposition} 
As $\lambda_{X}$ for $X \in \{A,B,C\}$ are central, we can move them freely around inside $\varphi(\cdots)$. Then, due to invariance of the product, $A$ and $C$ are fully symmetric, and $B^i_{j,k}$ is fully antisymmetric under permutations of $(i,j,k)$ and equal to $\tfrac{1}{2}f_{i,j}^k$. We remark that, if $\lambda_{B} = \tfrac{1}{2}$, the Lie algebra structure on $V = \mathbb{A}$ determined by the quantum Airy structure coincides with the Lie algebra structure of $\mathbb{A}$ given by the commutator. 

\noindent \textbf{Proof.}  Using the property $\varphi(xe_a)e_a = x$ of the pairing and the orthonormality of the  basis $(e_i)_{i \in I}$, we compute
$$
B^i_{j,a}B^a_{k,\ell} = \varphi\big(\lambda_{B}[e_i,e_j]e_a\big)\varphi\big(\lambda_{B}[e_a,e_k]e_{\ell}\big) = \varphi\big(\lambda_{B}^2[[e_i,e_j],e_k]e_{\ell}\big).
$$
With similar manipulations, expanding (anti)commutators and using cyclic invariance of $\varphi$ to move $e_{\ell}$ to the last position, we find that
\bea
&& B^i_{j,a}B^a_{k,\ell} + B^i_{k,a}B^j_{a,\ell} + C^i_{\ell,a}A^j_{a,k} - (i \leftrightarrow j) \nonumber \\
& = & \varphi\big(2\lambda_{B}^2(e_ie_je_k - e_je_ie_k - e_ke_ie_j + e_ke_je_i)e_{\ell}\big) \nonumber \\
&& + \varphi\big(\lambda_{B}^2(e_je_ie_k - e_ie_ke_j - e_je_ke_i + e_ke_ie_j - e_ie_je_k + e_je_ke_i + e_ie_ke_j - e_ke_je_i)e_{\ell}\big) \nonumber \\
&& + \varphi\big(\lambda_{A}\lambda_{C}(e_je_ke_i + e_ke_je_i + e_ie_je_k + e_ie_ke_j - e_ie_ke_j - e_ke_ie_j - e_je_ie_k - e_je_ke_i)e_{\ell}\big) \nonumber \\
& = & \varphi\big(  (\lambda_B^2 + \lambda_{A}\lambda_{C})(e_ie_je_k - e_je_ie_k - e_ke_ie_j + e_ke_je_i)e_{\ell}\big). \nonumber
\eea
Likewise, we compute that
\bea
&& B^i_{j,a}C^a_{k,\ell} + C^i_{k,a}B^j_{a,\ell} + C^i_{\ell,a}B^j_{a,k} - (i \leftrightarrow j) \nonumber \\
& = & \varphi\big(2\lambda_{B}\lambda_{C}(e_ie_je_k - e_je_ie_k + e_ke_ie_j - e_ke_je_i)e_{\ell}\big) \nonumber \\
&& + \varphi\big(\lambda_{B}\lambda_{C}(e_je_ie_k - e_ie_ke_j + e_je_ke_i - e_ke_ie_j - e_ie_je_k + e_je_ke_i - e_ie_ke_j + e_ke_je_i)e_{\ell}\big) \nonumber \\
&& + \varphi\big(\lambda_{B}\lambda_{C}( e_ke_je_i - e_je_ke_i + e_ie_ke_j - e_ie_je_k - e_ke_ie_j + e_ie_ke_j - e_je_ke_i + e_je_ie_k)e_{\ell}\big) \nonumber \\
& = & 0,
\eea
and 
\bea
&& B^i_{j,a}A^a_{k,\ell} + B^i_{k,a}A^{j}_{a,\ell} + B^i_{\ell,a}A^j_{a,k} - (i \leftrightarrow j) \nonumber \\
& = & \varphi\big(2\lambda_{B}\lambda_{A}(e_ie_je_k - e_je_ie_k + e_ke_ie_j - e_ke_je_i)e_{\ell}\big) \nonumber \\
&& + \varphi\big(\lambda_{B}\lambda_{C}(e_je_ie_k - e_je_ke_i + e_ie_ke_j - e_ke_ie_j - e_ie_je_k + e_ie_ke_j - e_je_ke_i + e_ke_je_i)e_{\ell}\big) \nonumber \\
&& + \varphi\big(\lambda_{B}\lambda_{C}(e_ke_je_i - e_ie_ke_j + e_je_ke_{i} - e_ie_je_k - e_ke_ie_j + e_je_ke_i - e_ie_ke_j + e_je_ie_k)e_{\ell}\big) \nonumber \\
& = & 0. \nonumber
\eea
Therefore, when $\lambda_{B}^2 + \lambda_{A}\lambda_{C} = 0$, we have a classical Airy structure. The last statement about $D$ is a consequence of Lemma~\ref{trL}, noticing that $D^{{\rm ref}} = \tfrac{1}{2}{\rm tr}\,B^i = 0$ since $B^i_{j,k} = -B^i_{k,j}$. \hfill $\Box$

If $\lambda_{A}$ is invertible, the quantum Airy structures of Proposition~\ref{NCFrob} are transformed by the symmetries \eqref{SymU} with $u_{a,b} = \varphi\big(\tfrac{\lambda_{B}}{2\lambda_{A}}e_{a}e_{b}\big)$ into 
\bea
\tilde{A}^i_{j,k} & = & \varphi\big(\mu_{A}\{e_i,e_j\}e_k\big), \nonumber \\ 
\tilde{B}^i_{j,k} & = & \varphi\big(\mu_{B}e_ie_je_k\big),\nonumber \\
\label{CzeroAiry} \tilde{C} & = & 0, \\
\tilde{D}^i & = & D^i + \varphi(\lambda_B H e_i),
\eea
where the new parameters are related to the old ones by
$$
\mu_{A} = \lambda_{A},\qquad \mu_{B} = 2\lambda_{B}.
$$
 
If $\mathbb{A}$ happens to be commutative, we retrieve particular cases of the quantum Airy structure of Section~\ref{S4Frob}, namely the one with $(\theta_{A},\theta_{B},\theta_{C}) = (2\lambda_{A},0,2\lambda_{C})$ for the quantum Airy structure of Proposition~\ref{NCFrob}, and the one with $(\theta_{A},\theta_{B},\theta_{C}) = (2\mu_{A},\mu_B,0)$ for \eqref{CzeroAiry}. Generically they fit in the case $\theta_B^2 - \theta_{A}\theta_{C} \neq 0$.

If we assume $-\lambda_{A} = \lambda_{B} = \lambda_{C} = 1$, the associated  infinitesimal symplectomorphisms $({\cal L}_i)_{ i \in I}$ have a particularly nice form
$$
{\cal L}_i (e_j,0) = ([e_i,e_j],0) + (0,\{e_i,e_j\}),
$$
and
$$
{\cal L}_i (0,e_j) = (\{e_i,e_j\},0) - (0,[e_i,e_j]),
$$
relative to the orthonormal basis $(e_i)$. By direct computation, one can check that $[{\cal L}_i,{\cal L}_j] = 2 \phi([e_i,e_j] e_a) {\cal L}_a$.

We already computed in Proposition~\ref{Cequal0} the partition function for quantum Airy structures having $C = 0$, and we get the following expression using the pairing $\varphi$.

\begin{lemma}
The partition function $\tilde{Z}$ of the quantum Airy structure \eqref{CzeroAiry} is $\tilde{Z}(x) = \exp\big(\tfrac{\tilde{S}_0(x)}{\hbar} + \tilde{S}_1(x)\big)$ with
\bea 
\tilde{S}_0(x) & = & \varphi\Big(\tfrac{\mu_{A}}{\mu_{B}^3}\big(-\ln(1 - \mu_{B}x) - \mu_{B}x - \tfrac{\mu_B^2x^2}{2}\big)\Big), \nonumber  \\
\tilde{S}_1(x) & = & \varphi\big(-\tfrac{\tilde{D}}{\mu_{B}}\ln(1 - \mu_{B}x)\big). \nonumber 
\eea \hfill $\Box$
\end{lemma}

Transforming back to the initial quantum Airy structure, we deduce an integral formula for the expression for its partition function.
\begin{corollary}
\label{co89}The partition function of the quantum Airy structure of Proposition~\ref{NCFrob} is $Z(x) = \mathfrak{Z}(x)/\mathfrak{Z}(0)$ with
$$
\mathfrak{Z}(x) = \int \,\exp\Big\{\varphi\Big(\tfrac{\lambda_A}{\hbar\lambda_B}\big(-\tfrac{3t^2}{16} + tx - \tfrac{x^2}{2} - \tfrac{t}{\lambda_B}\big) - \tfrac{1}{2}\big(\tfrac{\lambda_A}{4\hbar \lambda_B^3} + \tfrac{D}{\lambda_B} + H\big)\ln(1 - \lambda_B t)\Big\}.
$$
\hfill $\Box$
\end{corollary}

\subsection{From loop spaces}
\label{S4Loop}

We consider $V = \mathbb{C}[\![z]\!]$. We index a basis of $V$ by non-negative integers $k$. If $f \in \mathbb{C}(\!(z)\!).\dd z$ (germ of meromorphic $1$-forms), we denote
$$
\varphi(f) = \Res_{z = 0} f(z).
$$
Let $(\xi_k)_{k \geq 0}$ be a linearly independent family of germs of meromorphic $1$-forms, and $(\xi^*_{k})_{k \geq 0}$ be a linearly independent family of germs of functions (elements of $V$), such that
$$
\varphi(\xi_k \xi^*_{\ell}) = \delta_{k,\ell}.
$$
Let $\theta \in \mathbb{C}(\!(z)\!).(\dd z)^{-1}$. Inspired by the Frobenius algebra example, we declare
\begin{eqnarray}
A^i_{j,k} & := & \varphi(\xi_i^*\,\dd\xi_j^*\,\dd\xi_k^*\,\theta), \nonumber \\
B^i_{j,k} & := & \varphi(\xi_i^*\,\dd\xi_j^*\,\xi_k\,\theta), \nonumber \\
\label{Cr}C^i_{j,k} & := & \varphi(\xi_i^*\,\xi_j\,\xi_k\,\theta),
\end{eqnarray}
for some $\theta$ yet to be fixed.

\begin{proposition}
\label{Loop1} Let $u_{k,\ell} = u_{\ell,k}$ be scalars indexed by integers $k,\ell \geq 0$, and choose
\beq
\label{xikforms} \xi_{k} = \Big(\frac{k + 1}{z^{k + 2}} + \sum_{\ell \geq 0} u_{k,\ell}\,z^{\ell}\Big)\dd z,\qquad \xi_k^* = \frac{z^{k + 1}}{k + 1},\qquad \theta = \sum_{r \in \mathbb{Z}} t_{r}z^{r}(\dd z)^{-1},
\eeq
where $t_r$ are formal parameters. Then, $A$ is fully symmetric. Further, the triple $(A,B,C)$ given by \eqref{Cr} defines a classical Airy structure if and only if $\theta \in z^{-1}\mathbb{C}[\![z]\!].(\dd z)^{-1}$. In this case, $(A,B,C,D)$ defines a quantum Airy structure iff
\beq
\label{Deqnloop}\forall i \geq 1,\qquad \sum_{r \geq -1} t_{r}D^{i + r} = 0,
\eeq
\end{proposition}

Let us make a few comments on the result.

First, for a given $r_0 \geq -1$, if $t_{k} = 0$ for $k \in \{-1,\ldots,r_0\}$, $(D^i)_{i = 0}^{r_0}$ does not appear in \eqref{epsconstraint} and can be chosen freely.

Second, specialising from formal parameters $t_r$ to complex-valued parameters, we obtain that $\theta(z) = \sum_{r \geq -1} t_{r}z^{r}(\dd z)^{-1}$ in Proposition~\ref{Loop1}. Conversely, if $\theta$ contains higher negative powers, $(A,B,C,D)$ cannot be a quantum Airy structure for generic parameters $(t_r)_{r \in \mathbb{Z}}$ --- but we do not rule out neither confirm the existence of non-generic $\theta$ for which these formulas define quantum Airy structures.

Third, the restriction on $\theta$ in this proposition implies that $A = 0$. Therefore, the partition function has $F_{0,n} = 0$ for all $n \geq 1$. The possibility to have a non-trivial partition function stems from the possible non-zero choices of $D$ satisfying \eqref{Deqnloop}. Lemma~\ref{Lemm1} in Appendix \ref{App4} gives vanishing rules for the $F_{g,n}$ of this quantum Airy structure. In particular, the case $t_{-1} \neq 0$ is not really interesting, as the partition function in this case is $Z = 1$, \textit{i.e.} all $F_{g,n}$ vanish. 

Fourth, we can identify the underyling Lie algebra for those Airy structures. Independently of $u$ we have:
$$
B^i_{j,k} = \sum_{r \geq -1}  \frac{k + 1}{i + 1}t_{r}\,\delta_{i + j + r,k}.
$$
So, we get the commutation relations
\beq
\label{Lieloop} [\tilde{L}_i,\tilde{L}_j] = \sum_{r \geq -1} (j - i)t_{r}\tilde{L}_{i + j + r},\qquad \tilde{L}_i := (i + 1)L_i.
\eeq
This is just another form of a Lie subalgebra of the Virasoro algebra $\mathfrak{Vir}_c$. Recall that $\mathfrak{Vir}_c$ is the Lie algebra defined by generators $(\mathcal{L}_{i})_{i \in \mathbb{Z}}$ satisfying the commutation relations
$$
[\mathcal{L}_i,\mathcal{L}_{j}] = (i - j)\mathcal{L}_{i + j} + \frac{c}{12}i(i^2 - 1)\delta_{i+j,0}
$$
The central charge $c \in \mathbb{C}$ will not play a role here as the property of being a sub-Lie algebra is true for any value of $c$.
\begin{lemma}
\label{VirL} If $\theta(z) = t_{r_0}z^{r_0} + O(z^{r_0 + 1})$ with $t_{r_0} \neq 0$, define $\tfrac{z^{r_0}}{\theta(z)} = \sum_{k \geq 0} \tau_{k}\,z^{k}$, and for $n \geq r_0$
$$
\hat{L}_n = -\sum_{k \geq 0} \tau_{k}\tilde{L}_{n + k - r_0}\,.
$$
Then, for all $m,n \geq r_0$, we have that $[\hat{L}_{m},\hat{L}_{n}] = (m - n)\hat{L}_{m + n}$.
\end{lemma}
The proof is a straightforward computation and is omitted.

\noindent \textbf{Proof of Proposition \ref{Loop1}.} First, we observe that the result for
\beq
\label{refxi} \xi_{k} = \xi_{k}^{(0)} := \frac{k + 1}{z^{k + 2}}\,\dd z
\eeq
implies the general result, since the $(A,B,C)$ for general
$$
\xi_{k} = \xi_{k}^{(0)} + \sum_{\ell \geq 0} u_{k,\ell} \dd \xi_{\ell}^*
$$
in \eqref{xikforms} is obtained from the $(A,B,C)$ for \eqref{refxi} by the symmetries \eqref{SymU}. For the choice \eqref{refxi}, unfortunately, we will proceed by direct computation.

We start with the warm-up case $\theta(z) = z^{r}(\dd z)^{-1}$ for some integer $r \in \mathbb{Z}$, which yields
\begin{eqnarray}
A^i_{j,k} & := & \frac{1}{i + 1}\,\delta_{i + j + k + r + 2,0}. \nonumber \\
B^i_{j,k} & := &  \frac{k + 1}{i + 1}\,\delta_{i + j + r,k}. \nonumber \\
C^i_{j,k} & := & \frac{(j + 1)(k + 1)}{i + 1}\,\delta_{i + r,j + k + 2}. \nonumber
\end{eqnarray}
In the following computations, it is implicit that the index $a$ is summed over, and it is important to keep in mind that indices are always $\geq 0$. We compute
\begin{eqnarray}
B^{i}_{j,a}B^{a}_{k,\ell} & = & \frac{\ell + 1}{(i + 1)(j + 1)}\,(j + 1)\,\delta_{i + j + r \geq 0} \delta_{i + j + k + 2r,\ell}. \nonumber \\
B^{i}_{k,a}B^{j}_{a,\ell} & = & \frac{\ell + 1}{(i + 1)(j + 1)}\,(i + k + r + 1)\,\delta_{i + k + r \geq 0}\delta_{i + j + k + 2r,\ell}. \nonumber \\
C^{i}_{\ell,a}A^{j}_{a,k} & = & \frac{\ell + 1}{(i + 1)(j + 1)}\,(i + r - \ell - 1)\,\delta_{i + r \geq \ell + 2}\,\delta_{i + j + k + 2r,\ell}. \nonumber
\end{eqnarray}
Therefore, the left-hand side of the \textbf{BA} relation reads
\begin{eqnarray}
& \frac{\ell + 1}{(i + 1)(j + 1)}\delta_{i + j + k + 2r,\ell}&\Big\{ (j - i)\delta_{i + j + r \geq 0} + (i + k + r  +1)\delta_{i + k + r \geq 0} - (j + k + r + 1)\delta_{j + k + r \geq 0} \nonumber \\
&& + (i + r - \ell - 1)\delta_{i + r \geq \ell + 2} - (j + r - \ell - 1)\delta_{i + r \geq k + 2}\Big\} \nonumber \\
=& \frac{\ell + 1}{(i + 1)(j + 1)}\delta_{i + j + k + 2r,\ell}&\Big\{(i + r)(2\delta_{i + j + r < 0} - \delta_{i + k + r + 1,0} - \delta_{\ell + 1,i + r}) \nonumber \\
&& + k(\delta_{i + j + r < 0} - \delta_{i + k + r + 1,0}) + \ell(-\delta_{i + j + r < 0} + \delta_{\ell + 1,i + r})\Big\},
\end{eqnarray}
where we exploited the constraint in the delta function prefactor to get rid of all $j$s, and we used $\delta_{A} = 1 - \delta_{\overline{A}}$ where $\overline{A}$ is the negation of $A$. The coefficients of $\delta_{i + k + r + 1,0}$ cancel each other, and likewise for $\delta_{\ell + 1,i + r}$. We are left with a non-zero multiple of
\begin{equation}
\label{BBend} \mathrm{L.H.S.}\,\,\mathrm{of}\,\,\textbf{BA} = (i - j)\delta_{i + j + r < 0}\delta_{i + j + k + 2r,\ell}.
\end{equation}
For  non-negative indices $i,j$, the set $\{i + j + r < 0\}$ is empty if and only if $r \geq 0$. For $r = -1$, it consists only of $(i,j) = (0,0)$, but the prefactor $(i - j)$ vanish in this case, so \eqref{BBend} also vanish identically if $r = -1$. When $r \leq -2$, one can find non-negative $(i,j,k,\ell)$ for which \eqref{BBend} is non-zero. Therefore, the \textbf{BB-CA} relation holds if and only if $r \geq -1$.

Next, we compute
\begin{eqnarray}
B^{i}_{j,a}A^{a}_{k,\ell} & = & \frac{\delta_{i + j + k + \ell + 2r + 2,0}}{(i + 1)(j + 1)}\,(j + 1)\delta_{i + j + r \geq 0}\,, \nonumber \\
B^{i}_{k,a}A^{j}_{a,\ell} & = & \frac{\delta_{i + j + k + \ell + 2r + 2,0}}{(i + 1)(j + 1)}\,(i + k + r  +1)\,\delta_{i + k + r \geq 0}\,, \nonumber \\
B^{i}_{\ell,a}A^{j}_{a,k} & = & \frac{\delta_{i + j + k + \ell + 2r + 2,0}}{(i + 1)(j + 1)}\,(i + \ell + r + 1)\,\delta_{i + \ell + r \geq 0}. \nonumber
\end{eqnarray}
Applying the same principles used in the previous computation, the left-hand side of the \textbf{BA} relation reads
\begin{equation}
\label{BAend} \mathrm{L.H.S.}\,\,\mathrm{of}\,\,\textbf{BA} = \frac{\delta_{i + j + k + \ell + 2r + 2,0}}{(i + 1)(j + 1)}\,(i-j)\,\delta_{i + j + r + 1 \leq 0}.
\end{equation}
A similar analysis shows that \eqref{BAend} is identically zero (\textit{i.e.} the \textbf{BA} relation is satisfied) if and only if $r \geq -1$.

We also compute
\begin{eqnarray}
B^{i}_{j,a}C^{a}_{k,\ell} & = & \frac{(k + 1)(\ell + 1)}{(i + 1)(j + 1)}\,(j + 1)\delta_{i + j + 2r,k + \ell + 2}\,\delta_{i + j + r \geq 0}, \nonumber \\
C^{i}_{k,a}B^{j}_{a,\ell} & = &  \frac{(k + 1)(\ell + 1)}{(i + 1)(j + 1)}\delta_{i + j + 2r,k + \ell + 2}\,\delta_{i + r \geq k +2}, \nonumber \\
C^{i}_{\ell,a}B^{j}_{a,k} & = & \frac{(k + 1)(\ell + 1)}{(i + 1)(j + 1)}\delta_{i + j + 2r,k + \ell + 2}\,\delta_{i + r \geq \ell + 2}, \nonumber
\end{eqnarray}
and we obtain that the left-hand side of the \textbf{BC} relation reads
\begin{equation}
\label{BCend} \mathrm{L.H.S.}\,\,\mathrm{of}\,\,\textbf{BC} = \frac{(k + 1)(\ell + 1)}{(i + 1)(j + 1)}\,\delta_{i + j + 2r,k + \ell + 2}\,(i - j)\delta_{i + j + r + 1 \leq 0}.
\end{equation}
Here, we see that under the condition $i + j + 2r = k + \ell + 2$, we have $i + j + r + 1 \geq 1 - r$. Therefore, if $r \leq 0$, \eqref{BCend} vanishes identically. But $r \geq 0$ also implies that $i + j + r + 1 \leq 0$ cannot be satisfied for non-negative $(i,j)$. Hence, \eqref{BCend} vanishes identically (\textit{i.e.} the \textbf{BC} relation is satisfied) for any $r \in \mathbb{Z}$.

Let us now assume $r \geq -1$ and analyse the \textbf{D} relation. As we have $A = 0$, it takes the form, for all $i > j \geq 0$.
$$
\sum_{a \geq 0} \frac{(j - i)(a + 1)}{(i + 1)(j + 1)}\,\delta_{i + j + r,a}\,D^{a} = 0,
$$
which is equivalent to the system of equations $D^{k + r} = 0$ for all $k \geq 1$.

Now we consider the case
$$
\theta = \sum_{r \geq -1} t_r\,z^{r}(\dd z)^{-1}.
$$
This results in a decomposition
$$
X = \sum_{r \geq -1} t_r\,{}^{r}X,\qquad X \in \{A,B,C\}.
$$
As $(t_{r})_{r \in \mathbb{Z}}$ are considered as formal parameters, while checking the three quadratic relations it is enough to check the coefficient of the monomial $t_{r}t_{s}$ with $r \neq s$. Indeed the coefficient of $t_{r}^2$ already displays the relations which $({}^{r}A,{}^{r}B,{}^{r}C)$ satisfy according to Step 1. We should therefore check that
$$
{}^{r}B^{i}_{j,a}\cdot {}^{s}B^{a}_{j,k} + {}^{s}B^i_{j,a}\cdot {}^{r}B^a_{j,k} + {}^{r} B^i_{k,a}\cdot {}^{s}B^{j}_{a,\ell} + {}^{s}B^i_{k,a}\cdot {}^{r} B^j_{a,\ell} + \big({}^{r}C^i_{\ell,a}\cdot {}^{s} A^j_{a,k} + {}^{s}C^{i}_{\ell,a}\cdot {}^{r} C^j_{a,k}\big) - (i \leftrightarrow j) = 0,
$$
and likewise for the two other relations. This is again checked by direct computations, which are very similar to the previous ones, thus omitted. Since $t_r = 0$ for $r < -1$, we have $A = 0$ and the \textbf{D} relation takes the form
$$
\forall i > j \geq 0,\qquad (i - j) \sum_{r \geq -1} t_{r} D^{i + j + r} = 0.
$$
It is then equivalent to 
\begin{equation}
\label{epsconstraint} \forall i \geq 1,\qquad \sum_{r \geq -1} t_r D^{r + i} = 0.
\end{equation}

\hfill $\Box$

\subsection{Loop space with \texorpdfstring{$\mathbb{Z}_2$}{Z2}-symmetry}

We can formulate a $\mathbb{Z}_{2}$-symmetric version of Proposition~\ref{Loop1}.
\begin{proposition}
\label{Loop2} Let $u_{k,\ell} = u_{\ell,k}$ be scalars indexed by integers $k,\ell \geq 0$, and specialise \eqref{Cr}:
\begin{eqnarray}
A^i_{j,k} & := & \varphi(\xi_i^*\,\dd\xi_j^*\,\dd\xi_k^*\,\theta), \nonumber \\
B^i_{j,k} & := & \varphi(\xi_i^*\,\dd\xi_j^*\,\xi_k\,\theta), \nonumber \\
\label{Cr2}C^i_{j,k} & := & \varphi(\xi_i^*\,\xi_j\,\xi_k\,\theta),
\end{eqnarray}
to
$$
\xi_{k} = \bigg(\frac{2k + 1}{z^{2k + 2}} + \sum_{\ell \geq 0} u_{k,\ell}\,z^{2\ell}\bigg)\,\dd z,\qquad \xi_{k}^* = \frac{z^{2k + 1}}{2k + 1},\qquad \theta = \sum_{s \in \mathbb{Z}} t_{s} z^{2s}(\dd z)^{-1}.
$$
while assuming $\theta \in \mathbb{C}(\!(z^2)\!).(\dd z)^{-1}$. Then, $A$ is fully symmetric. Further, the triple $(A,B,C)$ given by \eqref{Cr2} defines a classical Airy structure if and only if $\theta \in z^{-2}\mathbb{C}[\![z^2]\!].(\dd z)^{-1}$. In this case, $(A,B,C,D)$ defines a quantum Airy structure iff
\beq
\label{DDD}\forall i \geq 1,\qquad \sum_{s \geq -1} (2i + 2s + 1)\,t_sD^{i + s} = \frac{t_{-1}^2}{8}\, \delta_{i,2} + \frac{t_{-1}(t_0 + 2u_{0,0}t_{-1})}{4}\,\delta_{i,1}.
\eeq
\end{proposition}

As before, setting $t_s$ to complex values also give Airy structures. The restriction on $\theta$ in this proposition implies that
\bea
\label{Aexp} A^i_{j,k} & = & t_{-1}\delta_{i,j,k,0}.  \\
\label{Bexp}B^i_{j,k} & = & \sum_{s \geq -1} \frac{2k + 1}{2i + 1}t_{s}\,\delta_{i + j + s,k} + t_{-1}u_{0,0}\delta_{i,j,k,0}. 
\eea
In particular, if $t_{-1} = 0$, then $A = 0$ hence $F_{0,n} = 0$ for all $n$. Lemma~\ref{Lemm2} in Appendix  \ref{App4} gives more general vanishing rules for the $F_{g,n}$s of this quantum Airy structure. Note that, if we assume $D^k = 0$ for $k \geq 2$, \eqref{DDD} is satisfied iff
\begin{equation}
\label{DssoL}D^0 = \frac{t_{0}}{8} + \frac{u_{0,0}t_{-1}}{2},\qquad D^1 = \frac{t_{-1}}{24}\,.
\end{equation}
Note that, for fixed $a \geq 0$
$$
B^i_{a,a} = (2a + 1)\big(\delta_{i,0}t_{0} + \delta_{i,1}\tfrac{t_{-1}}{3}\big) + \delta_{i,a,0}t_{-1}u_{0,0}\,.
$$
Therefore, ${\rm Tr}\,B^i$ is not well-defined, and even after zeta regularisation of the sum $\sum_{a \geq 0} (2a + 1)$, the expression $\tfrac{1}{2} ``\sum_{a} B^i_{a,a}"$ does not reproduce \eqref{DssoL}. The commutation relations deduced from \eqref{Bexp} are
$$
[\tilde{L}_i,\tilde{L}_j] = \sum_{s \geq -1} (j - i)t_{s} \tilde{L}_{i + j + s},\qquad \tilde{L}_i := \frac{2i + 1}{2}\,L_i.
$$
They are the same as \eqref{Lieloop}, so we obtain again a sub-Lie algebra of $\mathfrak{Vir}_c$.

\noindent \textbf{Proof of Proposition~\ref{Loop2}} Due to the symmetry \eqref{SymU} it is sufficient to consider
$$
\xi_k := \frac{2k +1}{z^{2k + 2}}\,\dd z.
$$
If we denote $\eta_k$ the $1$-forms used in Proposition~\ref{Loop1}, note that $\xi_k = \eta_{2k}$ is just a subset of the $(\eta_k)_{k \geq 0}$. So, the case $\theta \in \mathbb{C}[\![z]\!](\dd z)^{-1}$ is covered by the previous Theorem with $r = 2s$ (excluding the statement about the \textbf{D} relation), by restricting to even indices. Indeed, although we have to sum over all indices $a$ to check the relations, the terms with odd $a$ are always zero since $X^{p}_{q,m} = 0$ whenever $p + q + m \neq 0\,\,{\rm mod}\,\,2$ for any $X \in \{A,B,C\}$, and two of the indices are not summed over and always even by our restriction. This is not so for the \textbf{D} relation, because the indices $a$ and $b$ which are summed over appear in the same symbol.

The case $\theta \in \big(t_{-1}z^{-2} + \mathbb{C}[\![z]\!]\big)(\dd z)^{-1}$ needs special care, as the relations \textbf{BA} and \textbf{BB-CA} failed in the proof of Proposition~\ref{Loop1} where indices of any parity were allowed. We first focus on the warm-up case $\theta(z) = z^{-2}(\dd z)^{-1}$, for which
\begin{eqnarray}
A_{j,k}^i & = & \delta_{i,j,k,0}. \nonumber \\
B_{j,k}^i & = & \frac{2k + 1}{2i + 1}\,\delta_{i + j,k + 1}. \nonumber \\
C_{j,k}^i  & = & \frac{(2j + 1)(2k + 1)}{2i + 1}\,\delta_{i,j + k + 2}. \nonumber
\end{eqnarray}
We just have to check that the following specialisation of \eqref{BBend} and \eqref{BAend} hold
$$
2(i - j)\delta_{i + j < 1}\delta_{i + j + k,\ell + 1} = 0,\qquad 2(i - j)\delta_{i + j + k + \ell + 1,0}\delta_{i + j < 1} = 0.
$$
This is obviously the case since $\delta_{i + j < 1}$ forces $i = j = 0$ and the prefactor $(i - j)$ makes them vanish in this case.

We now consider the $\mathbf{D}$ relation in the full generality of \ref{Loop2}. Since in any case $A^i_{j,k} = t_{-1}\delta_{i,j,k,0}$, it takes the form
$$
\frac{2(i - j)}{(2i + 1)(2j + 1)} \sum_{s \geq -1} (2i + 2j + 2s + 1)\,t_{s} D^{i + j + s} = \frac{t_{-1}}{2}(C^i_{0,0}\delta_{j,0} - C^j_{0,0} \delta_{i,0})
$$
We compute:
$$
C^i_{0,0} = \frac{t_{-1}\delta_{i,2} + (t_0 + 2u_{0,0}t_{-1})\delta_{i,1}}{2i + 1}
$$
We then see that we do not lose any relation by considering them only for $j = 0$ and $i \geq 1$, and they take the form:
$$
\sum_{s \geq -1} (2i + 2s + 1)\,t_sD^{i + s} = \frac{t_{-1}}{4i}\big(t_{-1}\delta_{i,2} + (t_0 + 2u_{0,0}t_{-1})\delta_{i,1}\big) = \frac{t_{-1}^2}{8}\, \delta_{i,2} + \frac{t_{-1}(t_0 + 2u_{0,0}t_{-1})}{4}\,\delta_{i,1}
$$
which entails the claim.
\hfill $\Box$

\subsection{Loop space of Frobenius algebras}
\label{S444}
Let $\mathbb{A}$ be a (commutative) Frobenius algebra and recall the notations of Section~\ref{S4Frob}. We choose an orthonormal basis $(e_{\alpha})_{\alpha}$ of $\mathbb{A}$, i.e $e_{\alpha} = e^*_{\alpha}$. Set $\mathcal{V} := \mathbb{A}[\![z]\!]$. The proofs of Propositions~\ref{FrobABCD} and \ref{Loop1}-\ref{Loop2} can easily be adapted to this setting. If $f \in \mathbb{A}(\!(z)\!).(\dd z)$, we define
$$
\Phi(f) := \Res_{z = 0}\, \phi(f(z))
$$
using the linear form $\varphi\,:\,\mathbb{A} \rightarrow \mathbb{C}$ provided by the Frobenius structure. Let $\xi_{i,\alpha}$ be linearly independent family of elements of $\mathcal{V}$, indexed by $\alpha$ and integers $i \geq 0$.

We declare
\begin{eqnarray}
A^{(i,\alpha)}_{(j,\beta),(k,\gamma)} & = & \Phi(\xi_{i,\alpha}^*\,\dd\xi_{j,\beta}^*\,\dd\xi_{k,\gamma}^*\,\theta), \nonumber  \\
B^{(i,\alpha)}_{(j,\beta),(k,\gamma)} & = & \Phi(\xi_{i,\alpha}^*\,\dd\xi_{j,\beta}^*\,\xi_{k,\gamma}\,\theta), \nonumber  \\
\label{CrcCC} C^{(i,\alpha)}_{(j,\beta),(k,\gamma)} & = & \Phi(\xi_{i,\alpha}^*\,\xi_{j,\beta}\,\xi_{k,\gamma} \theta),
\end{eqnarray}
where $\theta \in \mathbb{A}(\!(z)\!).(\dd z)^{-1}$.

\begin{proposition}
\label{ThmLoopTQFT}Let $v_{(k,\alpha),(\ell,\beta)} = v_{(\ell,\beta),(k,\alpha)}$ be scalars indexed by basis elements $\alpha,\beta$ of $\mathbb{A}$ and integers $k,\ell \geq 0$. Assume
$$
\theta = \sum_{\substack{r \geq -1 \\ \alpha}} t_{r,\alpha} z^{r}(\dd z)^{-1}e_{\alpha}
$$
for some scalars $t_{r,\alpha}$, and choose
$$
\xi_{k,\alpha} = \bigg(\frac{(k + 1)e_{\alpha}\dd z}{z^{k + 1}} + \sum_{\substack{\ell \geq 0 \\ \beta}} v_{(k,\alpha),(\ell,\beta)}\,z^{\ell}e_{\beta}\bigg)\,\dd z,\qquad \xi^*_{k,\alpha} = \frac{z^{k + 1}}{k + 1}\,e_{\alpha}.
$$
Then, the triple $(A,B,C)$ given by \eqref{CrcCC} defines a classical Airy structure. Further, $(A,B,C,D)$ is a quantum Airy structure iff
$$
\forall i \geq 1\,\,\forall \alpha_1,\alpha_2 ,\qquad \sum_{\substack{r \geq -1 \\ \alpha,\beta}} \varphi(e_{\alpha_1}e_{\alpha_2}e_{\alpha}e_{\beta})\,t_{r,\beta}D^{(r + i,\alpha)} = 0.
$$
\end{proposition}

\begin{proposition}
\label{ThmLoopZ2TQFT} Let $v_{(k,\alpha),(\ell,\beta)} = v_{(\ell,\beta),(k,\alpha)}$ be scalars indexed by basis elements $\alpha,\beta$ of $\mathbb{A}$ and integers $k,\ell \geq 0$. Choose
$$
\xi_{k,\alpha} = \bigg(\frac{(2k + 1)e_{\alpha}}{z^{2k + 2}} + \sum_{\substack{\ell \geq 0 \\ \beta}} v_{(k,\alpha),(\ell,\beta)}\,z^{2\ell}e_{\beta}\bigg)\,\dd z,\qquad \theta = \sum_{\substack{s \geq -1 \\ \alpha}} t_{s,\alpha} z^{2s}(\dd z)^{-1}e_{\alpha}.
$$
Then, $(A,B,C)$ given by \eqref{CrcCC} defines a classical Airy structure. Further, $(A,B,C,D)$ then defines a quantum Airy structure iff, for any $\alpha_1,\alpha_2$ and $i \geq 1$
\begin{eqnarray}
\sum_{\substack{s \geq -1 \\ \alpha,\beta}} \varphi(e_{\alpha_1}e_{\alpha_2}e_{\alpha}e_{\beta})(2s + 3)t_{s,\alpha}D^{(s + 1,\beta)} & = &  \sum_{\alpha,\beta} \bigg( \varphi(e_{\alpha_1}e_{\alpha_2}He_{\alpha}e_{\beta})\,\frac{t_{-1,\alpha}t_{0,\beta}}{4} \nonumber \\
&& + \sum_{\gamma,\epsilon} \varphi(e_{\alpha_1}e_{\alpha_2}e_{\alpha}e_{\beta} e_{\gamma}e_{\epsilon})\,u_{(0,\gamma),(0,\epsilon)} \,\frac{t_{-1,\alpha} t_{-1,\beta}}{2}\bigg), \nonumber \\
\sum_{\substack{s \geq -1 \\ \alpha,\beta}} \varphi(e_{\alpha_1}e_{\alpha_2}e_{\alpha}e_{\beta})\,(2s + 5)t_{s,\alpha}D^{(s + 2,\beta)} & = &\,\sum_{\alpha,\beta} \varphi(e_{\alpha_1}e_{\alpha_2}He_{\alpha}e_{\beta})\,\frac{t_{-1,\alpha}t_{-1,\beta}}{8}\,, \nonumber \\
\label{fdsugn}\sum_{\substack{s\geq -1 \\ \alpha,\beta}} \varphi(e_{\alpha_1}e_{\alpha_2}e_{\alpha}e_{\beta})(2s + 2i + 1)t_{s,\alpha}D^{(s + i,\beta)} & = & 0, \qquad (i \geq 3)\,,
\end{eqnarray}
where $H := \sum_{\alpha} e_{\alpha}^2$.
\end{proposition}

The proofs combine the two aspects of the proofs given in Sections~\ref{S4Frob} and \ref{S4Loop}, and hence are omitted. They rely on the fact that, each of the three terms in the three relations are already symmetric in $i$ and $j$ in the Frobenius algebra case. Note that, if we assume $\mathbb{A}$ is semi-simple and $(e_{\alpha})_{\alpha}$ is an orthonormal basis such that $e_{\alpha}e_{\beta} = \delta_{\alpha\beta}e_{\alpha}$, the constraints on $D$ can be rewritten, for all $\alpha$ as follows
\bea
\sum_{s \geq -1} (2s + 3)t_{s,\alpha}\,D^{(s + 1,\alpha)} & = & \frac{t_{-1,\alpha}t_{0,\alpha}}{4} + u_{(0,\alpha),(0,\alpha)}\frac{t_{-1,\alpha}^2}{2} \\ 
\sum_{s \geq -1} (2s + 5)t_{s,\alpha}\,D^{(s + 2,\alpha)} & = & \frac{t_{-1,\alpha}^2}{8}, \nonumber \\
\sum_{s \geq -1} (2s + 2i + 1)t_{s,\alpha}\,D^{(s + i,\alpha)} & = & 0 \qquad (i \geq 3). \nonumber
\eea

Let us describe this classical Airy structure in the language of Section~\ref{SSymp}. On $W = \mathbb{A}((z^{-1}))$ we consider the following symplectic form
$$
\omega(f(z),g(z)) = \Res_{z  = 0}   \varphi\big(f(z)\dd g(z)\big).
$$
Let $W = V^* \oplus V$ be a polarisation of $W$, where $V$ (resp. $V^*$) has basis $\left(\xi_{k,\alpha}(z)\right)_{k,\alpha}$, respectively $\big(\xi_{k,\alpha}^*(z) := \frac{z^{k+1}}{k+1}e_\alpha\big)_{k,\alpha}$, such that 
$$
\forall (k,l) \in \mathbb{Z}_{\geq 0}^2 \, , \; \omega(\xi_{k,\alpha}^*(z), \xi_{l,\beta}(z)) = \delta_{k,l} \delta_{\alpha,\beta}.
$$
One defines a classical Airy structure given by the set of the infinitesimal symplectomorphisms $\left({\cal L}_{k,\alpha}\right)_{k \in \mathbb{Z}_{\geq 0}}$, where
$$
\forall k \in \mathbb{Z}_{\geq 0} \, , \; \forall f \in W \, , \; {\cal L}_{k,\alpha} f(z) = \xi_{k,\alpha}^*(-z) \, {\theta(-z) \, \dd f(-z) } ,
$$
and $\theta(z) \in z^{-1} \mathbb{A}[\![z]\!] \cdot (\dd z)^{-1}$.

\begin{corollary}
The operators $ \left({\cal L}_k\right)_{k \geq 0}$ together with the above orthonormal basis of $V$ and $V^*$ define a classical Airy structure.
\end{corollary}

\section{Relation with Chekhov--Eynard--Orantin topological recursion}
\label{TRcomp}

\subsection{Comparison}
\label{comparTR}
The original setting of \cite{EOFg} for the topological recursion is the data of a spectral curve, \textit{i.e.}
\begin{itemize}
\item[$\bullet$] a branched cover with simple ramification points $x\,:\,\Sigma \rightarrow \Sigma'$, where $\Sigma'$ is an open subset of $\mathbb{P}^1$.
\item[$\bullet$] a meromorphic function $y$ on $\Sigma$, such that the zeroes of $\dd y$ are distinct from the zeroes of $\dd x$. We set $\omega_{0,1} := y\dd x$ and assume $\dd x$ has finitely many zeroes.
\item[$\bullet$] a symmetric bidifferential $\omega_{0,2}$ on $\Sigma^2$ with a double pole at coinciding points and the following behaviour in local coordinates
$$
\omega_{0,2}(p_1,p_2) = \frac{\dd z(p_1) \dd z(p_2)}{(z(p_1) - z(p_2))^2} + O(1).
$$
Such an object is sometimes called a \emph{fundamental bidifferential of the second kind} on $\Sigma$.
\end{itemize}

We denote $\mathfrak{r} \subset \Sigma$ the set of the ramification points, \textit{i.e.} the zeros of $\dd x$. As they are simple, we can find around each $r \in \mathfrak{r}$ a local coordinate $z$ such that
$$
x(p) = x(r) + \tfrac{z(p)^2}{2}.
$$
Let $U \subseteq \Sigma$ be the disjoint union of small enough neighbourhoods of the ramification points, in which $\iota\,:\,z \mapsto -z$ is a well-defined holomorphic involution. By construction $x \circ \iota = x$. We introduce the recursion kernel
$$
K(p_0,p) = \frac{1}{2}\,\frac{\int_{\iota(p)}^{p} \omega_{0,2}(\cdot,p_0)}{\omega_{0,1}(p) - \omega_{0,1}(\iota(p))}.
$$
It is defined for $(p_0,p) \in \Sigma \times U$ as a $1$-form in $p_0$ and the inverse of a $1$-form in $p$. For $2g - 2 + n > 0$, \cite{EOFg} defines by induction
\beq
\label{oldTR}\omega_{g,n}(p_1,I) := \sum_{r \in \mathfrak{r}} \Res_{p = r} K(p_1,p)\bigg\{\omega_{g - 1,n + 1}(p,\iota(p),I) + \sum_{\substack{g' + g'' = g \\ J' \sqcup J'' = I}}^{*} \omega_{g',1 + |J'|}(p,J')\omega_{g'',1+|J''|}(\iota(p),J'')\bigg\},
\eeq
where $I = \{p_2,\ldots,p_n\}$, and $\sum^*$ means that we exclude the terms of the form $\omega_{0,1}\omega_{g,n}$ from the sum. Although \eqref{oldTR} gives a special role to the variable $p_1$, \cite{EOFg} proves inductively that $\omega_{g,n}(p_1,\ldots,p_n)$ is symmetric under permutation of all the $p_i$s, therefore for $L = \{p_1,\ldots,p_{\ell}\}$ an unordered $\ell$-tuple of variables in $\Sigma$, the notation $\omega_{h,\ell}(L)$ makes sense. Manifestly \eqref{oldTR} produces differential forms whose only poles are located at the ramification points. In other words
$$ \omega_{g,n} \in H^0\big(\Sigma^{\times n}, (K(\star \mathfrak{r}))^{\boxtimes n}\big)^{\mathfrak{S}_n}.$$

In \cite{EInter}, it is shown that the $\omega_{g,n}$ can be decomposed on a suitable family of meromorphic $1$-forms. To be self-contained we review this proof, and make explicit the recursion following from \eqref{oldTR} for the coefficients of the decomposition.

\begin{definition}
For $k \geq 0$ and $r \in \mathfrak{r}$, we define for $p_0 \in \Sigma$ the meromorphic $1$-form 
\beq
\label{xiforms} \xi_{k,r}(p_0) := \Res_{p = r} \Big(\int_{r}^{p} \omega_{0,2}(p_0,\cdot)\Big)\,\frac{(2k + 1)\dd z(p)}{z(p)^{2k + 2}}.
\eeq
We also define, for $p_0$ in a neighbourhood of $r$ in $\Sigma$
$$
\xi_{k,r}^*(p_0) := \frac{z^{2k + 1}(p_0)}{2k + 1},\qquad \theta(p_0) := \frac{-2}{\omega_{0,1}(p_0) - \omega_{0,1}(\iota(p_0))},
$$
and if $p_0$ is in a neighbourhood of $r_0 \neq r$, we define $\xi_{k,r}^*(p_0) := 0$.
\end{definition}

\begin{lemma}
\label{LemmaFogi}
For $2g - 2 + n > 0$, there exists a unique decomposition with finitely many non-zero terms
\beq
\label{Fogi}\omega_{g,n}(p_1,\ldots,p_n) = \sum_{\substack{r_1,\ldots,r_n \in \mathfrak{r} \\ k_1,\ldots,k_n \geq 0}} W_{g,n}\left[\begin{smallmatrix} r_1 & \cdots & r_n \\ k_1 & \cdots & k_n \end{smallmatrix}\right]\,\prod_{i = 1}^n \xi_{k_i,r_i}(p_i).
\eeq
\end{lemma}
More precisely, one can show \cite{EInter} that the coefficients in \eqref{Fogi} with $\sum_{i} k_i > 3g - 3 + n$ vanish. For completeness, we also give a proof in Appendix~\ref{App4}.

We can now compare with the quantum Airy structure of Section~\ref{S444}. We take $\mathbb{A} = \bigoplus_{r \in \mathfrak{r}} \mathbb{C}$ as the sum of trivial $1$-dimensional Frobenius algebras, and we let $V = \mathbb{A}[\![z]\!]$ be the vector space with a basis indexed by $k \geq 0$ and $r \in \mathfrak{r}$. As we assumed $y$ is holomorphic and $\dd y$ has no zero at $\mathfrak{r}$, we deduce that $\theta(p)$ has an expansion for $p \rightarrow r$ of the form
\beq
\label{om10exp}\theta(p) = \sum_{m \geq -1} t_{m,r}\,z^{2m}(p)\,(\dd z(p))^{-1}.
\eeq 
According to Proposition~\ref{ThmLoopZ2TQFT},
\bea
A^{(k_1,r_1)}_{(k_2,r_2),(k_3,r_3)} & := & \Res_{p = r_1} \xi_{k_1,r_1}^*(p)\,\dd\xi_{k_2,r_2}^*(p)\,\dd \xi_{k_3,r_3}^*(p)\,\theta(p), \nonumber \\
B^{(k_1,r_1)}_{(k_2,r_2),(k_3,r_3)} & := & \Res_{p = r_1} \xi_{k_1,r_1}^*(p)\,\dd \xi_{k_2,r_2}^*(p)\,\xi_{k_3,r_3}(p)\,\theta(p), \nonumber \\
C^{(k_1,r_1)}_{(k_2,r_2),(k_3,r_3)} & := & \Res_{p = r_1} \xi_{k_1,r_1}^*(p)\,\xi_{k_2,r_2}(p)\,\xi_{k_3,r_3}(p)\,\theta(p), \nonumber \\
D^{(k,r)} & = & \delta_{k,0}\big(\tfrac{t_{-1,r}}{2}\,\phi_{0,2}\left[\begin{smallmatrix} r & r \\ 0 & 0 \end{smallmatrix}\right] + \tfrac{t_{0,r}}{8}\big)+ \delta_{k,1}\tfrac{t_{-1,r}}{24}
\label{qAIryTR}
\eea
is a quantum Airy structure. Here, $\varphi_{0,2}\left[\begin{smallmatrix} r & r \\ 0 & 0 \end{smallmatrix}\right]$ is a scalar, which corresponds to the constant term in the expansion of $\omega_{0,2}$ near $(p_1,p_2) = (r,r)$ in local coordinates $(z(p_1),z(p_2))$, see \eqref{om20exp}. One can then check that indeed $D$ is a solution of the \textbf{D} relation in the form \eqref{fdsugn}. Substituting the expansion (\ref{Fogi}) in the residue formula (\ref{oldTR}) gives a recursion for the $W_{g,n}$, which is identical to KS recursion \eqref{TRForm} for the Taylor coefficients $F_{g,n}$ of this quantum Airy structure. Since we can check (see the proof below) that the initial data are the same, this leads to

\begin{proposition}
\label{FWid} For $2g - 2 + n > 0$, $F_{g,n}\big((k_1,r_1),\ldots,(k_n,r_n)\big)$ computed by KS topological recursion for the quantum Airy structure \eqref{qAIryTR} and $W_{g,n}\left[\begin{smallmatrix} r_1 & \cdots & r_n \\ k_1 & \cdots & k_n \end{smallmatrix}\right]$ computed by the topological recursion of \cite{EOFg}, agree.
\end{proposition}

\noindent \textbf{Proof.}  We start by a preliminary study of the recursion kernel. If we expand $\omega_{0,2}(p_1,p_2)$ in local coordinates when $p_i$ is in a neighbourhood of $r_i \in \mathfrak{r}$ we get that
\beq
\label{om20exp} \omega_{0,2}(p_1,p_2) = \frac{\delta_{r_1,r_2}\dd z(p_1)\dd z(p_2)}{(z(p_1) - z(p_2))^2} + \sum_{\ell_1,\ell_2 \geq 0} \phi_{0,2}\left[\begin{smallmatrix} r_1 & r_2 \\ \ell_1 & \ell_2 \end{smallmatrix}\right]\,z^{\ell_1}(p_1)z^{\ell_2}(p_2) \dd z(p_1)\dd z(p_2).
\eeq
We find the following expansion for \eqref{xiforms} when $p_0 \rightarrow r_0$ for some $r_0 \in \mathfrak{r}$
$$
\xi_{k,r}(p_0) = \frac{(2k + 1)\delta_{r,r_0}}{z^{2k + 2}(p_0)} + (2k + 1) \sum_{\ell \geq 0}  \phi_{0,2}\left[\begin{smallmatrix} r & r_0 \\ 2k & \ell \end{smallmatrix}\right]\,z^{\ell}(p_0)\,\dd z(p_0).
$$
In particular, $\xi_{k,r}(p_0)$ has a pole of order $2k + 2$ at $p_0 = r$, and is holomorphic elsewhere. We also find for $p$ in a neighbourhood of $r$
$$
\frac{1}{2} \int_{\iota(p)}^{p}\omega_{0,2}(\cdot,p_0) = \sum_{\substack{k \geq 0 \\ r \in \mathfrak{r}}} \xi_{k,r}(p_0)\,\xi_{k,r}^*(p)
$$
under the condition $|z(p_0)| > |z(p)|$ when $p_0$ is in the neighbourhood of $r$. Here we have used that
$$
\frac{1}{(z(p_0) - z(p))^2} = \sum_{\ell \geq 0} \frac{(\ell + 1)\,z^{\ell}(p)}{z^{\ell + 2}(p_0)}
$$
for $p_0,p$ in the neighbourhood of the same $r$. To perform the residue computation in \eqref{oldTR}, we will need the expansion of the recursion kernel $K(p_0,p)$ around $p \rightarrow r$
\beq
\label{Klocal} K(p_0,p) = -\tfrac{1}{2}\sum_{k \geq 0} \xi_{k,r}(p_0)\,\xi_{k,r}^*(p)\,\theta(p),\qquad \xi_{k,r}^*(p) \in O(z(p)^{2k + 1}).
\eeq

\vspace{0.2cm}

Let us start by computing $\omega_{0,3}$.
$$
\omega_{0,3}(p_1,p_2,p_3) = \sum_{r \in \mathfrak{r}} \Res_{p = r} K(p_1,p)\{\omega_{0,2}(p,p_2)\omega_{0,2}(\iota(p),p_3) + \omega_{0,2}(\iota(p),p_2)\omega_{0,2}(p,p_3)\big\}. \nonumber \\
$$
Since $\theta(p) \in O\big(z^{-2}(p)(\dd z)^{-1}\big)$, $K(p_1,p)$ has a simple pole at $p = r$. So, the residue selects the coefficient of $(\dd z(p))^2$ in $\big\{\cdots \big\}$, and as $\dd \xi_{k,r}(p) = z^{2k}(p)\dd z(p)$, we find that
\beq
\label{ome03} \omega_{0,3}(p_1,p_2,p_3) = \delta_{k_1,k_2,k_3,0} \sum_{r \in \mathfrak{r}} \bigg(\prod_{i = 1}^3 \dd\xi_{0,r}(p_i)\bigg) \Res_{p = r} \xi_{0,r}^*(p)\dd\xi_{0,r}^*(p)\dd\xi_{0,r}^*(p)\theta(p).
\eeq
The factor $-\tfrac{1}{2}$ in front of \eqref{Klocal} disappeared as \eqref{ome03} has two terms with equal contribution, and the $\iota(p)$ in one of the factor $\omega_{0,2}$ results into a minus sign in the local coordinate $z$. We therefore have proved \eqref{Fogi} for $(g,n) = (0,3)$, and can identify the coefficients $W_{0,3}\left[\begin{smallmatrix} r_1 & r_2 & r_3 \\ k_1 & k_2 & k_3 \end{smallmatrix}\right]$ with $A^{(k_1,r_1)}_{(k_2,r_2),(k_3,r_3)}$ introduced in \eqref{qAIryTR} -- these coefficients vanish unless $k_1 = k_2 = k_3 = 0$ and $r_1 = r_2 = r_3$.

\vspace{0.2cm}

Likewise we compute $\omega_{1,1}$. Examining the local behaviour at ramification points, we find
\bea
\omega_{1,1}(p_1) & = & \sum_{r \in \mathfrak{r}} \Res_{p = r} K(p_1,p)\omega_{0,2}(p,\iota(p)) \nonumber \\
& = & \sum_{r \in \mathfrak{r}} \sum_{k = 0}^{1} \xi_{k,r}(p_1) \Res_{p = r} \tfrac{z(p)^{2k + 1}}{2(2k + 1)}\Big(\tfrac{t_{-1,r}}{z^2(p)} + t_{0} + O(z^2(p))\Big)\Big(\tfrac{1}{4z^2(p)} + \phi_{0,2}\left[\begin{smallmatrix} r & r \\ 0 & 0 \end{smallmatrix}\right] + O(z^2(p))\Big) \nonumber \\
& = & \sum_{r \in \mathfrak{r}} \big(\tfrac{t_{-1,r}}{2}\,\phi_{0,2}\left[\begin{smallmatrix} r & r \\ 0 & 0 \end{smallmatrix}\right] + \tfrac{t_{0,r}}{8}\big)\xi_{0,r}(p_1) + \tfrac{t_{-1,r}}{24}\,\xi_{1,r}(p_1),
\eea
which proves \eqref{Fogi} for $(g,n) = (1,1)$ with $W_{1,1}\left[\begin{smallmatrix} r \\ k \end{smallmatrix}\right] = D^{(k,r)}$ given in \eqref{qAIryTR}.

\vspace{0.2cm}

Now let $2g - 2 + n > 2$ and assume the claim of Lemma~\ref{LemmaFogi} has been established for all $(g',n')$ such that $2g' - 2 + n' < 2g - 2 + n$. Let $I = \{p_2,\ldots,p_n\}$ an unordered $(n - 1)$-uple of variables in $\Sigma$. In Equation \eqref{oldTR} for $\omega_{g,n}(p_1,I)$, we denote $\omega_{g,n}^{B}$ the sum of terms in the right-hand side involving $\omega_{0,2}\omega_{g,n -  1}$, and $\omega_{g,n}^{C}$ the sum of all the other terms. We have that
\beq
\label{omegagnB}\omega_{g,n}^{B}(p_1,I) = \sum_{i = 2}^n \Res_{p = \mathfrak{r}} K(p_1,p)\big(\omega_{0,2}(p,p_i)\omega_{g,n - 1}(\iota(p),I\setminus\{p_i\}) + \omega_{0,2}(\iota(p),p_i)\omega_{g,n - 1}(p,I\setminus\{p_i\})\big).
\eeq
As $K(p_1,p)$  is invariant under $p \rightarrow \iota(p)$, the two terms give an equal contribution. The form \eqref{Fogi} of $\omega_{g,n - 1}$ by the induction hypothesis implies that
$$
\omega_{g,n - 1}(p,I\setminus\{p_i\}) = \tfrac{1}{2}\big(\omega_{g,n - 1}(p,I\setminus\{p_i\}) - \omega_{g,n - 1}(\iota(p),I\setminus\{p_i\})\big) + O(z(p)\dd z(p)).
$$
As $\omega_{0,2}(p,p_i)$ is holomorphic near $p \rightarrow r$, we deduce that replacing it with its odd part in \eqref{omegagnB} does not change the residue
$$
\omega_{g,n}^B(p_1,I) = \sum_{i = 2}^n \Res_{p = \mathfrak{r}} K(p_1,p)\,\tfrac{1}{2}\big(\omega_{0,2}(\iota(p),p_i) - \omega_{0,2}(p,p_i)\big)\omega_{g,n - 1}(p,I\setminus\{p_i\}).
$$
We substitute in this formula, for $p$ in the neighbourhood of $r$
$$
\tfrac{1}{2}\big(\omega_{0,2}(\iota(p),p_i) - \omega_{0,2}(p,p_i)\big) = -\sum_{k \geq 0} \xi_{k,r}(p_i)\dd\xi_{k,r}^*(p),
$$
and the decomposition \eqref{Fogi} for $\omega_{g,n - 1}$. The result for $\omega_{g,n}^B$ decomposes like \eqref{Fogi} with coefficients
$$
W_{g,n}^{B}\left[\begin{smallmatrix} r_1 & \cdots & r_n \\ k_1 & \cdots &k_n \end{smallmatrix}\right] = \sum_{i = 2}^n \sum_{k',r'} B^{(k_1,r_1)}_{(k_i,r_i),(k',r')}\,W_{g,n - 1}\left[\begin{smallmatrix} r' & r_{2} & \cdots & \widehat{r_i} & \cdots & r_n \\ k' & k_2 & \cdots & \widehat{k_i} & \cdots & k_n \end{smallmatrix}\right],
$$
where
$$
B^{(k_1,r_1)}_{(k_i,r_i),(k',r')} = \Res_{p = r_1} \xi_{k_1,r_1}^*(p)\,\dd \xi_{k_i,r_i}^*(p)\,\xi_{k',r'}(p)\,\theta(p)
$$
as given in \eqref{qAIryTR}. Due to the local behaviour of the integrand, $B^{(k_1,r_1)}_{(k_2,r_2),(k_3,r_3)}$ vanishes when $r_1 \neq r_2$, or when $r_1 = r_2 \neq r_3$ and $k_1 + k_2 > 0$, or when $r_1 = r_2 = r_3$ and $k_1 + k_2 \geq k_3 + 1$. In particular these selection rules imply that there are finitely many non-zero $W_{g,n}^B$s.

Let us turn to
$$
\omega_{g,n}^{C}(p_1,I) = \Res_{p = \mathfrak{r}} K(p_1,p)\bigg\{\omega_{g - 1,n + 1}(p,\iota(p),I) + \sum_{\substack{g' + g'' = g \\ J' \sqcup J'' = I}}^{**} \omega_{g',1+|J'|}(p,J')\omega_{g'',1 + |J''|}(\iota(p),J'')\bigg\},
$$ 
where $\sum^{**}$ excludes the terms of the form $\omega_{0,1}\omega_{g,n}$ or $\omega_{0,2}\omega_{g,n - 1}$. By induction hypothesis, we can directly substitute the decomposition \eqref{Fogi} for all the $\omega$s involved in the left-hand side. We find that $\omega_{g,n}^{C}$ has a decomposition again of the form \eqref{Fogi}, with coefficients
\bea
W_{g,n}^{C}\left[\begin{smallmatrix} r_1 & \cdots & r_n \\ k_1 & \cdots & k_n \end{smallmatrix}\right] & = & \tfrac{1}{2} \sum_{\substack{k',k'' \geq 0 \\ r',r'' \in \mathfrak{r}}} C^{(k_1,r_1)}_{(k',r'),(k'',r'')}\bigg(W_{g - 1,n + 1}\left[\begin{smallmatrix} r' & r'' & r_2 & \cdots & r_n \\ k' & k'' & k_2 & \cdots & k_n \end{smallmatrix}\right] \nonumber \\
\label{WgnC} && + \sum_{\substack{g' + g'' = g \\ J' \sqcup J'' = \{2,\ldots,n\}}}^{**} W_{g',1+|J'|}\left[\begin{smallmatrix} r' & (r_j)_{j \in J'} \\ k' & (k_j)_{j \in J'} \end{smallmatrix}\right]W_{g'',1+ |J''|}\left[\begin{smallmatrix} r'' & (r_j)_{j \in J''} \\ k'' & (k_j)_{j \in J''} \end{smallmatrix}\right] \bigg),
\eea
where
$$
C^{(k_1,r_1)}_{(k_2,r_2),(k_3,r_3)} = \Res_{p = r_1} \xi_{k_1,r_1}^*(p)\,\xi_{k_2,r_2}(p)\,\xi_{k_3,r_3}(p)\,\theta(p)
$$
as given in \eqref{qAIryTR}. Due to the local behaviour of the integrand, $C^{(k_1,r_1)}_{(k_2,r_2),(k_3,r_3)} = C^{(k_1,r_1)}_{(k_3,r_3),(k_2,r_2)}$ vanishes when $r_2,r_3 \neq r_1$ and $k_1 > 0$, or when $r_2 = r_1 \neq r_3$ and $k_1 \geq k_2 + 2$, or when $r_1 = r_2 = r_3$ and $k_1 \geq k_2 + k_3 + 3$. In particular, this implies that \eqref{WgnC} contains only finitely many non-zero terms. We therefore have justified that $\omega_{g,n} = \omega_{g,n}^{B} + \omega_{g,n}^{C}$ has the form \eqref{Fogi}, and proved Lemma~\ref{LemmaFogi} by induction.

Since we have checked $F_{0,3} = A = W_{0,3}$ and $F_{1,1} = D = W_{1,1}$, and the recursive rules to build the $W_{g,n}$s agree with the KS topological recursion \eqref{TRForm} for the $F_{g,n}$s, this entails Proposition~\ref{FWid}. \hfill $\Box$

More explicitly, in terms of coefficients of expansion of $\omega_{0,1}$ in \eqref{om10exp} and $\omega_{0,2}$ in \eqref{om20exp}, the relevant quantum Airy structure is
\bea
A^{(k_1,r_1)}_{(k_2,r_2),(k_3,r_3)} & = & \delta_{k_1,k_2,k_3,0}\delta_{r_1,r_2,r_3}t_{-1,r_1}, \nonumber \\
B^{(k_1,r_1)}_{(k_2,r_2),(k_3,r_3)} & = & \frac{2k_3 + 1}{2k_1 + 1}\,\delta_{r_1,r_2}\bigg(\delta_{r_2,r_3}t_{k_3 - k_2 - k_1,r_1} + \delta_{k_1,k_2,0}\,\varphi_{0,2}\left[\begin{smallmatrix} r_3 & r_1 \\ 2k_3 & 0 \end{smallmatrix}\right]\bigg),\nonumber\\ 
C^{(k_1,r_1)}_{(k_2,r_2),(k_3,r_3)} & = & \frac{(2k_3 + 1)(2k_2 + 1)}{2k_1 + 1}\bigg(\delta_{r_1,r_2,r_3}\,t_{1 + k_2 + k_3 - k_1,r_1} + \sum_{m = 0}^{1 + k_3 - k_1}  \delta_{r_1,r_3}\,\varphi_{0,2}\left[\begin{smallmatrix} r_2 & r_1 \\ 2k_2 & 2m \end{smallmatrix}\right] t_{k_3 - k_1 - m,r_1} \nonumber \\
&& + \sum_{m = 0}^{1 + k_2 - k_1} \delta_{r_1,r_2}\,\varphi_{0,2}\left[\begin{smallmatrix} r_3 & r_1 \\ 2k_3 & 2m \end{smallmatrix}\right] t_{k_2 - k_1 - m,r_1} + \delta_{k_1,0}\,\varphi_{0,2}\left[\begin{smallmatrix} r_2 & r_1 \\ 2k_2 & 0 \end{smallmatrix}\right]\varphi_{0,2}\left[\begin{smallmatrix} r_3 & r_1 \\ 2k_3 & 0 \end{smallmatrix}\right] t_{-1,r_1}\bigg). \nonumber
\eea

\subsection{The point of view of Givental quantisation of Lagrangian cones}
\label{SJUING}
One of the applications of the original topological recursion formalism is the study of Frobenius manifolds/cohomological field theories. In this setup, \cite{DBOSS} established that the topological recursion applied to a specific local spectral curve is equivalent to Givental's quantisation formalism \cite{GiventalQuad} for computing the ancestor potential of a semi-simple cohomological field theory. This correspondence was obtained by a direct comparison of the result of the topological recursion and of Givental reconstruction procedure. In this section, we revisit this equivalence from the point of view of quantisation of Givental's Lagrangian cone \cite{CoatesGivental,Giventalcone}, giving it a stronger geometric explanation. We first review the Lagrangian cone formalism, following Coates and Givental.

Let $V$ be a finite dimensional vector space equipped with a bilinear form $(\cdot,\cdot)$ and a distinguished vector $\mathbf{1}$, and let $\mathcal{W}:=V(\!(z^{-1})\!)$ be the corresponding loop space equipped with the symplectic form\footnote{This is not the same symplectic form as in the example of Section~\ref{S444}. One can go from one to the other by a Laplace transform.} $\tilde{\omega}$ defined by
$$
\forall (f,g) \in \mathcal{W}^2,\qquad \tilde{\omega}(f,g) := \frac{1}{2 {\rm i}\pi} \oint  \left(f(-z),g(z)\right) \, \dd\,z.
$$
Consider the polarisation $\mathcal{W} = \mathcal{V}_+ \oplus \mathcal{V}_-$ where $\mathcal{V}_+ := V[\![z]\!]$ and $\mathcal{V}_-:= z^{-1} V[\!(z)\!]$. Then the symplectic form gives an identification $({\cal W},\tilde{\omega}) \simeq (T^*{\cal V}_+,\tilde{\omega})$.

Parametrising elements $q$ of ${\cal V}_+$ by an infinite dimensional vector ${\bf t} :=(t_k)_{k \geq 0}$,
$$
q(z) := \sum_{k=0}^\infty (t_k - \delta_{k,1}\mathbf{1}) z^k,
$$
one defines the graph ${\cal L}_{\cal F}$ of the derivative of a function ${\cal F}({\bf t})$ on ${\cal V}_+$ by
$$
{\cal L}_{\cal F}:=\{(p,q) \in T^*{\cal V}_+\,\,:\,\,\, p = \dd_q {\cal F}({\bf t})\}.
$$
As a formal germ around $q = -z$, this defines a Lagrangian submanifold of $T^*{\cal V}_+$ and hence of $({\cal{W}},\tilde{\omega})$. 


An interesting choice for such functions are genus $0$ free energies coming from a CohFT, \textit{i.e.} functions $F_0(\mathbf{t})$ satisfying the following three axioms. It is convenient to state them by choosing a basis $(e_{\nu})_{\nu = 1}^{d}$ of $V$ and denoting $g_{\mu,\nu} := (e_{\mu},e_{\nu})$ and by $(g^{\mu,\nu})_{\mu,\nu}$ its inverse matrix. For $k \geq 0$, we denote $t_k := \sum_{\nu = 1}^{d} t_k^{\nu} e_{\nu}$. With these notations, the three axioms defining a genus 0 free energy read as follows.

\begin{itemize}

\item The \emph{dilaton equation}, which states that $F_0$ is homogenous of degree 2
\beq
\label{DE}
2 F_0({\bf t}) = \sum_{k \geq 0} \sum_{\nu = 1}^d t_{k}^\nu \frac{\partial F_0({\bf t})}{\partial t_{k}^\nu}.
\eeq

\item The \emph{string equation}, which decomposes the action of $\frac{\partial}{\partial t_{0}^1}$ (the unit vector field)
\beq
\label{SE}
\frac{\partial F_0}{\partial t_{0}^1} = \frac{1}{2} (t_0,t_0) + \sum_{k \geq 0}\sum_{\nu = 1}^d  t_{k+1}^\nu \frac{\partial F_0}{\partial t_k^\nu}.
\eeq

\item The \emph{topological recursion relations}\footnote{This set of equations is different from the topological recursion of \cite{EOFg}. It is unfortunate that both names coincide.}
\beq
\label{TRR}
\forall (\alpha,\beta,\gamma) \in \{1,\ldots,d\}^3\,\,\,\forall (k,l,m) \in \mathbb{Z}^3_{\geq 0} \, , \; \frac{\partial^3 F_0}{\partial t_{k+1}^\alpha \, \partial t_{l}^\beta \, \partial t_{m}^\gamma} = \sum_{\mu,\nu=1}^d  \frac{\partial^2 F_0}{\partial t_{k}^\alpha \, \partial t_{0}^\mu } g^{\mu,\nu } \frac{\partial^3 F_0}{\partial t_{0}^\nu \, \partial t_{l}^\beta \, \partial t_{m}^\gamma} .
\eeq

\end{itemize}

The Lagrangians ${\cal L}_{F_0}$ defined by such functions have a very nice characterisation.

\begin{theorem} \cite{Giventalcone}
\label{th.Givental.cone}
$F_0$ satisfies \eq{DE}, \eq{SE} and \eq{TRR} if and only if ${\cal L}_{F_0}$ is a Lagrangian cone with vertex at the origin and such that its tangent spaces $L$ satisfy $zL = L$. 
\end{theorem}
In addition, Givental described a large group of symmetries of the set of such cones. 
\begin{theorem}\cite{Giventalcone}
The twisted loop group $G_{{\rm tw}}$, consisting of elements $M(z) \in {\rm End}(V)[\![z^{-1}]\!]$ such that $M^*(-z) M(z) = {\rm Id}$ preserves the class of cones of Theorem \ref{th.Givental.cone}.
\end{theorem}
Note that the condition $M^*(-z) M(z) = {\rm Id}$ implies that $M(z)$ defines a symplectomorphism of ${\cal W}$. The set of tangent spaces to such a cone carries the structure of a Frobenius manifold $\mathcal{M}$. For instance, this applies to (and was motivated by the application to) the genus $0$ descendent or ancestor potentials of Gromov--Witten theory of a complex projective variety \cite{CoatesGivental}, to genus $0$ correlation functions of cohomological field theories \cite{KMCohFT} and to quantum $K$-theory \cite{CoatesGivental}.

Conversely, if $\mathcal{M}$ is a semi-simple Frobenius manifold, there is a notion of a descendent (resp. an ancestor) potential $F_0^{\mathcal{M}}(\mathbf{t})$, which satisfy the axioms above. Here $(t^{\nu})_{\nu}$ are local flat coordinates on $\mathcal{M}$, and we fix a point $m \in \mathcal{M}$ to identify $V := T_{m}\mathcal{M}$. We denote as before $(t^{\nu}_{k})_{k \geq 0}$ the linear coordinates on $V[\![z]\!]$. Dubrovin --- see e.g. \cite{Dubro0} --- associates to $\mathcal{M}$ a Riemann--Hilbert problem on $\mathbb{P}^1$. Its solution is an element $M_{v}(z)$ of the associated loop space depending on a point $m$ of the Frobenius manifold\footnote{This element and its factorisation are unique if the Frobenius manifold admits an Euler vector field. Otherwise, one needs to fix the diagonal ambiguity by some other geometric condition.}, and it admits a Birkhoff factorisation
$$
M_{v}(z) =  M_{v,\infty}(z)^{-1} M_{v,0}(z),
$$
where $M_{v,0}(z)$ (resp. $M_{v,\infty}(z)$) is analytic and invertible for $|z|<1$ (resp. $1<|z|\leq\infty$). Combining Givental's analysis of the action of the twisted loop group  \cite{Giventalcone} and Teleman's classification of semi-simple Frobenius manifolds \cite{Teleman}, one can conclude that the cone defined by the graph of the genus zero descendent (resp. ancestor) potential of a ${\rm dim}\,V = N$ semi-simple Frobenius manifold is obtained by the action of the symplectomorphism $\gamma(v) M_{v}(z)$ (resp. $M_{v,0}(z)$) on the cone ${\cal L}_N$ corresponding to the trivial theory of type $A_1^{\times N}$ where $\gamma(v)$ is a suitably chosen normalisation factor.

Finally, Givental reconstruction procedure proved by Teleman through its classification can be expressed as the following quantisation result.

\begin{theorem}\cite{Giventalcone,Teleman}
If ${\rm dim}\,V = N$, then the descendant (resp. ancestor) potential of a semi-simple Frobenius manifold is obtained by quantising the cone obtained by the action of the symplectomorphism $\gamma(v) M_{v}(z)$ (resp. $M_{v,0}(z)$) on the cone ${\cal L}_N$ corresponding to the trivial theory of type $A_1^{\times N}$ where $\gamma(v)$ is a suitably chosen normalisation factor.
\end{theorem}

In order to be more explicit, let us describe ${\cal L}_N$. Let $F_0^{{\rm KdV}}(\mathbf{t})$ be the genus 0 potential of the Gromov--Witten theory of a point, \textit{i.e.} the genus 0 part of the logarithm of the partition function of the quantum Airy structure of Proposition~\ref{ThmLoopZ2TQFT} with all $v_{(k,i),(l,j)}$ vanishing and $\theta(z) = z^{-2}\cdot 1_{V}$. Then, after the identification by the dilaton shift $q_{k,i} = t_{k}^i-\delta_{k,1}$, one has indeed
$$
{\cal L}_N:= \Big\{(p,q) \in T^*{\cal V}\,\,:\,\,\quad p = \sum_{i=1}^N \dd_{q}F_0^{{\rm KdV}}({\bf t}^{i})\Big\}.
$$
The full partition function of this quantum Airy structure is the matrix Airy function of \cite{Konts}.

Because $M_{v,0}(z)$ is analytic for $|z|>1$, the quantum structure corresponding to the ancestor's Lagrangian cone is obtained by the action of the operator $\exp\big(\tfrac{\hbar}{2}\sum_{k,\ell,i,j} u_{(k,i),(\ell,j)} \partial_{(k,i)} \partial_{(\ell,j)} \big)$ on the quantum Airy structure built from ${\cal L}_N$ where
$$ 
\frac{M_{v,0}^*(z_1) M_{v,0}(z_2) - {\rm Id}}{z_1+z_2} := \sum_{k,\ell\geq 0} (-1)^{k+\ell} U_{k,l} z_1^k z_2^{\ell},\qquad U_{k,\ell} e_i := \sum_{j=1}^N u_{(k,i),(\ell,j)} e_j.
$$ 
This transformation preserves Airy structures and the topological recursion gives a way to compute the ancestor potential. Further, the local spectral curve is fixed by the symplectomorphism $M_{v,0}(z)$.

The action of the Givental's twisted loop group on Lagragian cones is easily seen to coincide with the action of symplectomorphisms on Airy structures defined on ${\cal W}$. This leads to the equivalence of Givental quantisation procedure and quantisation of the corresponding Airy structure.

\begin{corollary}
The ancestor potential of a semi-simple Frobenius manifold is obtained from the quantum Airy structure defined on the loop space ${\cal W}$ from the action of the symplectomorphism $M_{v,0}(z)$ on the trivial quantum Airy structure whose partition function is the matrix Airy function.

This quantisation procedure is equivalent to Givental's quantisation of the corresponding Lagrangian cone.

\end{corollary}

On the other hand, the action of the quantisation of $M_{v,\infty}(z)^{-1}$ taking the ancestor potential to the descendant potential is just the multiplication of the partition function by $\exp\big( \tfrac{\hbar}{2} \sum_{k,\ell,i,j} s_{(k,i),(\ell,j)} x_{(k,i)} x_{(\ell,j)}\big)$ where
$$ 
\frac{M_{v,\infty}^*(z_1) M_{v,\infty}(z_2) - {\rm Id}}{1/ z_1+1/z_2} := \sum_{k,\ell\geq 0}  S_{k,\ell} z_1^{-k} z_2^{-\ell},\qquad S_{k,\ell} e_i := \sum_{j=1}^N s_{(k,i),(\ell,j)} e_j.
$$
Even though it is a simple transformation, let us remark that this does not preserve the Airy structure property, since it adds a linear term in the $x$s. This explains why the topological recursion does not directly provide the descendant potential, except in the cases where $M_{v,\infty}$ is trivial.

\section{A topological recursion without branched covers}
\label{SWithout}
In this section, we explain the simple observation that the quantum Airy structures of Proposition~\ref{ThmLoopTQFT} can be realised by a new variant of the topological recursion of \cite{EOFg}. We take as initial data
\begin{itemize}
\item[$\bullet$] a Riemann surface $\Sigma$.
\item[$\bullet$] a meromorphic $1$-form $\omega_{0,1}$ on $\Sigma$.
\item[$\bullet$] a fundamental bidifferential of the second kind $\omega_{0,2}$ on $\Sigma$.
\item[$\bullet$] a finite subset $\mathfrak{r} \subset \Sigma$, such that $\omega_{0,1}$ has at most simple zeroes at $\mathfrak{r}$ --- this allows poles of $\omega_{0,1}$ at $\mathfrak{r}$.
\item[$\bullet$] a meromorphic $1$-form $\omega_{1,1}$ on $\Sigma$, such that, for any $r \in \mathfrak{r}$, $(x(p) - x(r))^2\frac{\omega_{1,1}(p)}{\omega_{0,1}(p)}$ is holomorphic around $p \rightarrow r$. 
\end{itemize}
We define a recursion kernel
$$
\tilde{K}(p_0,p) = \frac{\int_{r}^{p} \omega_{0,2}(\cdot,p_0)}{\omega_{0,1}(p)},
$$
and for $2g - 2 + n > 0$ and $(g,n) \neq (1,1)$, we make the inductive definition, with $I = \{p_2,\ldots,p_n\}$:
\beq
\label{TRsans} \omega_{g,n}(p_1,I) = \sum_{r \in \mathfrak{r}} \Res_{p  = r} \tilde{K}(p_0,p)\bigg\{\omega_{g - 1,n + 1}(p,p,I) + \sum_{\substack{g' + g'' = g \\ J' \sqcup J'' = I}}^* \omega_{g',1+|J'|}(p,J') \omega_{g'',1 + |J''|}(p,J'')\bigg\}.
\eeq 
We have to include $\omega_{1,1}$ in the initial data since $\omega_{0,2}(p,p)$, which would appear in \eqref{TRsans} for $(g,n) = (1,1)$, does not make sense due to the double pole at coinciding point. As in Section~\ref{comparTR}, one can prove that $\omega_{g,n}$ decomposes on a basis of $1$-forms.
\begin{definition}
For $k \geq 0$ and $r \in \mathfrak{r}$, we define for $p_0 \in \Sigma$ the meromorphic $1$-form
$$
\xi_{k,r}(p_0) := \Big(\Res_{p = r} \int_{r}^{p_0} \omega_{0,2}(\cdot,p_0)\Big)\frac{(k + 1)\dd x(p)}{(x(p) - x(r))^{k + 1}}.
$$
We also define, for $p_0$ in a neighbourhood of $r$ in $\Sigma$
$$
\xi_{k,r}^*(p_0) := \frac{(x(p) - x(r))^{k + 1}}{k + 1},\qquad \theta(p) := \frac{1}{\omega_{0,1}(p)},
$$
and if $p_0$ is in a neighbourhood of $r_0 \neq r$, we define $\xi_{k,r}^*(p_0) := 0$.
\end{definition}

\begin{lemma}
\label{PPP0}For $2g - 2 + n > 0$, there exists a unique decomposition with finitely many non-zero terms
\beq
\label{Fogisans}\omega_{g,n}(p_1,\ldots,p_n) = \sum_{\substack{r_1,\ldots,r_n \in \mathfrak{r} \\ k_1,\ldots,k_n \geq 0}} W_{g,n}\left[\begin{smallmatrix} r_1 & \cdots & r_n \\ k_1 & \cdots & k_n \end{smallmatrix}\right]\,\prod_{i = 1}^n \xi_{k_i,r_i}(p_i).
\eeq \hfill $\Box$
\end{lemma}
The assumption on $\omega_{0,1}$ guarantees that $\theta(p)$ for $p \rightarrow r$ has an expansion of the form
$$
\theta(p) = \sum_{m \geq -1} t_{m,r}z^{m}(p)\,(\dd x(p))^{-1}.
$$
According to Proposition~\ref{ThmLoopTQFT}, we have a quantum Airy structure given by
\bea
A^{(k_1,r_1)}_{(k_2,r_2),(k_3,r_3)} & := & 0, \nonumber \\
B^{(k_1,r_1)}_{(k_2,r_2),(k_3,r_3)} & := & \Res_{p = r_1} \xi_{k_1}^*(p)\,\dd \xi_{k_2,r_2}^*(p)\,\xi_{k_3,r_3}(p)\,\theta(p), \nonumber \\
C^{(k_1,r_1)}_{(k_2,r_2),(k_3,r_3)} & := & \Res_{p = r_1} \xi_{k_1}^*(p)\,\xi_{k_2,r_2}(p)\,\xi_{k_3,r_3}(p)\,\theta(p), \nonumber \\
\label{KSsf} D^{(k_1,r_1)} & := & W_{1,1}\left[\begin{smallmatrix} r_1 \\ k_1 \end{smallmatrix}\right],
\eea
and by comparison of KS-TR and the recursive relation for $W_{g,n}$s ensuing from \eqref{TRsans} we obtain that
\begin{proposition}
\label{PPP1}For $2g - 2 + n > 0$, $F_{g,n}\big((k_1,r_1),\ldots,(k_n,r_n)\big)$ computed by KS topological recursion for the quantum Airy structure \eqref{KSsf}, and $W_{g,n}\left[\begin{smallmatrix} r_1 & \cdots & r_n \\ k_1 & \cdots & k_n \end{smallmatrix}\right]$, agree. \hfill $\Box$
\end{proposition}

We omit the proof of Lemma~\ref{PPP0} and Proposition~\ref{PPP1}, as it is similar to Lemma~\ref{LemmaFogi} and Proposition~\ref{FWid}, in fact simpler due to the absence of the involution. Note that the assumption made on $\omega_{1,1}$ is equivalent to the \textbf{D} relation.

\section{Dynamics on (coloured) Young diagrams}\label{s:TR}
\label{SYoung}

\def\Col#1{
\begin{pspicture}(0.2,1)
\multiput(0,0)(0,0.2){#1}{\psframe(0,0)(0.2,0.2)}
\end{pspicture}
}

\def\ColOne#1{
\begin{pspicture}(0.2,1)
\multiput(0,0)(0,0.2){#1}{\psframe(0,0)(0.2,0.2)}
\rput(0,-0.2){\makebox(0,0)[lb]{\tiny$1$}}
\end{pspicture}
}

\def\ColP#1{
\begin{pspicture}(0.2,1)
\multiput(0,0)(0,0.2){#1}{\psframe(0,0)(0.2,0.2)}
\rput(0,-0.2){\makebox(0,0)[lb]{\tiny $p$}}
\end{pspicture}
}

\def\Colg#1{
\begin{pspicture}(0.2,1)
\multiput(0,0)(0,0.2){#1}{\psframe[fillstyle=solid,fillcolor=lightgray](0,0)(0.2,0.2)}
\end{pspicture}
}

\def\ColgOne#1{
\begin{pspicture}(0.2,1)
\multiput(0,0)(0,0.2){#1}{\psframe[fillstyle=solid,fillcolor=lightgray](0,0)(0.2,0.2)}
\rput(0,-0.2){\makebox(0,0)[lb]{\tiny$1$}}
\end{pspicture}
}

\def\ColgP#1{
\begin{pspicture}(0.2,1)
\multiput(0,0)(0,0.2){#1}{\psframe[fillstyle=solid,fillcolor=lightgray](0,0)(0.2,0.2)}
\rput(0,-0.2){\makebox(0,0)[lb]{\tiny $p$}}
\end{pspicture}
}

\def\Cols#1{
\begin{pspicture}(0.2,0.6)
\multiput(0,0)(0,0.2){#1}{\psframe(0,0)(0.2,0.2)}
\end{pspicture}
}

\def\ColsOne#1{
\begin{pspicture}(0.2,0.6)
\multiput(0,0)(0,0.2){#1}{\psframe(0,0)(0.2,0.2)}
\rput(0,-0.2){\makebox(0,0)[lb]{\tiny$1$}}
\end{pspicture}
}

\def\ColsP#1{
\begin{pspicture}(0.2,0.6)
\multiput(0,0)(0,0.2){#1}{\psframe(0,0)(0.2,0.2)}
\rput(0,-0.2){\makebox(0,0)[lb]{\tiny $p$}}
\end{pspicture}
}

\def\Colsg#1{
\begin{pspicture}(0.2,0.6)
\multiput(0,0)(0,0.2){#1}{\psframe[fillstyle=solid,fillcolor=lightgray](0,0)(0.2,0.2)}
\end{pspicture}
}

\def\ColsgOne#1{
\begin{pspicture}(0.2,0.6)
\multiput(0,0)(0,0.2){#1}{\psframe[fillstyle=solid,fillcolor=lightgray](0,0)(0.2,0.2)}
\rput(0,-0.2){\makebox(0,0)[lb]{\tiny$1$}}
\end{pspicture}
}

\def\ColsgP#1{
\begin{pspicture}(0.2,0.6)
\multiput(0,0)(0,0.2){#1}{\psframe[fillstyle=solid,fillcolor=lightgray](0,0)(0.2,0.2)}
\rput(0,-0.2){\makebox(0,0)[lb]{\tiny $p$}}
\end{pspicture}
}

\subsection{Setting}

In this section we show that a quantum Airy structure on $V = \mathbb{C}[\![z]\!]$ (or $V = \mathbb{C}^{d}[\![z]\!]$, or $z\mathbb{C}^{d}[\![z^2]\!]$, etc.) gives a recursion on (coloured) Young diagrams, which are in correspondence with the monomials that can appear in the Taylor expansion of the partition function. We first formulate abstractly the recursion on Young diagrams, and relate it to quantum Airy structure in Proposition~\ref{PYoung} below. Please see also \cite{ACNP}, where this dynamics on Young diagrams in some special cases was given.

Let $d$ be a positive integer. For a Young diagram $\lambda = (\lambda_1 \geq \cdots \geq \lambda_{\ell(\lambda)})$, we consider $\lambda_i$ as column heights. We denote ${\rm Col}(\lambda)$ the set of columns, and  $|\lambda|$ be the number of boxes.
\begin{definition}
A \emph{$d$-colouring} of a Young diagram $\lambda$ is a map ${\rm Col}(\lambda) \rightarrow \{1,\ldots,d\}$. A \emph{column type} is an ordered pair $(k,\alpha) \in \mathbb{Z}_{> 0} \times \{1,\ldots,d\}$, where $k$ is the height and $\alpha$ is the colour. We denote ${\rm Aut}\,\lambda$ the group of permutations of columns respecting column types.
\end{definition}

We use the notation $N_{k,\alpha}(\lambda)$ for the number of columns of type $(k,\alpha)$, hence
$$
|\lambda| = \sum_{k \geq 1} \sum_{\alpha = 1}^d N_{k,\alpha}(\lambda) k,\qquad |{\rm Aut}\,\lambda| = \prod_{\substack{k \geq 1 \\ 1 \leq \alpha \leq d}} N_{k,\alpha}(\lambda)!.
$$

\begin{definition}
Let $\mathcal{Y}_{g,n}^{(d)}$ be the set of $d$-coloured Young diagrams $\lambda$ such that
\beq
\label{boundsY}\ell(\lambda) = n \qquad {\rm and}\qquad |\lambda| \leq (2g - 2 + n)r,
\eeq
and $\tilde{\mathcal{Y}}_{g,n}^{(d)}$ the set of such $d$-coloured Young diagrams together with the choice of a column type --- remembered by a label ``1".
\end{definition}
We denote $s\,:\,\tilde{\mathcal{Y}}_{g,n}^{(d)} \rightarrow \mathcal{Y}_{g,n}^{(d)}$ the map which forgets the label ``1''. We have an injective linear map
$$
\begin{array}{rcccc}
S & : & {\mathbb C}[\mathcal{Y}_{g,n}^{(d)}] & \longrightarrow & {\mathbb C}[\tilde{\mathcal{Y}}_{g,n}^{(d)}] \\
 & & \lambda & \longmapsto & \sum_{\tilde{\lambda} \in s^{-1}(\lambda)} \tilde{\lambda},
\end{array}
$$
that is, each Young diagram is mapped to the sum (linear combination with unit coefficients) of the same Young diagrams differing only by placing the label ``1'' on all types of columns present in this diagram. We now define two unary operations on diagrams from $\mathcal{Y}_{g,n}^{(d)}$ which results in two linear maps
$$
\Delta_{{\rm B}}\,:\, {\mathbb C}[\mathcal{Y}_{g,n}^{(d)}] \longrightarrow {\mathbb C}[\tilde{\mathcal{Y}}_{g,n+1}^{(d)}],\qquad \Delta_{{\rm C}}\,:\,{\mathbb C}[\mathcal{Y}_{g-1,n+1}^{(d)}] \longrightarrow {\mathbb C}[\tilde{\mathcal{Y}}_{g,n}^{(d)}],
$$
and a binary operation on ordered pairs of coloured Young diagrams, which results in the bilinear map
$$ \Delta_{{\rm C}}^{(2)}\,:\, {\mathbb C}[\mathcal{Y}_{g_1,n_1}^{(d)}] \otimes {\mathbb C}[\mathcal{Y}_{g_2,n_2}^{(d)}]  \longrightarrow {\mathbb C}[\tilde{\mathcal{Y}}_{g_1+g_2,n_1+n_2-1}^{(d)}].$$

The data for our recursion will be either finite or semi-infinite complex tensors $B = (B^{(k_1,\alpha_1)}_{(k_2,\alpha_2),(k_3,\alpha_3)})$ and $C = (C^{(k_1,\alpha_1)}_{(k_2,\alpha_2),(k_3,\alpha_3)})$, where $(k_i,\alpha_i) \in \mathbb{Z}_{> 0} \times \{1,\ldots,d\}$ characterise the possible column types. We assume  in the semi-infinite case that the entries of $B$ vanish whenever $k_1 + k_2 > k_3 + r$, and the entries of $C$ vanish whenever $k_1 > k_2 + k_3 + r$. This guarantees that all sums appearing below are finite. The bound \eqref{boundsY} on the number of boxes of our Young diagrams are tailored to this property of $B$ and $C$.

\subsection{The operations}

The first unary operation $\Delta_{{\rm B}}\,:\,{\mathbb C}[\mathcal{Y}_{g,n}^{(d)}]\to {\mathbb C}[\tilde{\mathcal{Y}}_{g,n + 1}^{(d)}]$
is defined by the following rule. It is a sum over $\alpha_3 \in \{1,\ldots,d\}$ followed by a sum over all possible column types  in $\lambda$ of colour $\alpha_3$. The terms of this sum are obtained by replacing a column of the selected type with two new columns of colours $\alpha_1$ and $\alpha_2$
\beq
\begin{pspicture}(0.2,0)
\rput(0.1,0.4){$\Delta_{{\rm B}}\biggl(\,\,\,\,\,\,\,\,\,\,$}
\end{pspicture}
\Col4
\begin{pspicture}(0.5,0)
\rput(-.1,0.9){\makebox(0,0)[cb]{\tiny $\alpha_3$}}
\put(0,0){\psline{<->}(0.1,0)(0.1,0.8)}
\rput(0.2,0.4){\makebox(0,0)[lc]{$k_3$}}
\end{pspicture}
\begin{pspicture}(0.5,0)
\rput(0.1,0.4){\makebox(0,0)[lc]{$\biggr)$}}
\end{pspicture}
=\sum_{\substack{k_1,k_2 \geq 1\\ k_1+k_2\le k_3+r}} \sum_{\alpha_1,\alpha_2 = 1}^{d} B_{(k_2,\alpha_2),(k_3,\alpha_3)}^{(k_1,\alpha_1)} N_{k_2,\alpha_2}(\lambda)
\begin{pspicture}(1,1)
\put(0,0){\psline{<->}(0.9,0)(0.9,0.4)}
\rput(0.8,0.2){\makebox(0,0)[rc]{$k_1$}}
\rput(1.1,0.5){\makebox(0,0)[cb]{\tiny $\alpha_1$}}
\end{pspicture}
\ColOne2\,\,
\Col3
\begin{pspicture}(1,1)
\put(0,0){\psline{<->}(0.1,0)(0.1,0.6)}
\rput(0.2,0.3){\makebox(0,0)[lc]{$k_2$}}
\rput(-.1,0.65){\makebox(0,0)[cb]{\tiny $\alpha_2$}}
\end{pspicture}.
\label{rule1}
\eeq
In this operation, we place a label ``1'' on the column type $(k_1,\alpha_1)$ in the resulting Young diagram.

For the second unary operation $\Delta_{{\rm C}}\,:\,{\mathbb C}[\mathcal{Y}_{g-1,n+1} ]\to {\mathbb C}[\tilde{\mathcal{Y}}_{g,n}]$, we proceed analogously, except now we sum over ordered pairs $((k_2,\alpha_2),(k_3,\alpha_3))$ of column types in $\lambda$. The terms are obtained by replacing the ordered pair of column of each type
\be
\begin{pspicture}(0.2,1)
\rput(0.1,0.1){$\Delta_{{\rm C}}\biggl(\,\,\,\,\,\,\,\,\,\,\,$}
\end{pspicture}
\begin{pspicture}(0.5,1)
\put(0,0){\psline{<->}(0.4,0)(0.4,0.4)}
\rput(0.3,0.2){\makebox(0,0)[rc]{\tiny$k_2$}}
\end{pspicture}
\underbracket[1pt]{\Col2\,\,\Col3}
\begin{pspicture}(0.5,1)
\put(0,0){\psline{<->}(0.1,0)(0.1,0.6)}
\rput(-.1,0.7){\makebox(0,0)[cb]{\tiny $\alpha_3$}}
\rput(-.45,0.7){\makebox(0,0)[cb]{\tiny $\alpha_2$}}
\rput(0.2,0.3){\makebox(0,0)[lc]{\tiny$k_3$}}
\end{pspicture}
\rput(0.1,0.1){\makebox(0,0)[lc]{$\biggr)$}}
\quad =\sum_{1 \leq k_1 \leq k_2 + k_3 + r} \sum_{\alpha_1 = 1}^d \tfrac{1}{2}\,C_{(k_2,\alpha_2),(k_3,\alpha_3)}^{(k_1,\alpha_1)}
\ColOne6
\begin{pspicture}(1.4,1)
\rput(-.1,1.3){\makebox(0,0)[cb]{\tiny $\alpha_1$}}
\put(0,0){\psline{<->}(0.1,0)(0.1,1.2)}
\rput(0.2,0.6){\makebox(0,0)[lc]{$k_1$}}
\end{pspicture},
\label{PP1}
\ee
with a new column of colour $\alpha_1$.

The binary operation $ \Delta_{{\rm C}}^{(2)}\,: \,{\mathbb C}[\mathcal{Y}_{g_1,n_1}] \times {\mathbb C}[\mathcal{Y}_{g_2,n_2}]  \ra{\mathbb C}[\tilde{\mathcal{Y}}_{g_1+g_2,n_1+n_2-1}]$  is obtained in a similar way to $\Delta_{{\rm C}}$, but fusing the two Young diagrams. More precisely, $ \Delta_{{\rm C}}^{(2)}(\lambda , \lambda')$ is the sum over all column types $(k_2,\alpha_2)$ in $\lambda$ and column types $(k_3,\alpha_3)$ in $\lambda'$, of the following contribution. We erase one column of the selected type in $\lambda$ and $\lambda'$, fuse the two Young diagrams, and insert a column of type $(k_1,\alpha_1)$ with a label ``1'' and a weight
$$
\tfrac{1}{2}C^{(k_1,\alpha_1)}_{(k_2,\alpha_2),(k_3,\alpha_3)}.
$$
These terms are then summed over $(k_1,\alpha_1) \in \mathbb{Z}_{> 0} \times \{1,\ldots,d\}$ such that $k_1 \leq k_2 + k_3 + r$ to give $\Delta_{{\rm C}}^{(2)}(\lambda , \lambda')$.

\subsection{Evaluation and relation to quantum Airy structures}

Let $\mathcal{S}^{(d)}$ be the $\mathbb{C}$-algebra of symmetric functions in infinite number of variables $x_{\alpha,i}$, $i \in \mathbb{Z}_{\geq 0} $ and $\alpha \in \{1,\ldots,d\}$. The power sums $p_{k,\alpha} := \sum_{i \geq 0} x_{\alpha,i}^{k}$ give a linear basis for $\mathcal{S}^{(d)}$. If $\lambda$ is a Young diagram with a $d$-colouring we denote
$$
P_{\lambda} := \prod_{\substack{k \geq 1 \\ 1 \leq \alpha \leq d}} p_{k,\alpha}^{N_{k,\alpha}(\lambda)}\,.
$$

We define the linear evaluation map
$$
\begin{array}{rcccc} {\rm ev} & \,:\, & \mathbb{C}[\mathcal{Y}_{g,n}^{(d)}] &  \longrightarrow  & \mathcal{S}^{(d)} \\ && \lambda & \longmapsto & |{\rm Aut}\,\lambda|^{-1}\,P_{\lambda} \end{array}\,.
$$
The map ${\rm ev}$ is obviously injective. Let $\underline{k_1,\alpha_1|\,\cdots\,|k_{\ell},\alpha_{\ell}}$ be the Young diagram with columns of height $k_i$ and colour $\alpha_i$. We remark that $\mathcal{S}^{(d)}$ is isomorphic --- \textit{via} Taylor expansions --- to the algebra of polynomial functions on
\beq
\label{Vyoung} V:= z\mathbb{C}^{d}[\![z]\!] \cong \bigoplus_{\substack{k \geq 1 \\ 1 \leq \alpha \leq d}} \mathbb{C}.\underline{k,\alpha}.
\eeq

\begin{proposition}
\label{PYoung} Assume that $(A,B,C,D)$ defines a quantum Airy structure on $V$ given by \eqref{Vyoung}, and assume that $A^{(k_1,\alpha_1)}_{(k_2,\alpha_2),(k_3,\alpha_3)} = 0$ whenever $k_1 + k_2 + k_3 > r$, and $D^{(k,\alpha)}$ vanishes whenever $k > r$. Then there exists unique $\Omega_{g,n} \in {\mathbb C}[Y_{g,n}]$ indexed by $g \geq 0$ and $n \geq 1$ satisfying $2g-2+n>0$, such that
\bea
\label{imfsg}\Omega_{0,3} & = & \sum_{(k_i,\alpha_i)_{i = 1}^3} A^{(k_1,\alpha_1)}_{(k_2,\alpha_2),(k_3,\alpha_3)}\,\,\underline{k_1,\alpha_1|k_2,\alpha_2|k_3,\alpha_3}\,,\label{A3} \\
\Omega_{1,1} & = & \sum_{k,\alpha} D^{(k,\alpha)}\,\,\underline{k,\alpha}\,, \label{EPS}
\eea
and for $2g + n \geq 0$
\beq
\label{imfsg2} S(\Omega_{g,n}) = \Delta_{{\rm B}}(\Omega_{g,n-1}) + \Delta_{{\rm C}}(\Omega_{g-1,n+1}) + \sum_{\substack{g_1+g_2 = g\\ n_1+n_2 = n}} \Delta_{{\rm C}}^{(2)}(\Omega_{g_1,n_1} ,\Omega_{g_2,n_2}).
\eeq
Moreover the coefficients of the partition function of the quantum Airy structure are $F_{g,n} = {\rm ev}(\Omega_{g,n})$.

Vice versa, if the dynamics of Young diagrams is governed by relations \eqref{rule1}, \eqref{PP1} endowed with the initial conditions
\eqref{A3} and \eqref{EPS} and we require the result of this action
to have the form $S(\Omega_{g,n})$ for some $\Omega_{g,n}\in {\mathbb C}[Y_{g,n}]$ for all $g$ and $n$, \textit{i.e.}, this result must belong to the image of the mapping $S$ for all $g$ and $n$, then the partition function of these correlation functions exists and is annihilated by the $L_i$ given by \eqref{Lform}.
\end{proposition}

Let us comment on this formalism. The operations $\Delta_{{\rm B}},\Delta_{{\rm C}}$ and $\Delta_{{\rm C}}^{(2)}$ introduce some asymmetry in the treatment of the column types, tracked by the label ``1''. The linear map $S$ discards this label by summing over all underlying coloured Young diagrams. For given $(A,B,C,D)$, we would like to define a dynamic on (coloured) Young diagrams by the formulae \eqref{imfsg}--\eqref{imfsg2} --- note that $B$ and $C$ enter in the definition of $\Delta_{{\rm B}}$, $\Delta_{{\rm C}}$ and $\Delta_{{\rm C}}^{(2)}$. However, at each step the right-hand side of \eqref{imfsg2} is an expression in terms of Young diagrams with a label ``1''. As $S$ is an injection, there is at most one expression in coloured Young diagrams $\Omega_{g,n}$ satisfying \eqref{imfsg2}. Such a $\Omega_{g,n}$ does exist if and only if the right-hand side of \eqref{imfsg2} produces a function on labelled coloured Young diagrams which lies in the image of the linear map $S$ --- \textit{i.e.} it gives a symmetric function on $V$ when evaluated. This is true only if the quadruple $(A,B,C,D)$ satisfies some conditions. The first part of Proposition~\ref{Explicitsym} shows that a sufficient condition for the right-hand side to be in the image of $S$ is that $(A,B,C,D)$ defines a quantum Airy structure. In this case, \eqref{imfsg}--\eqref{imfsg2} just mimicks, at the level of functions on Young diagrams, the recursive computation of the partition function of the quantum Airy structure. The proof is straightforward and thus omitted. 
Let us prove the inverse statement of Proposition \ref{PYoung}.

\noindent {\bf Proof}.
For simplicity, we replace the multi-index $(k,\alpha)$ merely by $k$. The consideration below is general and does not depend on
details of the model. It is also insensitive to whether we are in a  finite or infinite-dimensional situation.

We first identify the coefficients of $F_{g,n}$ with the Taylor coefficients at $\bs{\xi} = 0$ of a function $S_{g}(\bs{\xi})$
$$
F_{g,n}(k_1,\dots,k_n)=\left.\frac{\partial^n S_{g}}{\partial \xi_{k_1}\partial \xi_{k_2}\cdots\, \partial \xi_{k_n}}\right|_{\xi_i=0}.
$$ 
It is convenient to interpret $\Omega_{g,n}$ as symmetric differentials
$$
\Omega_{g,n}:=\sum_{\{k\}}F_{g,n}(k_1,\dots,k_n)\,\dd\xi_{k_1}\cdots\,\dd\xi_{k_n}.
$$

We introduce an auxiliary object
$$
\overline{\Omega}:=\sum_{2g - 2 + n >0}\hbar^{2g - 2 + n} \sum_{\{k_i\}_{i=1}^n} \frac{\partial^n S_g}{\partial \xi_{k_1}\cdots\, \partial \xi_{k_n}}
\dd\xi_{k_1}\cdots\,\dd\xi_{k_n},
$$
where \emph{we do not impose the constraint} $\xi_i=0$. We segregate the term proportional to $\dd\xi_s \dd\xi_{k_1}\cdots\,\dd\xi_{k_n}$ without a priori symmetrisation with respect to the index $s$. The condition that the right-hand side is actually fully symmetric with respect to permutations of all indices including $s$ implies that it must be of the form
$$
\dd\overline{\Omega}=\sum_{s} \sum_{2g - 2 + n >0}\hbar^{2g - 2 + n} \sum_{\{k_i\}_{i=1}^n} \frac{\partial^{n+1} S_{g}}{\partial \xi_s\partial \xi_{k_1}\cdots \,\partial \xi_{k_n}}\dd\xi_s \dd\xi_{k_1}\cdots\,\dd\xi_{k_n},
$$
which is fully symmetric by construction.

In the right-hand side we have several terms. Let us segregate the coefficients of $\dd\xi_s \dd\xi_{k_1}\cdots\,\dd\xi_{k_n}$. Our strategy is to push the whole collection of partial derivatives $\frac {\partial^n} {\partial \xi_{k_1}\cdots\, \partial \xi_{k_n}}$ outside the action of the other operators. At the end of the calculations we set all $\xi_i=0$, thus obtaining the original TR relations.
\begin{itemize}
\item [(1)] In the term corresponding to $C^s_{q,p}$ we remove two differentials $\dd\xi_{q}$ and $\dd\xi_{p}$ and replace them by $\dd\xi_s$. The corresponding term proportional to $\dd\xi_s \dd\xi_{k_1}\cdots\,\dd\xi_{k_n}$ reads
\begin{align*}
&C^s_{q,p}\sum_{\substack{g_1+g_2=g \\ I\sqcup J=\{1,\dots,n\}}}\frac{\partial}{\partial \xi_{q}}\frac{\partial^{|I|}S_{g_1}}{\partial \xi_{k_{\alpha_1}}\cdots \,\partial \xi_{k_{\alpha_{|I|}}}} \frac{\partial}{\partial \xi_{p}}\frac{\partial^{|J|}S_{g_2}}{\partial \xi_{k_{\beta_1}}\cdots\, \partial \xi_{k_{\beta_{|J|}}}} +
\frac{\partial^2}{\partial \xi_{q}\partial \xi_{p}}\frac{\partial^{n}S_{g}}{\partial \xi_{k_1}\cdots\, \partial \xi_{k_n}}
\cr
&
=\frac {\partial^n} {\partial \xi_{k_1}\cdots\, \partial \xi_{k_n}}\left[ C^s_{q,p}\sum_{g_1+g_2=g} \frac{\partial S_{g_1}}{\partial \xi_{q}}
\frac{\partial S_{g_2}}{\partial \xi_{p}} + C^s_{q,p} \frac{\partial^2 S_{g-1}}{\partial \xi_{q}\partial \xi_{p}} \right].
\end{align*}
\item[(2)] In the term corresponding to $B^s_{q,p}$ we erase $\dd\xi_{p}$ and add $\dd\xi_s$ and $\dd\xi_{q}$ and multiply the result by $N_{q}$ --- the number of times the index $q$ appears in the set $\{k_1,\dots, k_n\}$. For the action of this operation not to vanish, the index $q$ has to be found at least once in the set
$k_1,\dots,k_n$, say, $q=k_a$ and the corresponding coefficient is $\tfrac{\partial^{n-1}S_{g}}{\partial \xi_{k_1}\cdots\, \widehat{\partial \xi_{k_a}}\cdots\, \partial \xi_{k_n}}$. Here, the hat denotes the omission of the corresponding term. In order to collect the set of partial derivatives with respect to all $\xi_{k_i}$, $i=1,\ldots, n$ we use the following trick. We write
\beq
N_{k_a}\frac{\partial^{n}S_{g}}{\partial \xi_{p} \partial \xi_{k_1}\cdots\, \widehat{\partial \xi_{k_a}}\cdots\, \partial \xi_{k_n}}
=\frac {\partial^n} {\partial \xi_{k_1}\cdots\, \partial \xi_{k_n}}\left[\xi_{k_a}\frac {\partial S_{g}}{\partial \xi_{p}}\right] + N_{k_a} \xi_{k_a}
\frac{\partial^{n+1}S_{g}}{\partial \xi_{p} \partial \xi_{k_1}\cdots\, \partial \xi_{k_n}},
\label{TR-B1}
\eeq
where $N_{k_a}=N_q$ is exactly the proper coefficient appearing in the
TR relations (\ref{rule1}). The second term in the right-hand side of (\ref{TR-B1}) vanishes, when we impose the condition $\xi_i=0$ at the end of calculations. The remaining term reads
\beq
\frac {\partial^n} {\partial \xi_{k_1}\cdots \partial \xi_{k_n}}\left[ \sum_{q,p} B^s_{q,p}\xi_{q}\frac {\partial S_{g}}{\partial \xi_{p}}\right].
\label{TR-B2}
\eeq
\item[(3)] The last two terms describe the lowest order terms of TR, not determined by recursion formulae, namely $F_{0,3}$ and $F_{1,1}$. For the first term, we just use that
$$
\frac{\partial^3 S_{0}}{\partial \xi_s\partial \xi_{q}\partial \xi_{p}}= A^s_{q,p} \frac{\partial^2 }{\partial \xi_{q}\partial \xi_{p}}\bigl[ \xi_{q}\xi_{p}\bigr] +O(\xi_i), \ \hbox{where}\ A^s_{q,p} =\left. \frac{\partial^3 S_{0}}{\partial \xi_s\partial \xi_{q}\partial \xi_{p}}\right|_{\xi_i = 0}
$$
and for the second we get that
$$
\frac{\partial S_{1}}{\partial \xi_s}= D_s +O(\xi_s) \ \hbox{where}\  D_s=\left. \frac{\partial S_{1}}{\partial \xi_s}\right|_{\xi_s=0}.
$$
\end{itemize}
Combining all terms together, we obtain the following statement.

The coefficient of $\hbar^{2g+n-2} \dd\xi_s \dd\xi_{k_1}\cdots\,\dd\xi_{k_n}$ in the TR relations is given by the following expression
\begin{align}
&\frac {\partial^n} {\partial \xi_{k_1}\cdots\, \partial \xi_{k_n}}\left[\hbar^{-1} \frac{\partial S_{g}}{\partial \xi_s}-\sum_{q,p} B^s_{q,p}\xi_{q}\frac{\partial S_{g}}{\partial \xi_{p}}-\sum_{q,p} C^s_{q,p}\biggl( \sum_{g_1+g_2=g}\frac{\partial S_{g_1}}{\partial \xi_{q}}
\frac{\partial S_{g_2}}{\partial \xi_{p}}+\frac{\partial^2 S_{g - 1}}{\partial \xi_{q}\partial \xi_{p}}\biggr)\right.\nonumber\\
&\qquad \left.\left.
-\delta_{2g+n,3}\Bigl(\sum_{q,p}A^s_{q,p}\xi_{q}\xi_{p} + D_s\Bigr)
\right] \right|_{\xi_i=0}=0.
\label{XXX}
\end{align}

Because relation (\ref{XXX}) holds for all sets of external partial derivatives  $\frac {\partial^n} {\partial \xi_{k_1}\cdots \,\partial \xi_{k_n}}$ and the expression in square brackets depends neither on $n$ nor on the set $\{k_i\}_{i=1}^n$, whereas the quantities $S_{g}$ are defined to be formal power series in $\xi_i$, we conclude that this expression is identically zero for all $\xi_i$ in an open neighbourhood of the set of initial values $\xi_i=0$. Thus the set of TR relations is equivalent to the set of $(A,B,C,D)$-differential constraints $L_s$ imposed on the partition function ${\mathcal Z}$. \hfill $\Box$

This proposition applies in particular to the quantum Airy structure of Proposition~\ref{ThmLoopTQFT} and \ref{ThmLoopZ2TQFT}, and \textit{a fortiori} to the one underlying the topological recursion of \cite{EOFg} according to Proposition~\ref{FWid} and its new, branched cover-free version Proposition~\ref{PPP1}.

\section{A list of problems}
\label{SConclu}

By way of conclusion, we collect a few problems opened throughout tshe article --- a disjoint list of problems was put forward in \cite{KSTR}.

\begin{problem}
Complete the classification of finite-dimensional quantum Airy structures based on semi-simple Lie algebras.
\end{problem}
This is likely to be a case study of the candidate modules listed in Proposition~\ref{calsssls}. One may wonder in particular whether $\mathfrak{sl}_{2}(\mathbb{C})$ is the only simple Lie algebra supporting a quantum Airy structure, and if not, if the resulting classification has a geometric meaning. The case of simple Lie algebras has been solved in \cite{RHAiry} (see Remark~\ref{rem61}), but there seems to be much more flexibility in the semi-simple case.

\begin{problem}
What is the field theoretic meaning of the quantum Airy structure/partition function attached to a non-commutative Frobenius manifold?
\end{problem}
In the commutative case, we have found that the partition function computes the 2d TQFT amplitudes (Lemma~\ref{TQFTpart}). It is desirable to have a similar interpretation, maybe involving open-closed 2d TQFTs \cite{Lazaroiu,Mooreopen} --- these are indeed in correspondence with pairs of commutative and non-commutative Frobenius algebras together with some morphisms between them \cite{Nataclass}. Independently, one may wonder if the amplitudes of open-closed 2d TQFTs can be computed from quantum Airy structures.

We gave in Section~\ref{SS4} the basic definitions of moduli spaces of classical and quantum Airy structures. The geometry of these spaces is worth studying. In particular, the translations define commuting flows (although maybe not independent) on them. In this direction, one may wonder if those spaces carry integrable systems.
\begin{problem}
Study the algebraic geometry of the moduli spaces of classical and quantum Airy structures.
\end{problem}

The Lagrangian cones studied by Givental and Coates are quadratic Lagrangians in a symplectic space $T^*V[\![z]\!]$, therefore seem to be close to the setting for classical Airy structures. For those cones related to semi-simple Frobenius manifolds, one could use the action of the twisted loop group to bring such Lagrangians in the form of a standard Lagrangian cone which is known to describe an Airy structure (whose partition function is a product of the matrix Airy function of \cite{Konts}), and therefore the original cone does correspond to a classical Airy structure. This is however rather indirect.

\begin{problem}
Can one associate directly to Givental's Lagrangian cones a classical Airy structure (or some generalisation, \textit{e.g.} dropping the assumption that the differential operators are at most quadratic), without the semi-simplicity assumption?
\end{problem}

\begin{problem}
Given a quantum Airy structure on $V$, when does $S_0$ defines the structure of a (germ of a) Frobenius manifolds at $0$ on $V_0$? Or, more generally, if $V$ is infinite-dimensional and contains a distinguished finite-dimensional subspace $V_0$, when does the restriction of $S_0$ to $V_0$ defines a (germ of a) Frobenius manifolds at $0$ on $V_0$?
\end{problem}
We have checked that the $S_0$ of the quantum Airy structure of Proposition~\ref{FrobABCD} on $V =$ a Frobenius algebra, does define the prepotential of a germ of a Frobenius manifold at $0$ in $V$. We also know that this is true for the quantum Airy structure corresponding (see Section~\ref{comparTR}) to TR for compact spectral curves. In this case $S_0$ is the prepotential of the Hurwitz space equipped with its usual Frobenius structure \cite{Dubrovin,TRFROBN}. It would be interesting to know whether this is still true for the loop space examples of Section~\ref{S444} especially when $\mathbb{A}$ is non semi-simple.

The setting of quantum Airy structure is very much restricted to the case of a symplectic space isomorphic to $T^*V$. Yet, some works indicate that TR should be related to quantisation of moduli spaces, which are curved K\"ahler manifolds.
\begin{problem}
Can one construct interesting families of quantum Airy structure from (curved) symplectic manifolds, or from K\"ahler manifolds?
\end{problem}

\newpage

\appendix

\section{Asymptotics of special functions}

\subsection{Cardinality of \texorpdfstring{$\mathfrak{T}_{g}$}{Tg}}

In Proposition~\ref{ABequal0}, the partition function for quantum Airy structures with $A = B = 0$ was computed as a sum over the set $\mathfrak{T}_{g}$ of rooted trivalent trees with $g$ leaves. Let
$$
N_g := \sum_{T \in \mathfrak{T}_{g}} \frac{1}{|{\rm Aut}\,T|}\,.
$$
Applying Proposition~\ref{ABequal0} results in the one-dimensional quantum Airy structure
$$
L = \hbar\partial_{x} - \tfrac{\hbar^2}{2}\partial_{x}^2 - \hbar
$$
yields $Z = \exp\big(\sum_{g \geq 1} \hbar^{g - 1}N_{g} x\big)$. The differential equation $L\cdot Z = 0$ is easy to solve
$$
Z = \exp\big(\tfrac{x}{\hbar}(1 - \sqrt{1 - 2\hbar})\big)\,.
$$
We deduce
$$
N_{g} = \frac{(2g - 3)!!}{g!} = \frac{{2g \choose g}}{2^g (2g - 1)}\,.
$$
with convention $(-1)!! = 1$.

\subsection{The Bairy function}
\label{Bairyapp}
The full asymptotic expansion for the Bairy function is known in closed form
$$
{\rm Bi}(z) = \frac{e^{\frac{2}{3}z^{\frac{3}{2}}}}{\pi^{\frac{1}{2}}z^{\frac{1}{4}}}\bigg(\sum_{m \geq 0} \alpha_m\,z^{-\frac{3m}{2}}\bigg),\qquad \alpha_m = \frac{(3/4)^m\Gamma(m + \tfrac{1}{6})\Gamma(m + \tfrac{5}{6})}{2\pi m!}\,.
$$
The first few values are $\alpha_0 = 1$, $\alpha_1 = \tfrac{5}{48}$, $\alpha_2 = \tfrac{385}{4608}$, etc. There is no simple closed formula for the asymptotic expansion of $\ln\,{\rm Bi}(z)$, which are more directly related to the value of $F_{g,n}$. We only know it is of the form
$$
\ln\,{\rm Bi}(z) = \tfrac{2}{3}z^{\frac{3}{2}} - \tfrac{\ln z}{4} - \tfrac{\ln \pi}{2} + \sum_{g \geq 1} \frac{(3/4)^{g}\beta_{g}}{g}\,z^{-\frac{3g}{2}}
$$
where $(\beta_g)_{g \geq 1}$ is the integer sequence defined by the initial data $\beta_1 = \frac{5}{36}$ and the recurrence relation
$$
\forall g \geq 1,\qquad \beta_{g + 1} = (g + 1)\beta_{g} + \sum_{h = 1}^{g - 1} \beta_h\beta_{g - h}
$$ 
which can easily be obtained from the Airy differential equation. The first values are
\begin{center}
\begin{figure}[h!]
\centering
\begin{tabular}{|c||c|c|c|c|c|c|c|c|c|}
\hline
$g$ & $1$ & $2$ & $3$ & $4$ & $5$ & $6$ & $7$ & $8$ & $9$ \\ \hline
$\beta_{g}$ & $\tfrac{5}{36}$ & $\tfrac{5}{18}$ & $\tfrac{1105}{1296}$ & $\tfrac{565}{162}$ & $\tfrac{414125}{23328}$ & $\tfrac{78700}{729}$ & $\tfrac{1282031525}{1679616}$ & $\tfrac{80727925}{13122}$ & $\tfrac{1683480621875}{30233088}$ \\
\hline
\end{tabular}
\end{figure}
\end{center}

\subsection{The Whittaker function}
\label{Whitapp}

The partition function for the general Airy structure in dimension $1$ (Equation~\ref{Whity}) has as Taylor coefficients
\bea
F_{0,3} & = & \theta_{A} \nonumber \\
F_{0,4} & = & 3\theta_{A}\theta_{B}  \nonumber \\
F_{0,5} & = & 3\theta_{A}(\theta_{A}\theta_{C} + 4\theta_{B}^2) \nonumber \\
F_{0,6} & = & 15\theta_{A}\theta_{B}(3\theta_{A}\theta_{C} + 4\theta_{B}^2) \nonumber \\
F_{0,7} & = & 45\theta_{A}(\theta_{A}^2\theta_{C}^2 + 12\theta_{A}\theta_{B}^2\theta_{C} + 8\theta_{B}^4) \nonumber \\
F_{1,1} & = & D \nonumber \\
F_{1,2} & = & \tfrac{1}{2}\theta_{A}\theta_{C} + \theta_{B}D \nonumber \\
F_{1,3} & = & \tfrac{5}{2}\theta_{A}\theta_{B}\theta_{C} + \theta_{A}\theta_{C}D + 2\theta_{B}^2D \nonumber \\
F_{1,4} & = & 3\theta_{A}^2\theta_{C}^2 + \tfrac{27}{2}\theta_{A}\theta_{B}^2\theta_{C} + 9\theta_{A}\theta_{B}\theta_{C}D + 6\theta_{B}^3D \nonumber \\
F_{1,5} & = & \tfrac{111}{2}\theta_{A}^2\theta_{B}\theta_{C}^2 + 9\theta_{A}^2\theta_{C}^2D + 84\theta_{A}\theta_{B}^3\theta_{C} + 24\theta_{B}^4D \nonumber \\
F_{2,1} & = & \theta_{C}(\tfrac{1}{4}\theta_{A}\theta_{C} + \tfrac{1}{2}\theta_{B}D + \tfrac{1}{2}D^2) \nonumber \\
F_{2,2} & = & \theta_C(\tfrac{3}{2}\theta_{A}\theta_{B}\theta_{C} + \theta_{A}\theta_{C}D + \tfrac{3}{2}\theta_{B}^2D + \tfrac{3}{2}\theta_{B}D^2) \nonumber \\
F_{2,3} & = & \theta_C(2\theta_{A}^2\theta_{C}^2 +\tfrac{39}{4}\theta_{A}\theta_{B}^2\theta_{C} + \tfrac{21}{2}\theta_{A}\theta_{B}\theta_{C}D + \tfrac{3}{2}\theta_{A}\theta_{C}D^2 + 6 \theta_{B}^3D + 6\theta_{B}^2D^2) \nonumber \\
F_{3,1} & = & (\theta_{B} + D)\theta_{C}^2(\tfrac{3}{4}\theta_{A}\theta_{C} + \tfrac{3}{4}\theta_{B}D + \tfrac{1}{2}D^2) \nonumber \\
F_{3,2} & = & \theta_{C}^2(\tfrac{9}{8}\theta_{A}^2\theta_{C}^2 + \tfrac{45}{8}\theta_{A}\theta_{B}^2\theta_C + 8\theta_{A}\theta_{B}\theta_{C}D + 2\theta_{A}\theta_{C}D^2 + \tfrac{15}{4}\theta_{B}^3D + \tfrac{25}{4}\theta_{B}^2D^2 + \tfrac{5}{2}\theta_{B}D^3)  \nonumber 
\eea
They are enumerating the trivalent graphs appearing when unfolding the topological recursion formula \eqref{TRForm}. For low values of $(g,n)$ this can be directly checked in Figure~\ref{TRGraph}.

\section{Vanishing results for quantum Airy structures on loop spaces}
\label{App4}
Let $\mathbb{A}$ be a Frobenius manifold. We consider the quantum Airy structure on $\mathbb{A}[\![z]\!] = \mathbb{A} \otimes \mathbb{C}[\![z]\!]$ given by Proposition~\ref{Loop1}. The label $\alpha$ will index a basis of $\mathbb{A}$, and $i \geq 0$ the natural basis of $\mathbb{C}[\![z]\!]$.
\begin{lemma}
\label{Lemm1}Assume $\theta(z) = t_{r_0}z^{r_0} + O(z^{r_0 + 1})$ with $t_{r_0} \neq 0$ for some $r_0 \geq -1$.
\begin{itemize}
\item[$\bullet$] If $r_0 = -1$, the partition function is $Z = 1$, \textit{i.e.} all $F_{g,n}$ vanish.
\item[$\bullet$] If $r_0 \geq 0$, for any $g,n$ such that $2g - 2 + n > 0$, we have $F_{g,n}\big((i_1,\alpha_1),\ldots,(i_n,\alpha_n)\big) = 0$ whenever $\sum_{m = 1}^n i_m > 2g - 2 + r_0(2 - n)$.
\end{itemize}
\end{lemma}
 
Now we consider the $\mathbb{Z}_{2}$-symmetric version of a quantum Airy structure on $\mathbb{A}[\![z^2]\!]$ given by Proposition~\ref{Loop2}. Here the label $i \geq 0$ indexes the natural basis of $\mathbb{C}[\![z^2]\!]$.

\begin{lemma}
\label{Lemm2}Assume $\theta(z) = t_{s_0}z^{2s_0} + O(z^{2(s_0 + 1)})$ with $t_{s_0} \neq 0$ for some $s_0 \geq -1$.
\begin{itemize}
\item[$\bullet$] If $s_0 = -1$, for any $g,n$ such that $2g - 2 + n > 0$, we have $F_{g,n}\big((i_1,\alpha_1),\ldots,(i_n,\alpha_n)\big) = 0$ whenever $\sum_{m = 1}^n i_m > 3g - 3 + n$.
\item[$\bullet$] If $s_0 \geq 0$, for any $g,n$ such that $2g - 2 + n > 0$, we have $F_{g,n}\big((i_1,\alpha_1),\ldots,(i_n,\alpha_n)\big) = 0$ whenever $\sum_{m = 1}^n i_m > g - 1 + s_0(2 - n)$.
\end{itemize}
\end{lemma}

\noindent \textbf{Proof of Proposition~\ref{Lemm1}.} We give the proof in detail for $\mathbb{A} = \mathbb{C}$, as the argument and the result are similar for general Frobenius algebras. We also remark that the conjugation by $\exp\big(\tfrac{\hbar}{2}u_{a,b}\partial_{a}\partial_{b}\big)$ preserves the vanishing property, so it is sufficient to prove it when
$$
\xi_{k} = \frac{k + 1}{z^{k + 2}}\,\dd z\,.
$$
The corresponding quantum Airy structure has $A = 0$ and
\bea
B^i_{j,k} & = & \sum_{r \geq r_0} \frac{k + 1}{i + 1} t_{r} \delta_{i + j + r,k}\,, \\
C^i_{j,k} & = & \sum_{r \geq r_0} \frac{(k + 1)(j + 1)}{i + 1} t_{r} \delta_{i + r,j + k + 2}\,.
\eea
So, $B^i_{j,k} = 0$ unless $i + j + r_0 \leq k$ and $C^i_{j,k} = 0$ unless $i + r_0 \leq j + k + 2$. We also know from Section~\ref{S4Loop} that $D^k = 0$ unless $k \leq r_0$.

The fact that $A = 0$ already guarantees that $F_{0,n} = 0$ for all $n$. If $r_0 = -1$, we furthermore have $D = 0$. Since $A$ and $D$ are initial data for the recursion formula \eqref{TRForm}, it therefore can only produce $F_{g,n} = 0$ for all $(g,n)$.

We now assume $r_0 \geq 0$, and examine the conditions under which $F_{g,n}$ for $2g - 2 + n > 0$ are possibly non-zero. We already know that $F_{0,3}(i,j,k) = A^i_{j,k} = 0$, and that $F_{1,1}(i) = D^i$ vanishes unless $i \leq r_0$. We have 
$$
F_{1,2}(i,j) = B^i_{j,a}D^a\,,
$$
and it will vanish unless the two conditions $i + j + r_0 \leq a$ and $a \leq r_0$ are satisfied. This imposes $i + j \leq 0$, hence $i = j = 0$. We also have
$$
F_{2,1}(i) = \tfrac{1}{2}C^i_{a,b}\big(F_{1,2}(a,b) + D^aD^b\big)\,.
$$
The first term vanishes unless the two conditions $i + r_0 \leq a + b + 2$ and $a = b = 0$ are satisfied, hence unless $i \leq 2 - r_0$. The second term vanishes unless the three conditions $i + r_0 \leq a + b + 2$, $a \leq r_0$ and $b \leq r_0$ are satisfied, hence unless $i \leq r_0 + 2$. As $r_0 \geq 0$, we deduce that $F_{2,1}(i)$ vanishes unless $i \leq r_0 + 2$. These three cases $(g,n) \in \{(1,1),(1,2),(2,1)\}$ are sharply compatible with a linear bound
$$
\sum_{m = 1}^n i_m \leq d_{g,n} = \alpha n + \beta g + \gamma,\qquad {\rm  fixing} \quad (\alpha,\beta,\gamma) = (-r_0,2,2r_0 - 2).
$$
Note that this bound also holds for $g = 0$ and all $n \geq 3$, as then $d_{g,n} < 0$. Now that we have guessed a candidate for $d_{g,n}$, let us prove by induction on $2g - 2 + n > 0$ that indeed
$$
\sum_{m = 1}^n i_m > d_{g,n} \quad \Longrightarrow \quad F_{g,n}(i_1,\ldots,i_n) = 0\,.
$$

We already know it is true for $2g - 2 + n \leq 2$. Assume it is true for all $(g',n')$ such that $0 < 2g' - 2 + n' < \chi_0$, and pick $(g,n)$ such that $\chi_0 = 2g - 2 + n$. The recursion gives
\bea
\label{RHG}F_{g,n}(i_1,\ldots,i_n) & = & \sum_{m = 2}^n B^{i_1}_{i_m,a}F_{g,n - 1}(a,i_2,\ldots,\widehat{i_m},\ldots,i_n) \\
&& + \tfrac{1}{2}C^i_{a,b}\bigg(F_{g - 1,n + 1}(a,b,i_2,\ldots,i_m) + \sum_{\substack{g' + g'' = g \\ n' + n'' = n + 1}} F_{g'',n'}(a,\cdots)F_{g'',n''}(b,\cdots)\bigg)\,. \nonumber
\eea
We have not written the details of the sum, but the only thing to know is that in each term, all indices $(i_m)_{m = 2}^n$ appear once. Denote $[I] = \sum_{m = 1}^n i_m$. Using the induction hypothesis --- which is valid for all terms in the right-hand side of \eqref{RHG} --- and the vanishing rules for $B$ and $C$, we find that $F_{g,n}(i_1,\ldots,i_n)$ vanish unless one of the following three conditions is satisfied
\begin{itemize}
\item[$\bullet$] $[I] \leq d_{g,n - 1} - r_0 = d_{g,n}$,
\item[$\bullet$] $[I] \leq d_{g - 1,n + 1} + 2 - r_0 = d_{g,n}$,
\item[$\bullet$] $[I] \leq d_{g',n'} + d_{g'',n''} + 2 - r_0 = d_{g,n}$.
\end{itemize}
This proves the claim by induction. We also see that, as the bound was sharp for the three basic cases, it remains sharp throughout the induction.

\noindent \textbf{Proof of Proposition~\ref{Lemm2}.} As before we can restrict to the case $\mathbb{A} = \mathbb{C}$ and 
$$
\xi_{k} = \frac{(2k + 1)}{z^{2k + 2}}\,\dd z\,,
$$
which correspond to
\bea
A^i_{j,k} & = & \sum_{s \geq s_0} \frac{1}{2i + 1}\delta_{i + j + k + s + 1,0}t_{s} \,,\nonumber \\
B^i_{j,k} & = & \sum_{s \geq s_0} \frac{2k + 1}{2i + 1} \delta_{i + j + s,k} t_{s} \,,\nonumber \\
C^i_{j,k} & = & \sum_{s \geq s_0} \frac{(2k + 1)(2j + 1)}{2i + 1}\delta_{i + s,j + k + 1} t_{s} \,.\nonumber
\eea
So, $B^i_{j,k} = 0$ unless $i + j + s_0 \leq k$, and $C^i_{j,k} = 0$ unless $i + s_0 \leq j + k + 1$.

For $s_0 = -1$, we have $F_{0,3}(i,j,k) = A^i_{j,k} = 0$ unless $i = j = k = 0$, and we have $D^i = 0$ unless $i \geq 2$. We compute
\bea
F_{1,2}(i,j) = B^i_{j,a}D^a + \tfrac{1}{2}C^i_{a,b}A^j_{a,b}\,.
\eea
The first term vanishes unless $i + j - 1 \leq a$ and $a \leq 1$, that is unless $i + j \leq 2$. The second term vanishes unless $i - 1 \leq a + b + 1$ and $j = a = b = 0$, that is unless $i + j \leq 2$. Hence, $F_{1,2}(i,j)$ vanishes unless $i + j \leq 2$. The three cases $(g,n) \in \{(0,3),(1,1),(1,2)\}$ are sharply compatible with a linear bound
\beq
\label{indoc}\sum_{m = 1}^n i_m \leq d_{g,n} = \alpha n + \beta g + \gamma,\qquad {\rm fixing}\quad (\alpha,\beta,\gamma) = (1,3,-3).
\eeq
An induction similar to the proof of Lemma~\ref{Lemm1} proves \eqref{indoc} for any $(g,n)$ such that $2g - 2 + n > 0$.

Now consider the case $s_0 \geq 0$. We then have $A = 0$, hence $F_{0,n} = 0$ for all $n$. From Section~\ref{S4Loop} we already know that $F_{1,1}(i) = D^i$ vanishes unless $i \leq s_0$. We compute
$$
F_{1,2}(i,j) = B^i_{j,a}D^a\,,
$$
and it vanishes unless $i + j + s_0 \leq a$ and $a \leq s_0$, hence unless $i + j \leq 0$. We also compute
$$
F_{2,1}(i) = \tfrac{1}{2}C^i_{a,b}\big(F_{1,2}(a,b) + D^aD^b\big)\,.
$$
The first term vanishes unless the two conditions $i + s_0 \leq a + b + 1$ and $a + b \leq 0$ are satisfied, hence unless $i \leq 1 - s_0$. The second term vanishes unless the three conditions $i + s_0 \leq a + b + 1$ and $a \leq s_0$ and $b \leq s_0$ are satisfied, hence unless $i \leq s_0 + 1$. As $s_0 \geq 0$, we deduce that $F_{2,1}(i)$ vanishes unless $i \leq s_0 + 1$. The three cases $(g,n) \in \{(1,1),(1,2),(2,1)\}$ are compatible with the linear bound
\beq
\label{indoc2}\sum_{m = 1}^n i_m \leq d_{g,n} = \alpha n + \beta g + \gamma,\qquad {\rm fixing}\quad (\alpha,\beta,\gamma) = (-s_0,1,2s_0 - 1)\,.
\eeq
Again, an induction similar to the proof of Lemma~\ref{Lemm1} proves \eqref{indoc2} for any $(g,n)$ such that $2g - 2 + n > 0$. \hfill $\Box$

\providecommand{\bysame}{\leavevmode\hbox to3em{\hrulefill}\thinspace}
\providecommand{\MR}{\relax\ifhmode\unskip\space\fi MR }
\providecommand{\MRhref}[2]{%
  \href{http://www.ams.org/mathscinet-getitem?mr=#1}{#2}
}
\providecommand{\href}[2]{#2}

%
%
%
%
%

\end{document}